\documentclass[onecolumn]{emulateapj}
\usepackage{multirow}
\usepackage{graphics}
\usepackage{amsmath, amsthm, amssymb}
\usepackage{epsfig}

\begin{document}

\title{On the stability criteria for equatorial circular orbits\\ in Galactic Dynamics I:
Newtonian Thin Disks}

\author{Ronaldo S. S. Vieira}
\affil{Instituto de F\'{i}sica ``Gleb Wataghin'', Universidade Estadual de
Campinas, 13083-777, Campinas, SP, Brazil}
\email{ronssv@ifi.unicamp.br}

\author{Javier Ramos-Caro}
\affil{Departamento de Física, Universidade
Federal de S\~{a}o Carlos, 13565-905, SP, Brazil}
\email{javier@ufscar.br}

\begin{abstract}
We make a revision of the stability criteria for equatorial circular orbits, obtained
from the epicyclic approximation, which is widely used in Newtonian models for axisymmetric galaxies. We find that,
for the case of thin disk models, the criterion for vertical stability must be reformulated, due to the
discontinuity in the gravitational field. We show that, for a model characterized by a surface mass density $\Sigma$,
 the necessary and sufficient condition to have
vertically stable circular orbits is that $\Sigma>0$. On the other hand, the criterion for radial stability is the same as in thick
diks, i.e. that the (radial) epicyclic frequency squared is positive. As an application, we present finite thin disk models for nine
galaxies, as superpositions of members of the Morgan \& Morgan family (in Newtonian version), which can be considered
as stable configurations in a first
approximation. Also, as an additional  product of this study, we show that any galactic model with a thin disk component admits a
wide variety of integrable disk-crossing orbits, which are determined approximately by a third integral of motion of the form
$Z\Sigma^{1/3}$, where $Z$ is the z-amplitude of the motion.
\end{abstract}


\section{Introduction}\label{sec:intro}

It is usually accepted that many galaxies in the universe are nearly axisymmetric, with a
mass distribution formed by several components: a thin disk, a central bulge and a
surrounding halo. In consequence, there is a number of mass models incorporating one or all of these
features, depending on the particular case (\cite{freeman, kent1986, kent1987, sofue2008}).
For example, there exists a number of galaxies in the Ursa Major cluster  which can be modeled, at the scale of the optical radius,
with only the thin disk component (\cite{ngc}), suggesting that they obey the
so-called \emph{maximum disk hypothesis} (\cite{GD}). If we decide to continue using Newtonian theory beyond
the optical radius, presumably we have to introduce a dark halo, but the disk component still provides a significant
contribution, taking into account that the main part of
the stellar population is located there. For this reason, thin disk  models
have been an issue of interest in galactic dynamics (see for example \cite{hunter2, morgan, gonzalez-reina, pedraza}
as well as \cite{GD} and references therein).

Thin disks have also been used to model self-gravitating rings (\cite{letelier1994, letelier2007}), with applications
in other branches of astrophysics. These models can be used to describe accretion disks, stellar structures with
central black holes  (see  \cite{sesu,LOP}) or planetary rings. For example, the last issue was partially encompassed in \cite{pn},
where a study of the linear stability of the monopole-ring system was performed by using superpositions of Morgan \& Morgan disks.
Similar studies were conducted by the authors in the Newtonian realm of galactic dynamics (\cite{javier, pedraza}).

A fundamental step in the formulation of galactic models is the stability analysis.
In fact, the stability analysis could suggest sometimes the introduction of new features
in a given model in order to, presumably, obtain a more realistic representation. Consider for example the study conducted
by \cite{ostriker} where it is shown that a flattened system of self-gravitating particles, initially supported
against gravity by rotation, does not maintain its discoidal form in the course of time. They suggested that
the introduction of a spherical halo with mass of the order of the disk mass (or greater)
 improves substantially the stability of the disk, as it was corroborated by simulations.
But, on the other hand, consider also the counter argument made by \cite{kalnajs2},
in which it is suggested  (i) that the stability problems can be overcome
by improving the features of the the inner parts of galactic models (for example, by considering hot centers or small bulges)
and (ii) that a halo with a scale length larger than the disk and more massive than it, does not contribute significantly
to the stability. This discussion has its roots in the fact that internal kinematics of self-gravitating disks determines
the stability of the system: cool disks, which are mainly formed by stars in circular motion (the ones considered by
\cite{ostriker}), require a prominent halo in order to avoid instabilities,
contrary to the case of hot disks (the ones considered by \cite{kalnajs2}) which can be supported by the random motions of the stars
(for a recent review of  disk instabilities see \cite{sellwood2011}).
A detailed knowledge of the orbital features associated with a given galactic model provides the basis for the stability analysis.
Usually this knowledge can be achieved once we have at hand the distribution function (DF) of the model.

In general, the obtention of the DF associated to a particular model is not an easy task (except for those models which are defined
by an analytical DF) and, in consequence, the corresponding stability analysis is far from trivial. However it is possible to perform a first test
of stability without knowing the DF, by using the so-called \textit{epicyclic approximation} (\cite{GD}), i.e.
by performing a linear stability analysis of fixed points of the effective potential. Once it is verified that the model is linearly stable,
it deserves to perform more conclusive stability tests based on  statistical mechanics.

The linear stability analysis can be thought  as a first test to evaluate how realistic
any particular model is. In disk galaxies many stars are on nearly circular orbits,
so we have to demand, as a basic requirement, that any galactic model must be characterized by
allowing the existence of stable circular motion, especially in regions where the stellar population is maximum,
i.e. the equatorial plane. The epicyclic approximation provides a formalism to study
motion in the equatorial plane and leads to the establishment of simple stability criteria. However, when one deals with
 models incorporating a razor-thin disk, this method needs to be reformulated in view of the mathematical features
introduced by the thin disk component.
Next, we will briefly illustrate this problem.

Consider an axisymmetric potential-density pair (APDP) with mass distribution $\rho(R,z)$ and gravitational potential $\Phi(R,z)$, where $R$ and $z$ are
the usual cylindrical coordinates.
The motion of test particles can be described
by the Hamiltonian (\cite{GD})
\begin{equation}
   H = \frac{P_R^2 + P_z^2}{2}  + \Phi_{eff}(R, z),
\end{equation}
where $P_R=\dot{R}$, $P_z=\dot{z}$ and $\Phi_{eff}$ is the so-called effective potential, defined as
  \begin{equation}
   \Phi_{eff}(R, z) = \Phi(R, z) + \frac{l^2}{2R^2}.
  \end{equation}
Here, $l$ represents the $z$-component of the angular momentum,
often called azimuthal angular momentum, which is a first integral of motion.
The resulting equations of motion can be written as
  \begin{equation}\label{Req}
   \ddot{R} = -\frac{\partial \Phi_{eff}}{\partial R},
  \end{equation}
\begin{equation}\label{zeq}
   \ddot{z} = - \frac{\partial \Phi_{eff}}{\partial z}.
  \end{equation}
In particular, the equatorial circular orbits (i.e., belonging to the plane $z = 0$)
correspond to the minimum of $\Phi_{eff}$, obtained by setting
$\partial \Phi_{eff}/\partial R=0$ and $z=0$, which lead to the relation
  \begin{equation}
   \left.\frac{\partial \Phi}{\partial R}\right|_{z=0} - \frac{l^2}{R^3}=0.
  \end{equation}
The value of the $R$-coordinate that is solution of the above equation
is the radius of the circular orbit with angular momentum $\ell$
(in this case $R$ is called the \emph{guiding-center radius}, which we will denote as $R_{o}$).
If this orbit (or any other in the equatorial plane) suffers a small perturbation, one would expect that the resulting motion is not very
different from the original one (say, a nearly circular orbit), in order to guarantee the stability of the entire configuration.

Strictly speaking, a nearly circular orbit is defined as a trajectory with  coordinates $(R,z)$ very close to $(R_{o},0)$, so we
can expand $\Phi_{eff}$ near this minimum and neglect cubic and higher  terms in the expansion, that is,
\begin{equation}\label{expansion}
    \Phi_{eff}\approx\Phi_{eff}(R_{o},0)+\frac{\kappa^{2}}{2}(R-R_{o})^{2}+\frac{\nu^{2}}{2}z^{2},
\end{equation}
where $\kappa$ and $\nu$ are defined as
\begin{equation}\label{frecuencias}
    \kappa^{2}\equiv\left.\frac{\partial^{2}\Phi_{eff}}{\partial R^{2}}\right|_{(R_{o},0)}, \qquad
    \nu^{2}\equiv\left.\frac{\partial^{2}\Phi_{eff}}{\partial z^{2}}\right|_{(R_{o},0)}.
\end{equation}
By introducing (\ref{expansion}) in the equations of motion (\ref{Req})-(\ref{zeq}), it  can be seen that
$R-R_{o}$ and $z$ evolve like the displacements of two harmonic oscillators with frequencies $\kappa$ (epicyclic
frequency) and
$\nu$ (vertical frequency), respectively, once it is guaranteed that $\kappa^2> 0$ and $\nu^2> 0$. In this case the corresponding circular
orbit is said to be \textit{stable} (\cite{GD}). So we say that, in a first approximation, the conditions for stability of the self-gravitating structure
located at the equatorial plane (which it is assumed to be composed principally by particles describing nearly circular motions)
is that the epicyclic and vertical frequencies squared are both positive.

Note that, whenever one  deals with razor-thin disks, it is not possible to define a Taylor expansion of the effective potential around the point
$(R_o, 0)$. This procedure, which leads to the analysis of vertical and epicyclic frequencies around the circular orbit (\cite{GD}),
is only valid for potentials that are smooth or that have at least continuous second derivatives. This means that the analysis
of vertical frequencies of thin disk potentials, as did in \cite{pedraza, javier, ngc, pn}, is not a reliable indicator of the
vertical stability of the corresponding circular orbits. The analysis must take into account the discontinuity in the partial
derivative of the potential with respect to $z$ due to the surface distribution of matter in the equatorial plane.

In the following sections we will show that, when considering a razor-thin disk,
it can be constructed a vertical stability criterion in terms of the first derivative of
the potential, whereas the criterion of the epicyclic frequency remains unchanged (secs. \ref{sec:thin} and \ref{sec:vert-stab}).
Also we will present some consequences of
this approach in the description of disk-crossing orbits. In particular we will show that many of them can be described
by an approximate third integral of motion depending on the vertical amplitude and the surface mass density (sec. \ref{sec:integrability}).
Finally we briefly address these issues in the realm of modified theories of gravity (sec. \ref{sec:MOND}).

\section{Galactic Models via Thin Disks}\label{sec:thin}

As it was pointed out in the Introduction, many galaxies are modeled as
 an axisymmetric thin disk surrounded by an axisymmetric 3D smooth density distribution, which is symmetric with respect to the
equatorial plane. The total density distribution can be written as
  \begin{equation}\label{totaldensity}
   \rho(R,z) = \Sigma(R)\delta(z) + \rho_s(R, z),
  \end{equation}
where $\delta$ is the Dirac delta, $\Sigma(R)$ is the surface density distribution of the thin disk and
$\rho_s(R, z)$ describes the surrounding matter. The
gravitational potential of the system is
  \begin{equation}\label{totalpotential}
   \Phi(R, z) = \Phi_\Sigma(R,z) + \Phi_s(R, z),
  \end{equation}
where $\Phi_s$ and $\Phi_\Sigma$ are the contributions due to $\rho_s$ and $\Sigma$, respectively, in
such a way that
$$\nabla^2\Phi_s = 4\pi G \rho_s,
\qquad \nabla^2\Phi_\Sigma = 4\pi G \Sigma(R)\delta(z).$$
The symmetry of $\rho_s$ with respect to the equatorial plane implies a reflection symmetry of the gravitational potential
with respect to the equatorial plane, i.e. $\Phi(R, z) = \Phi(R, -z)$.
This reflection symmetry  implies that the $z$-dependence of $\Phi$ actually is solely on $|z|$, so we can write
  \begin{eqnarray}\label{Heaviside}
   \frac{\partial \Phi}{\partial z} = \frac{\partial \Phi}{\partial |z|}\frac{\partial |z|}{\partial z}
    =[2 \Theta(z)-1]\frac{\partial \Phi}{\partial |z|},
  \end{eqnarray}
where $\Theta$ is the Heaviside step-function. Thus, the potential $\Phi$ in these models is  a
smooth function of $|z|$ with a discontinuity in the $z$-derivative  at the plane $z=0$.

It is suggestive to note that there is a close connection between $\partial \Phi/\partial |z|$ and
$\Sigma$. To see this, first note that a direct consequence of (\ref{Heaviside}) is
  \begin{equation}
   \frac{\partial \Phi}{\partial z}\bigg|_{0^+} - \frac{\partial \Phi}{\partial z}\bigg|_{0^-} =
  2 \frac{\partial \Phi}{\partial |z|}\bigg|_{z = 0}.\label{limits}
  \end{equation}
Now, from Gauss's theorem we have
  $$
   4 \pi G \Sigma(R) = \frac{\partial \Phi}{\partial z}\bigg|_{0^+} - \frac{\partial \Phi}{\partial z}\bigg|_{0^-},
  $$
and, according to (\ref{limits}), it follows that
   \begin{eqnarray}
   \Sigma(R) &=& \frac{1}{2\pi G}\frac{\partial \Phi}{\partial |z|}\bigg|_{z = 0},\label{sigma}
  \end{eqnarray}
which is equivalent to the expression appearing in \cite{gonzalez-reina} and references therein.
This fact will be relevant in the construction of a stability criterion for circular orbits
under the action of vertical perturbations.

\section{Vertical stability of circular orbits}\label{sec:vert-stab}

We pointed out in the Introduction that
the stability analysis of equatorial orbits must take into account the discontinuity in the partial
derivative of the potential with respect to $z$, due to the surface distribution of matter in the equatorial plane.
The aim of this section is to
analyze the behavior of the $z$-coordinate of a vertically perturbed orbit
in the vicinity of the discoidal distribution.

Consider an equatorial circular orbit of radius $R$ under the
action of a small vertical perturbation
\footnote{Here, the term ``small'' means \textit{small enough to neglect variations in the projection of the orbit on
the $z=0$ plane}.}. This perturbation can be seen as an instantaneous vertical increase, $v_{0z}$, in the
velocity of the particle, which does not affect the value of $l$.
In order to study the evolution of the perturbation in the course of time, we have to use the
$z$-equation of motion.
From eqs. (\ref{zeq}) and (\ref{Heaviside}), we have
  \begin{equation}
   \ddot{z} = - [2 \Theta(z)-1]\frac{\partial \Phi}{\partial |z|},
  \end{equation}
from which we can establish the following statements:
\begin{enumerate}
\item Suppose that the particle hits the disk from below. If $\frac{\partial \Phi}{\partial |z|}\bigg|_{z = 0}  > 0$
then $\ddot{z}\bigg|_{z\rightarrow 0^+} < \ddot{z}\bigg|_{z\rightarrow 0^-}$;
\newline therefore its vertical acceleration decreases and the particle tends to come back to the disk.
\item Suppose that the particle hits the disk from above. If $\frac{\partial \Phi}{\partial |z|}\bigg|_{z = 0} > 0$
then $\ddot{z}\bigg|_{z\rightarrow 0^-} >  \ddot{z}\bigg|_{z\rightarrow 0^+}$;
\newline therefore its vertical acceleration increases and the particle also tends to come back to the disk.
\item Consider one of the two above situations but with $$\frac{\partial \Phi}{\partial |z|}\bigg|_{z = 0} < 0.$$
In both cases, the change in acceleration due to the
discontinuity in the z-derivative of
the potential makes the particle to gain velocity and, intuitively, move out from the disk.
\end{enumerate}

From the above considerations, it is natural to establish that
 a \textbf{necessary} condition for vertical stability is given by the relation
  \begin{equation}\label{necessary}
   \frac{\partial \Phi}{\partial |z|}\bigg|_{z = 0} > 0.
  \end{equation}
But the vertical stability is guaranteed once we verify that
any particle that crosses the disk, with initial vertical velocity $v_{0z}$
(just after crossing), will come back to the disk.
We show in the appendix that, for $v_{0z}$ sufficiently small, the perturbed trajectory will
oscillate around the original one for large time. Then we can consider that
condition (\ref{necessary}) is also \textbf{sufficient}  to guarantee vertical stability.
If we start with
 a small $v_{0z}$, the particle will always cross the disk with a velocity whose vertical component
has modulus $|v_{0z}|$.
 This is a consequence of: (i) the conservation of mechanical energy;
(ii) the assumption that the $R$-coordinate does not change in the process and
 (iii) the fact that the discontinuity is only in acceleration. So, just after crossing the disk the particle will also
have a velocity with vertical component of magnitude $v_{0z}$, which means that the motion after crossing the disk will have the same
behavior as the motion before crossing it. Therefore, the perturbed orbit remains oscillatory around
the original one for sufficiently small initial vertical velocity $v_{0z}$.

According to (\ref{sigma}), condition (\ref{necessary}) can be written in terms of the
surface mass density.
Then we can state that \textit{in a Newtonian thin disk model, a necessary and sufficient condition for a circular orbit of radius R to
be stable under small vertical perturbations is}
  \begin{equation}\label{condition}
   \Sigma(R) > 0.
  \end{equation}
This condition has indeed the same status as the vertical stability condition for smooth potentials (that is, replacing $\Sigma$  by  $\nu^2$ in the
above statement, as was employed in \cite{ngc,pn}), in the sense that both conditions imply vertical stability of the circular orbit if we neglect
variations in the $R$-coordinate of the perturbed orbit (see appendix).

In particular, if there are regions where $\Sigma(R) = 0$, i.e. without a thin disk component, then in these
regions we must apply the criterion
of vertical frequencies to analyze the vertical stability of the corresponding circular orbits.

For the case in which the perturbation introduces small variations
in the $R$-coordinate, we must know something about the radial dependence of the effective potential to ensure
stability of the perturbed orbit. It turns out that a sufficient condition for the $R$-variations in the perturbed orbit to be
negligible is (see appendix)
  \begin{equation}\label{conditionkappa}
   \kappa^2(R) > 0.
  \end{equation}
We also point out that the simultaneous conditions (\ref{condition})-(\ref{conditionkappa}) imply (Liapunov) stability of the circular orbit under small
perturbations in an arbitrary direction of the meridional plane (by neglecting variations in $l$ due to these perturbations). We will address this issue
in detail in Sec. \ref{sec:integrability}.

By way of verification of condition (\ref{condition}), in the following subsection  we show that, for a
sufficiently small vertical perturbation,
it is possible to compute characteristic periods and amplitudes of the oscillations around the discoidal plane.

\subsection{Characteristic period and amplitude of the oscillations}

Without loss of generality, consider a circular orbit of radius $R$ which suffers a vertical perturbation
in the direction $z>0$ at time $t=0$ and neglect its radial variations. In this moment, the test particle has an initial vertical velocity
$v_{0z} > 0$ and starts to rise, but it is attracted to the equatorial plane by gravity. The equation of motion for the vertical
coordinate, restricted to the region $z>0$, is
  $$
   \dot{z}(t) - v_{0z} =  - \int_0^t \frac{\partial \Phi}{\partial z}(R, z(t'))dt'.
  $$
Suppose that after a time interval $\Delta t$ the particle
returns to the disk. Then we have
  $$
   (-v_{0z}) - v_{0z} =  - \int_0^{\Delta t} \frac{\partial \Phi}{\partial z}(R, z(t'))dt'.
  $$
For sufficiently small $v_{0z}$ we can approximate the integrand by a constant, obtaining
  $$
   2v_{0z} \approx  (\Delta t) \frac{\partial \Phi}{\partial z}(R, 0^{+}).
  $$
We can perform an analogous procedure for the region $z < 0$ and the result is
the above equation but with a minus sign. Then, we can extend the above relation
for any real value of $z$:
  \begin{equation}
   2v_{0z} \approx  (\Delta t) \frac{\partial \Phi}{\partial |z|}(R, 0).
  \end{equation}
Since the characteristic period of one oscillation is $T\equiv 2\Delta t $, eq. (\ref{sigma})
gives us
\begin{equation}\label{zperiod}
  T = \frac{2 v_{0z}}{\pi G \Sigma(R)}.
  \end{equation}

The amplitude of the oscillation is given by the maximum of the parabola described by the equation
  $$
   \dot{z}(t) \approx  v_{0z} - t \frac{\partial \Phi}{\partial |z|}\bigg|_{z = 0},
  $$
and, since $z(0)=0$, we have
  $$
  z(t) \approx v_{0z} t - \frac{t^2}{2}\frac{\partial \Phi}{\partial |z|}\bigg|_{z = 0}.
 $$
The maximum of $z$, which we will denote as $z_{max}$, occurs at $t = T/4$, and is given by
  \begin{equation}\label{zamplitude}
   z_{max} = \frac{v_{0z}^2}{4 \pi G \Sigma(R)}.
  \end{equation}
As expected, the amplitude of oscillations is inversely proportional to the surface mass density (but not to the volumetric mass density,
od the surrounding matter, if present), a natural consequence of the attractive gravitational force.
These estimates give the order of magnitude of the parameters. More rigorous and accurate estimates are discussed in the appendix.

\subsection{Thin disks as a limiting case of a high density thick disk}

The stability condition obtained makes sense physically. For a thick disk, the vertical stability of a circular orbit is given by
the sign of the square of the vertical frequency,
  \begin{equation}\label{vertical}
   \nu^2 = \nabla^2\Phi\bigg|_{z = 0} - \frac{1}{R}\frac{d v_c^2}{dR}.
  \end{equation}
In general, the density distribution is assumed to be very high around the plane $z = 0$ and to fall off quickly with $z$.
If we consider that the rotation curves are bounded, the last term in the above equation is also bounded
(remember that $v_c^2 = R \partial\Phi/ \partial R$), and the sharpening of the density distribution makes the first term on the
right-hand side of (\ref{vertical}) increase. Eventually, it will increase an amount such that $\nu^2$ is positive (this may
depend on $R$).
In the limiting case when the (positive) density distribution becomes very close to $\delta(z)$, $\nu^2$ will be positive everywhere on
the disk.

\subsection{Finite Thin Disk Models for Galaxies}\label{sec:FTDModels}

Some authors have demonstrated that it is possible to obtain galactic models with  the thin
disk component only (without invoking a halo), modeling with high precision the rotation curves
in the optical range and in accordance with reasonable mass density profiles (\cite{kalnajs,kent1986,ngc}).
Galaxies that can be described by these models, which usually have high surface brightness (HSB),
are said to obey the so-called maximum disk hypothesis (\cite{GD}). This property was confirmed by \cite{persic1996},
where they found by means of model-independent methods that the amount of dark matter in HSB galaxies is negligible
inside the optical radius.

In particular, \cite{ngc} show some examples
in the Ursa Major cluster, by using the Hunter method \footnote{This is the
same formalism to obtain the Generalized Kalnajs Disks (\cite{gonzalez-reina}), which are the Newtonian version
of the general relativistic Morgan \& Morgan solutions (\cite{morgan}).}. As it was pointed out, one of the conclusions stated by Gonz\'{a}lez et al.
is that the  models obtained are vertically unstable, due to the fact that they considered the stability criterion of the
quadratic vertical frequency. But this is not true, in light of the stability criterion constructed here and taking into account that
all of the mass density profiles obtained in the aforementioned reference are positive. In fact, we remark that
the formalism showed by Gonz\'{a}lez et al. is a powerful method to obtain stable maximum diks describing the optical region of
many HSB galaxies.

\begin{figure*}
$$
\begin{array}{ccc}
  \epsfig{width=5.4cm,file=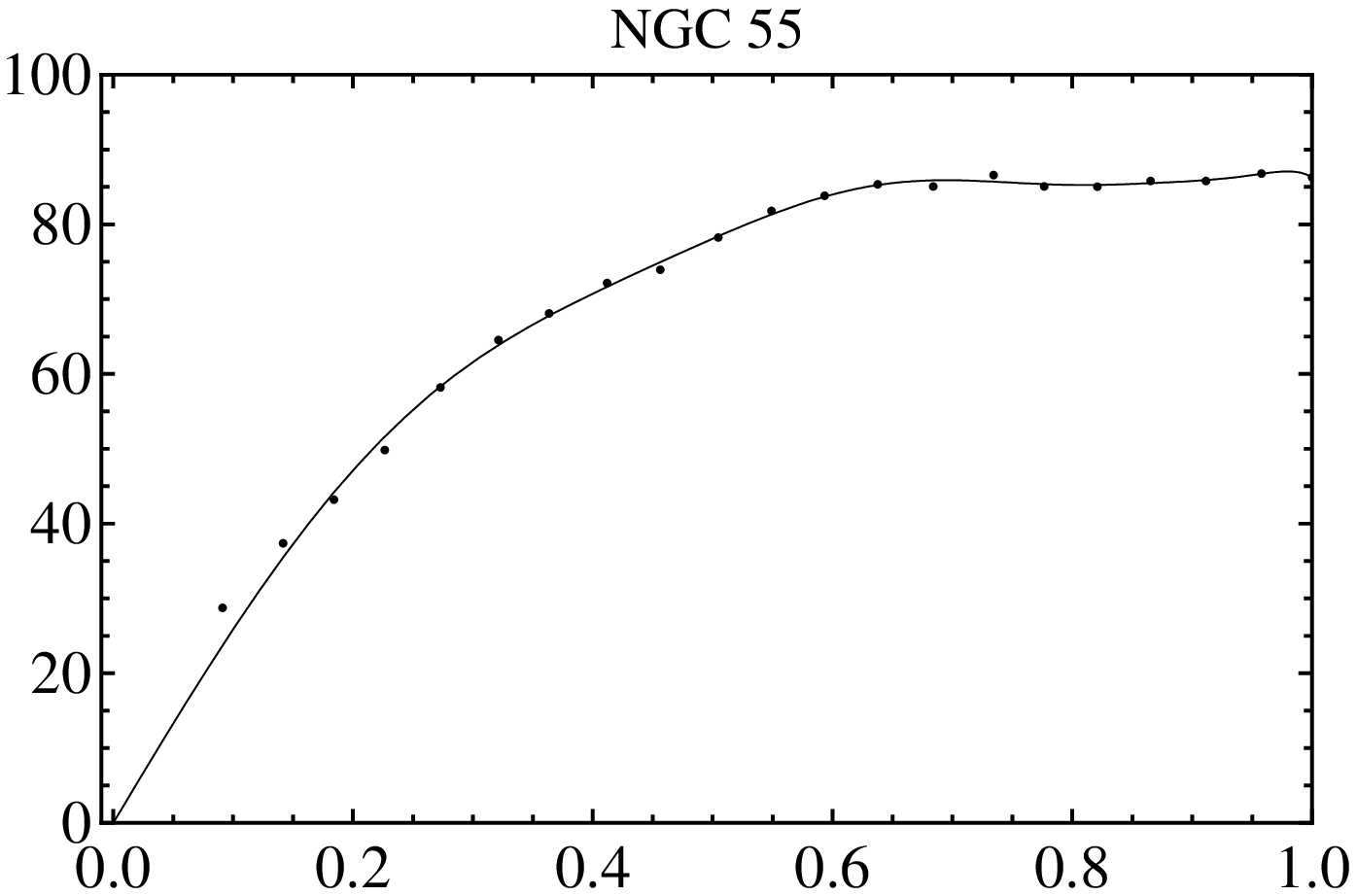} & \epsfig{width=5.4cm,file=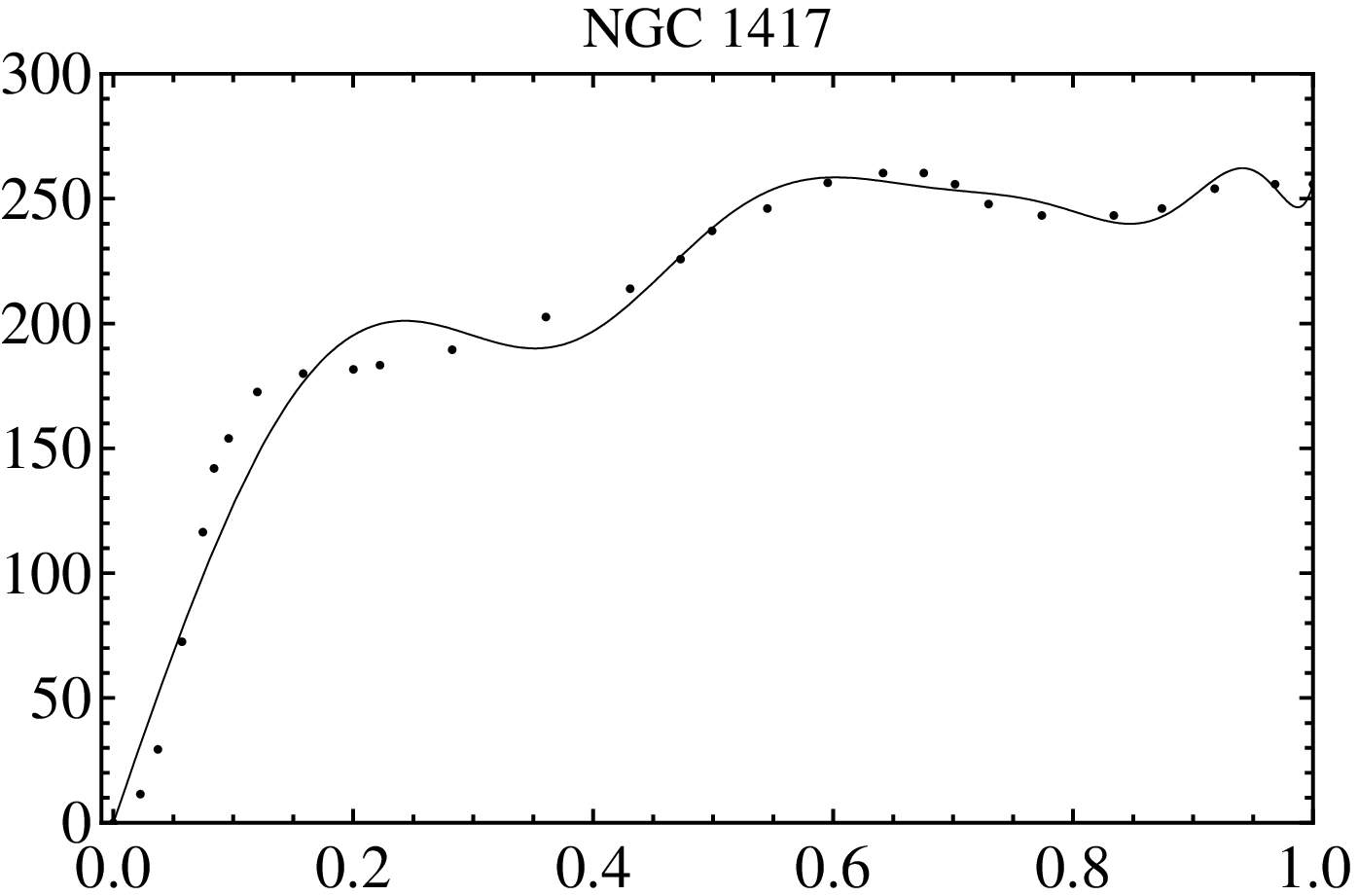} & \epsfig{width=5.4cm,file=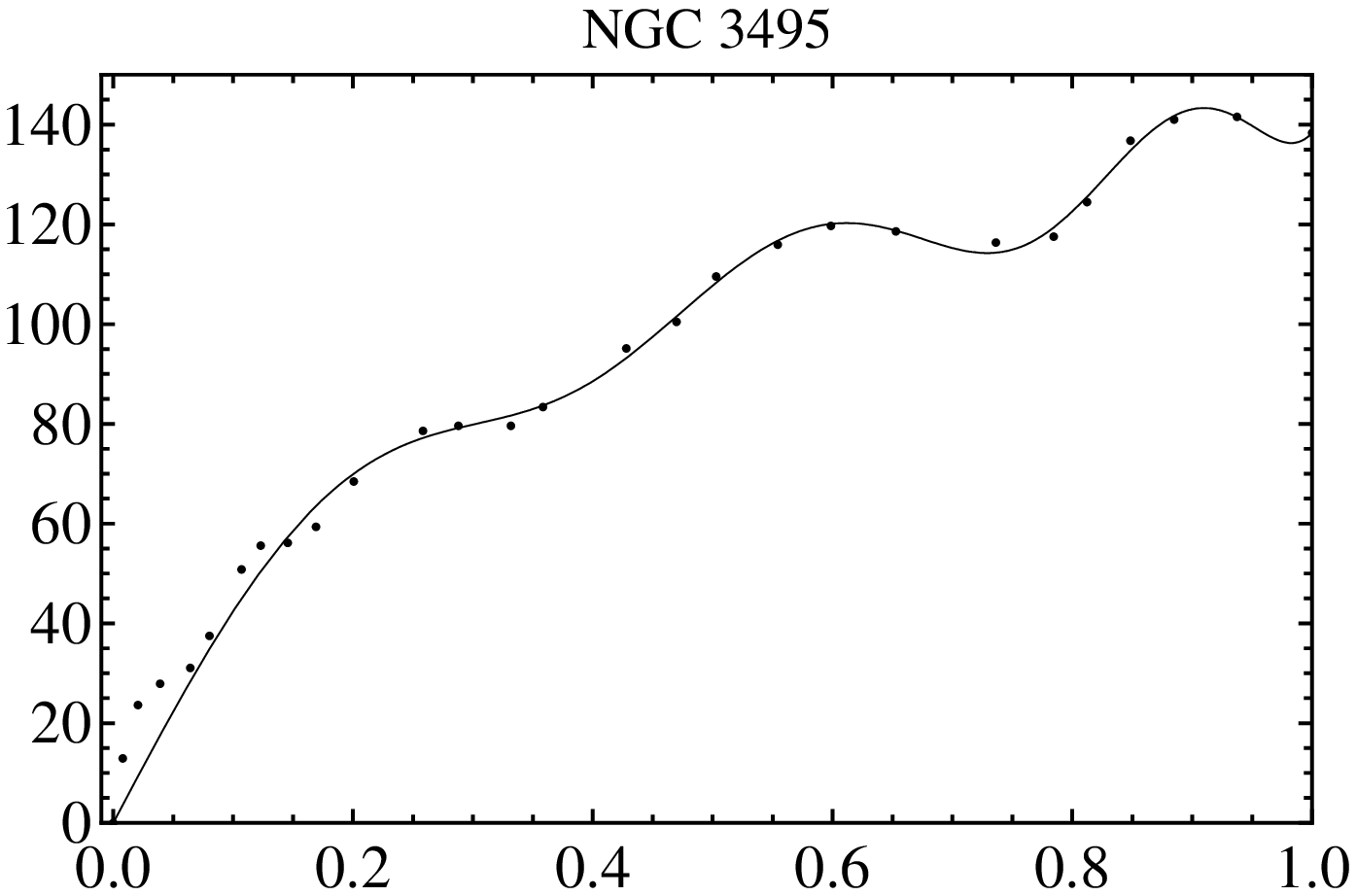} \\
  \epsfig{width=5.4cm,file=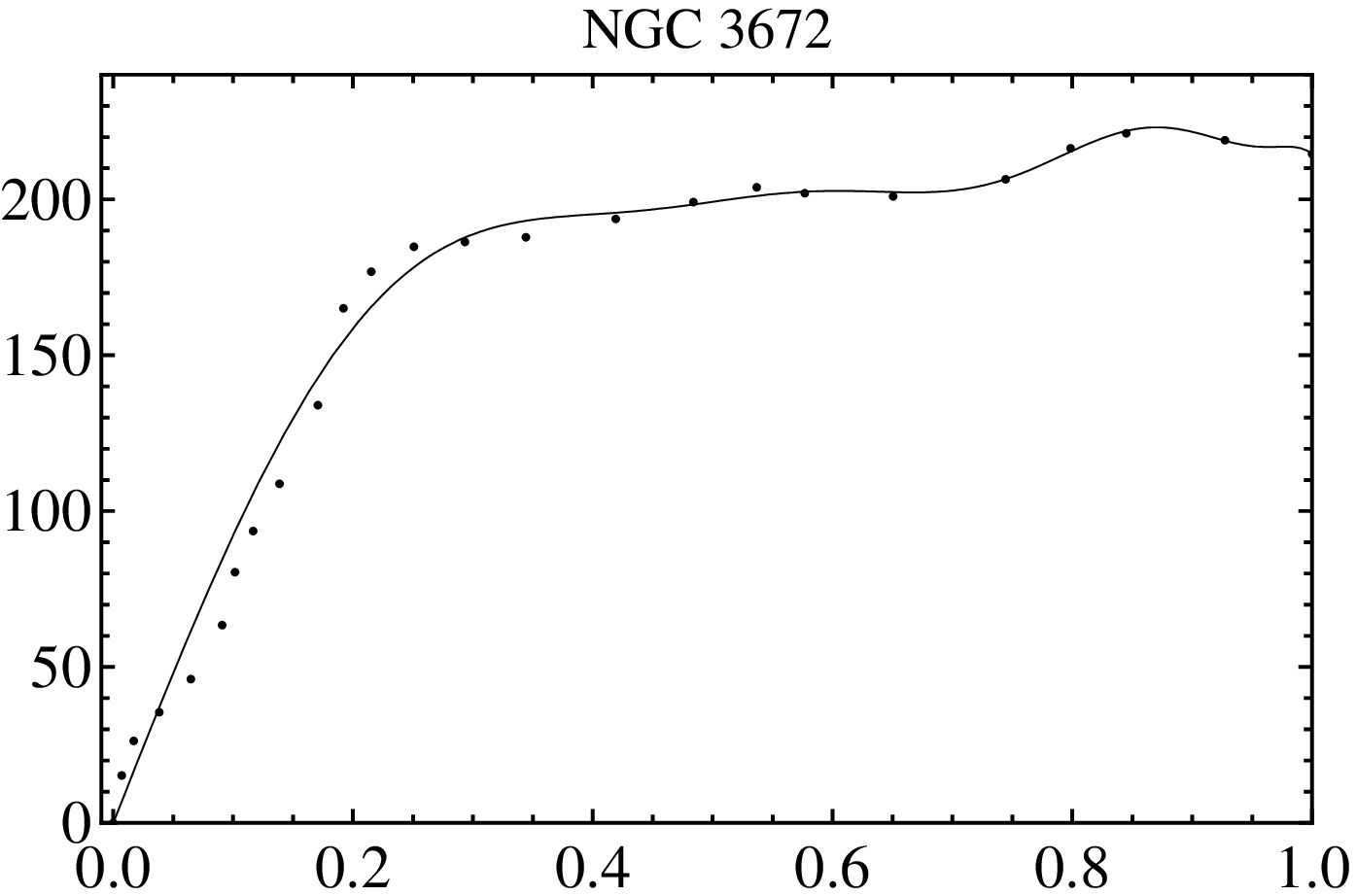} & \epsfig{width=5.4cm,file=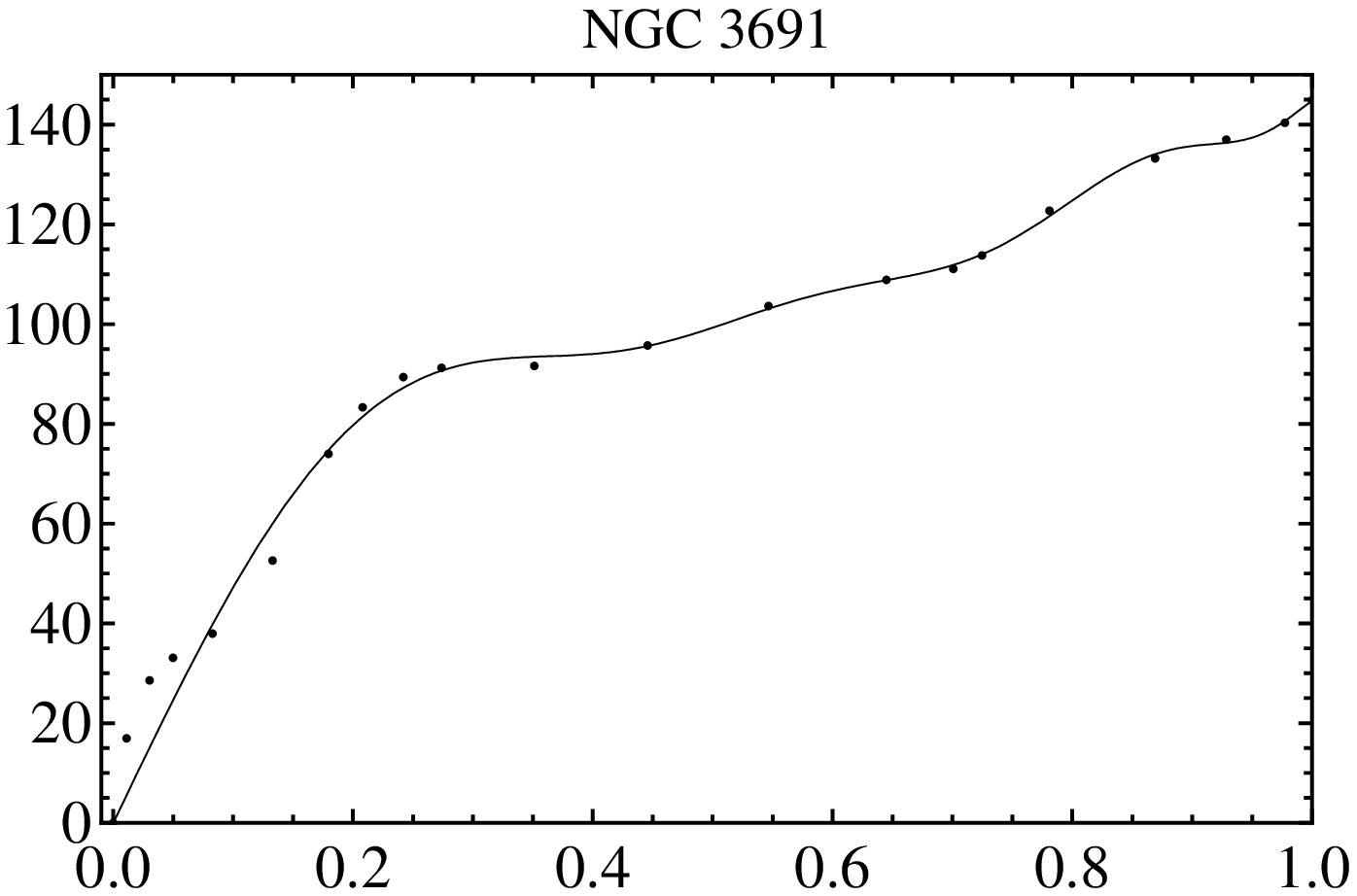} & \epsfig{width=5.4cm,file=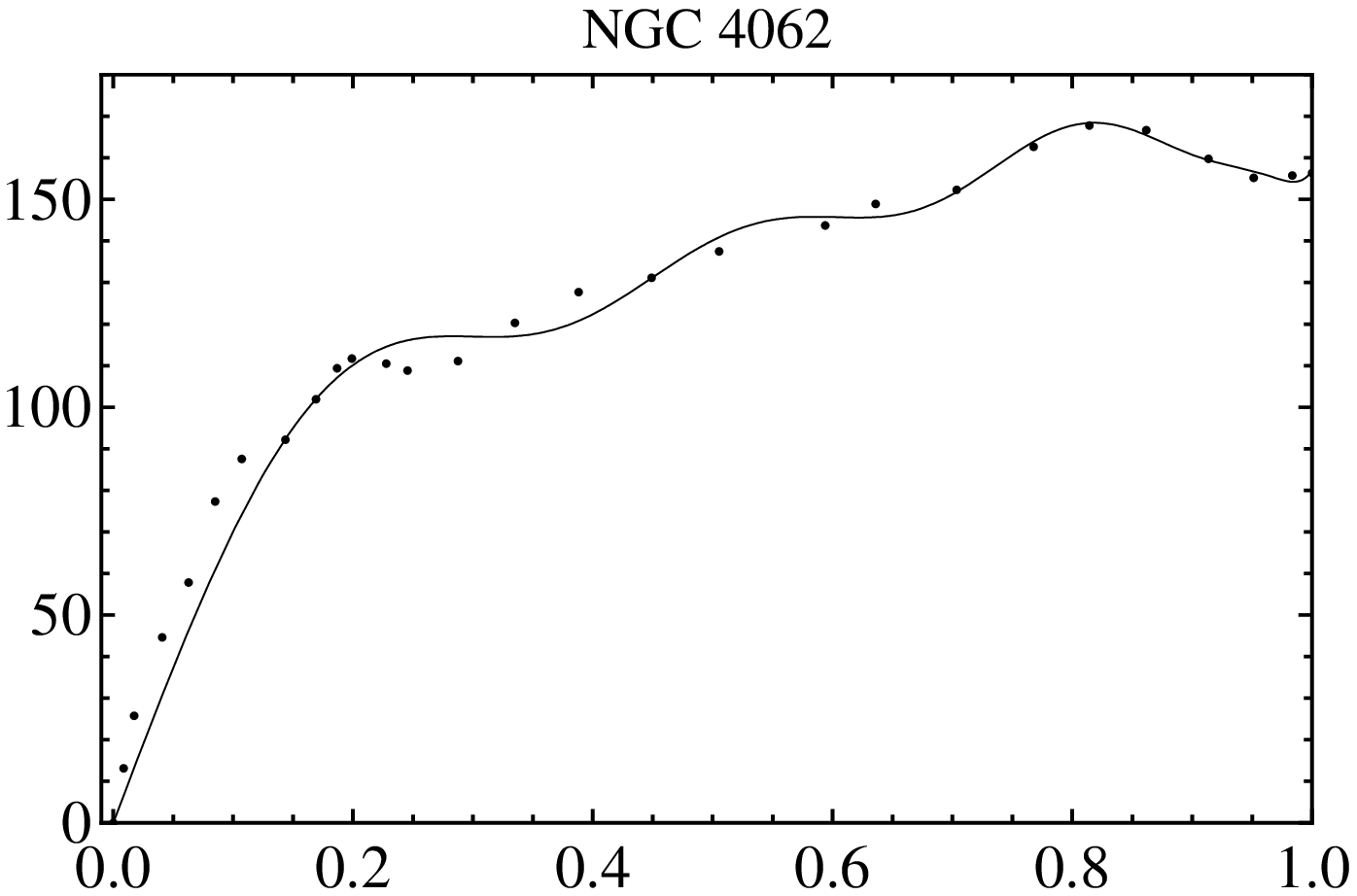} \\
  \epsfig{width=5.4cm,file=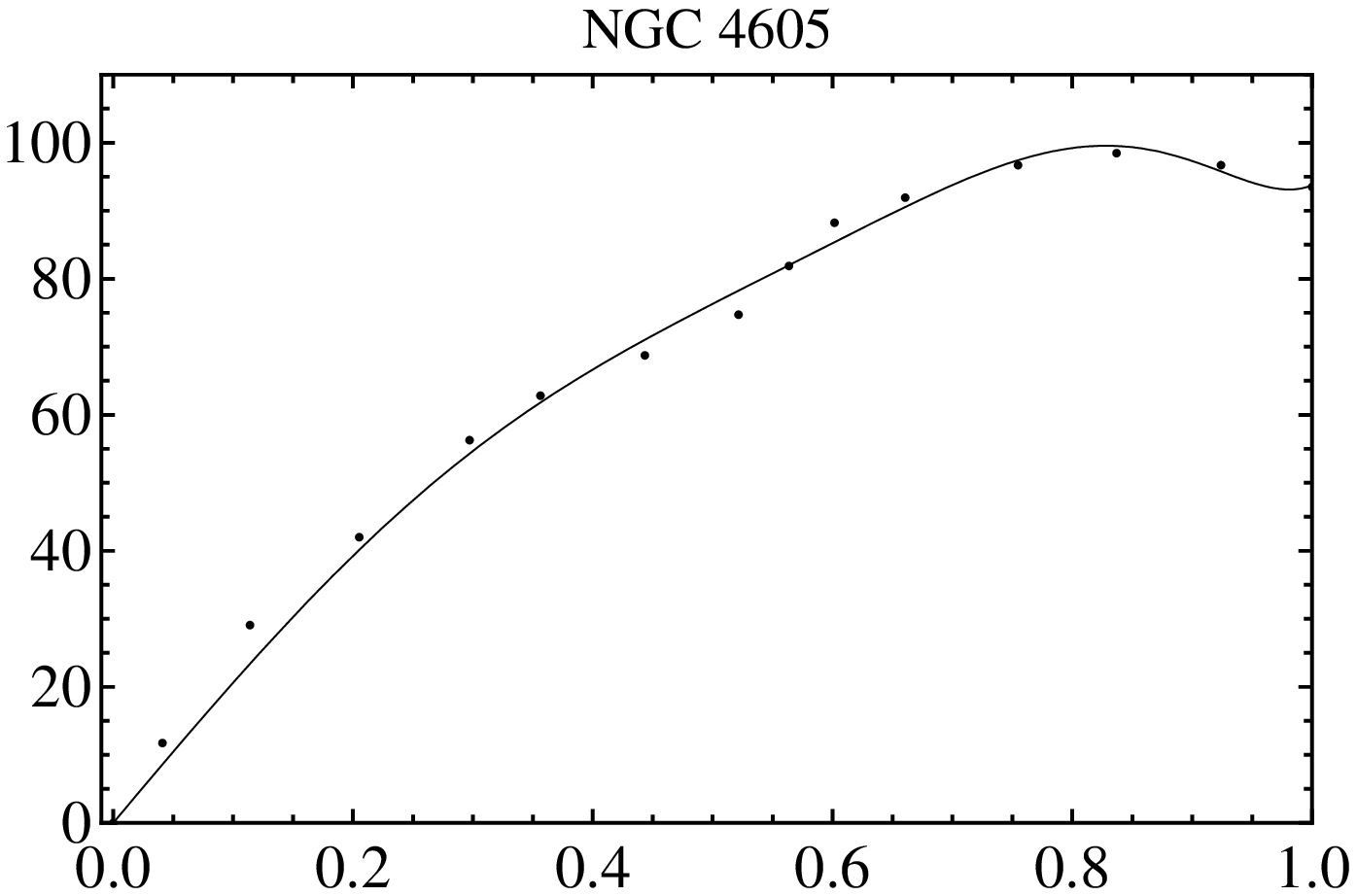} & \epsfig{width=5.4cm,file=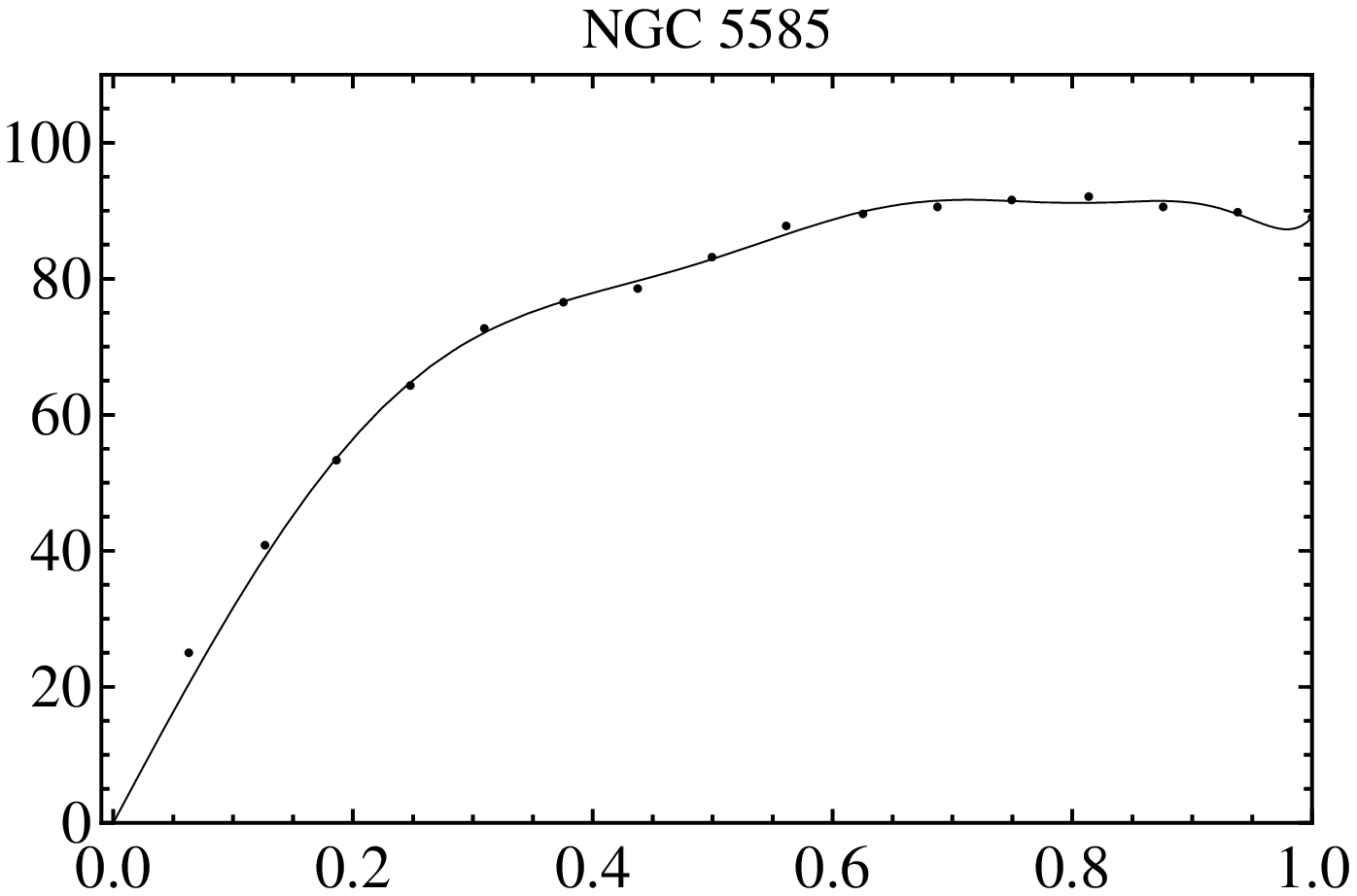} & \epsfig{width=5.4cm,file=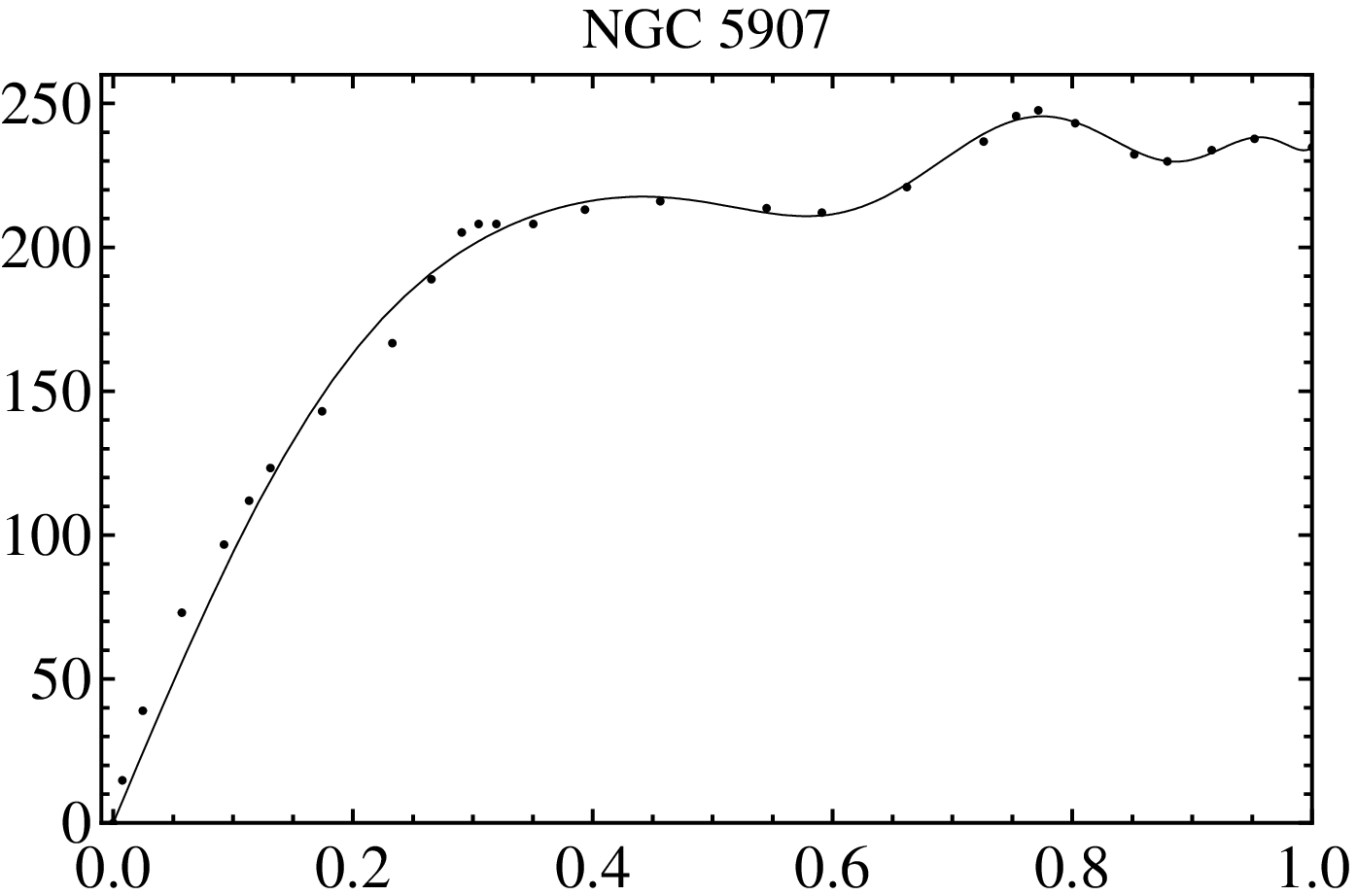}
\end{array}
$$
\caption{Circular velocity, $V_{c}$ (km s$^{-1}$), vs. $R/a$ (optical disk region) for a sample of nine spiral galaxies. The points correspond to
observational data from \cite{sofue1999} and the full line is the best fit using relation (\ref{velcirchunter}).} \label{fig:Vc}
\end{figure*}

\begin{figure*}
$$
\begin{array}{ccc}
  \epsfig{width=5.4cm,file=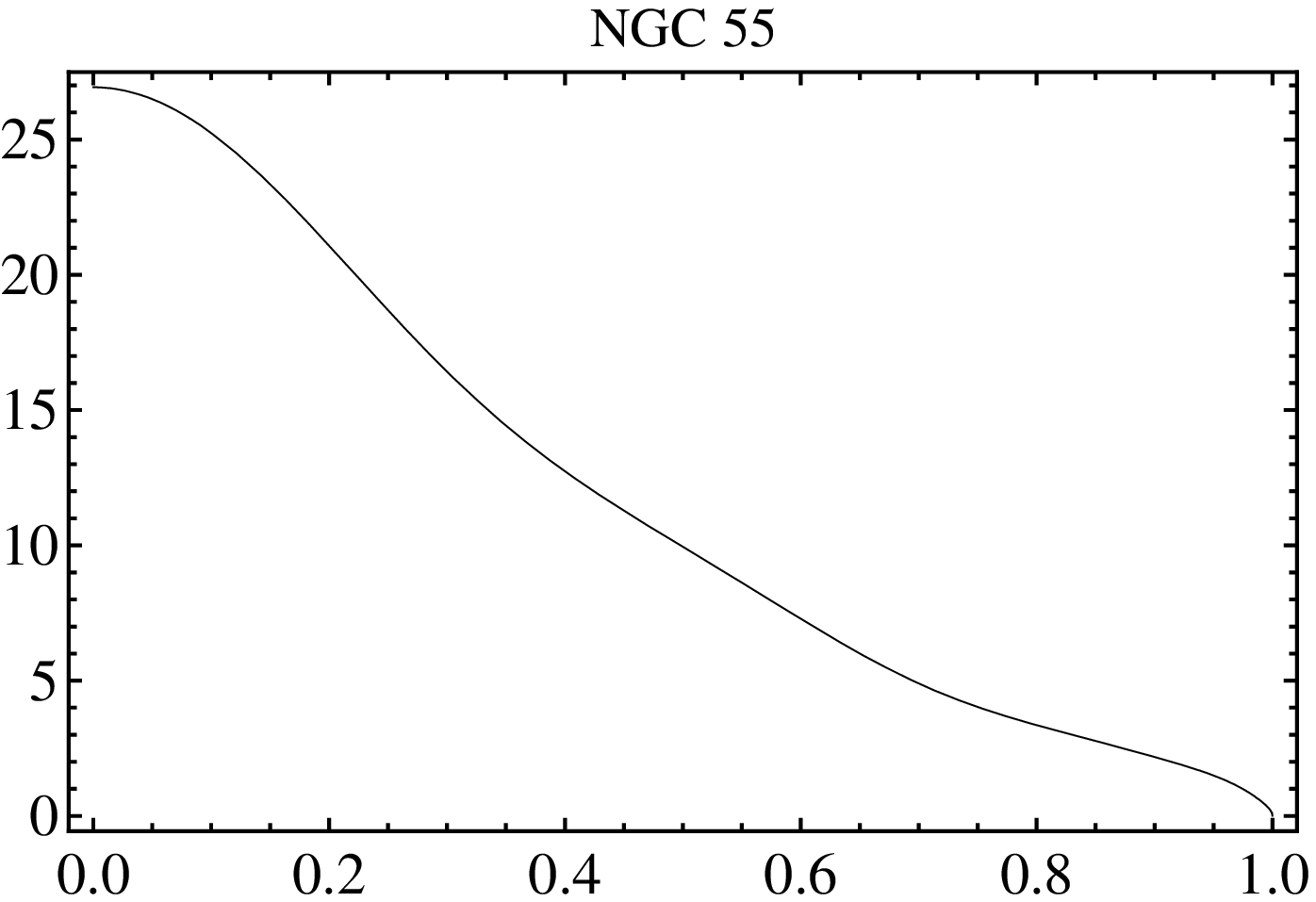} & \epsfig{width=5.4cm,file=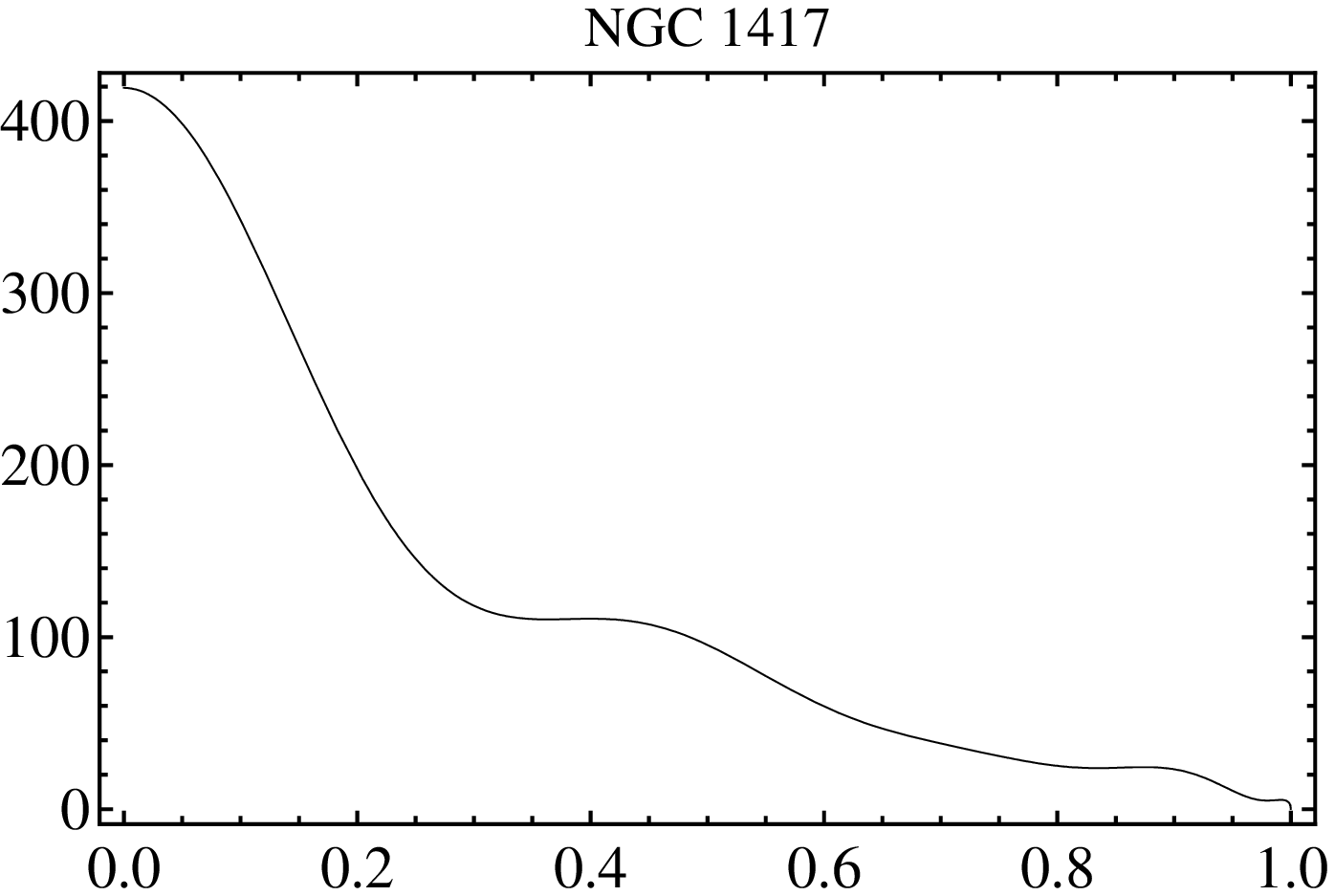} & \epsfig{width=5.4cm,file=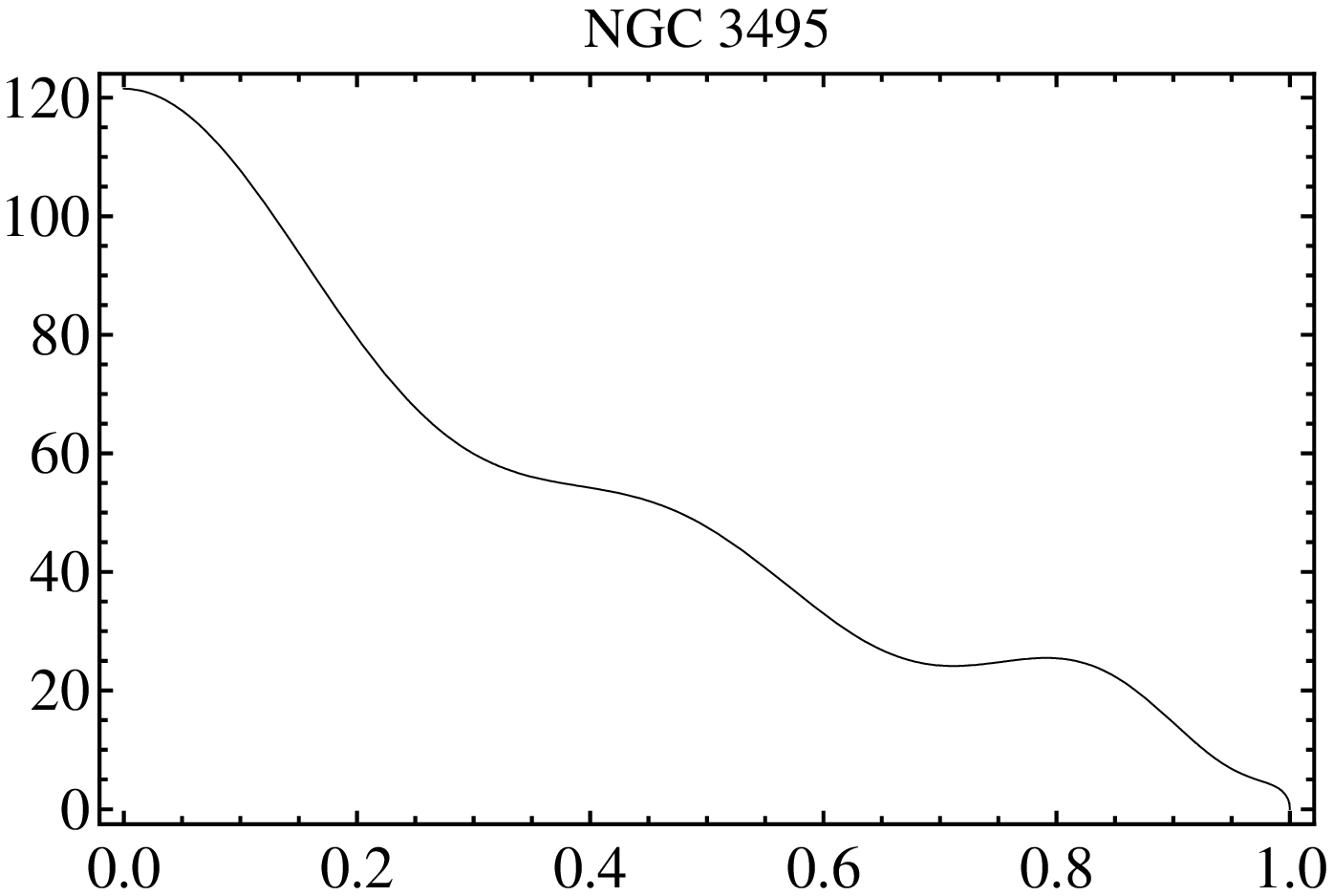} \\
  \epsfig{width=5.4cm,file=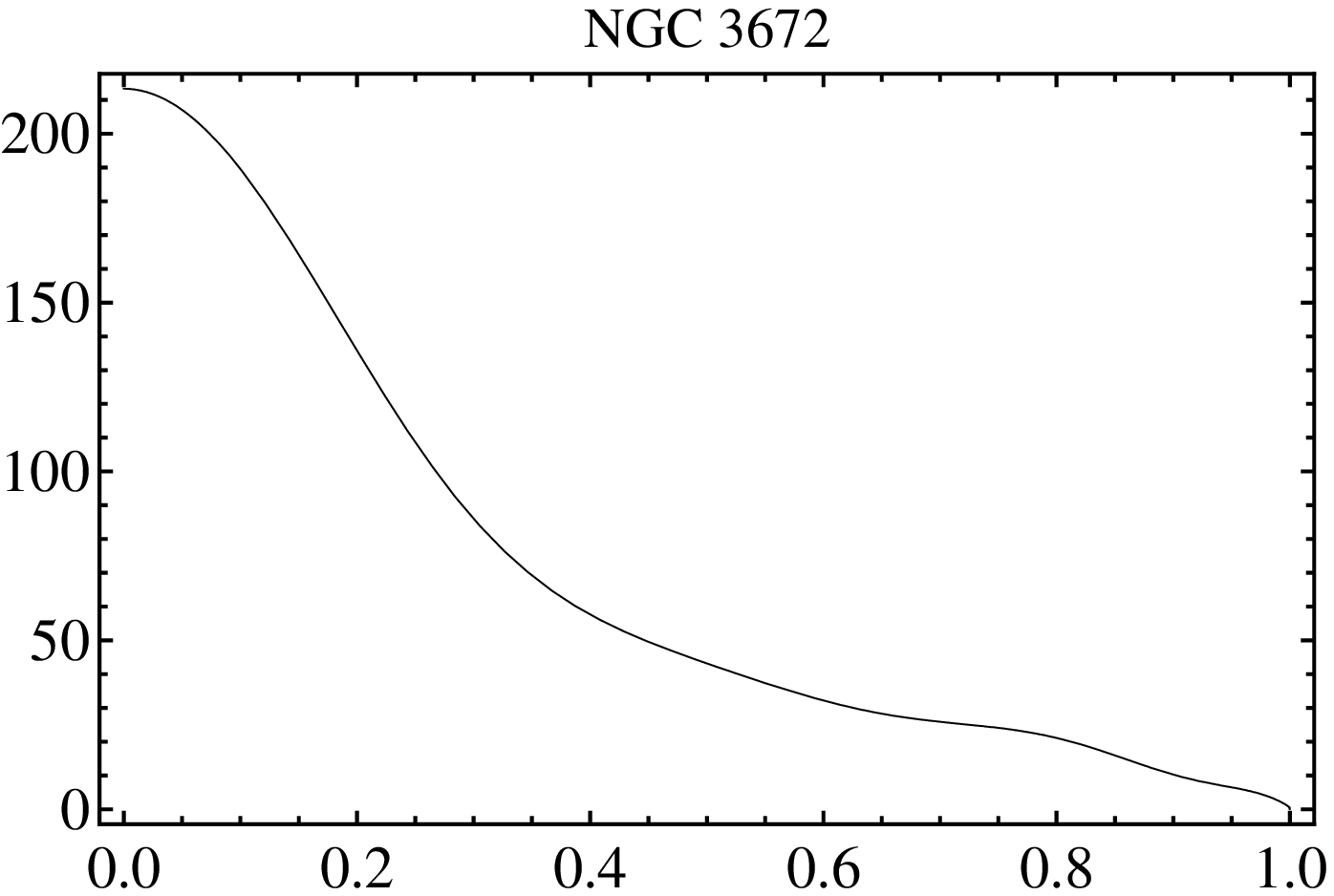} & \epsfig{width=5.4cm,file=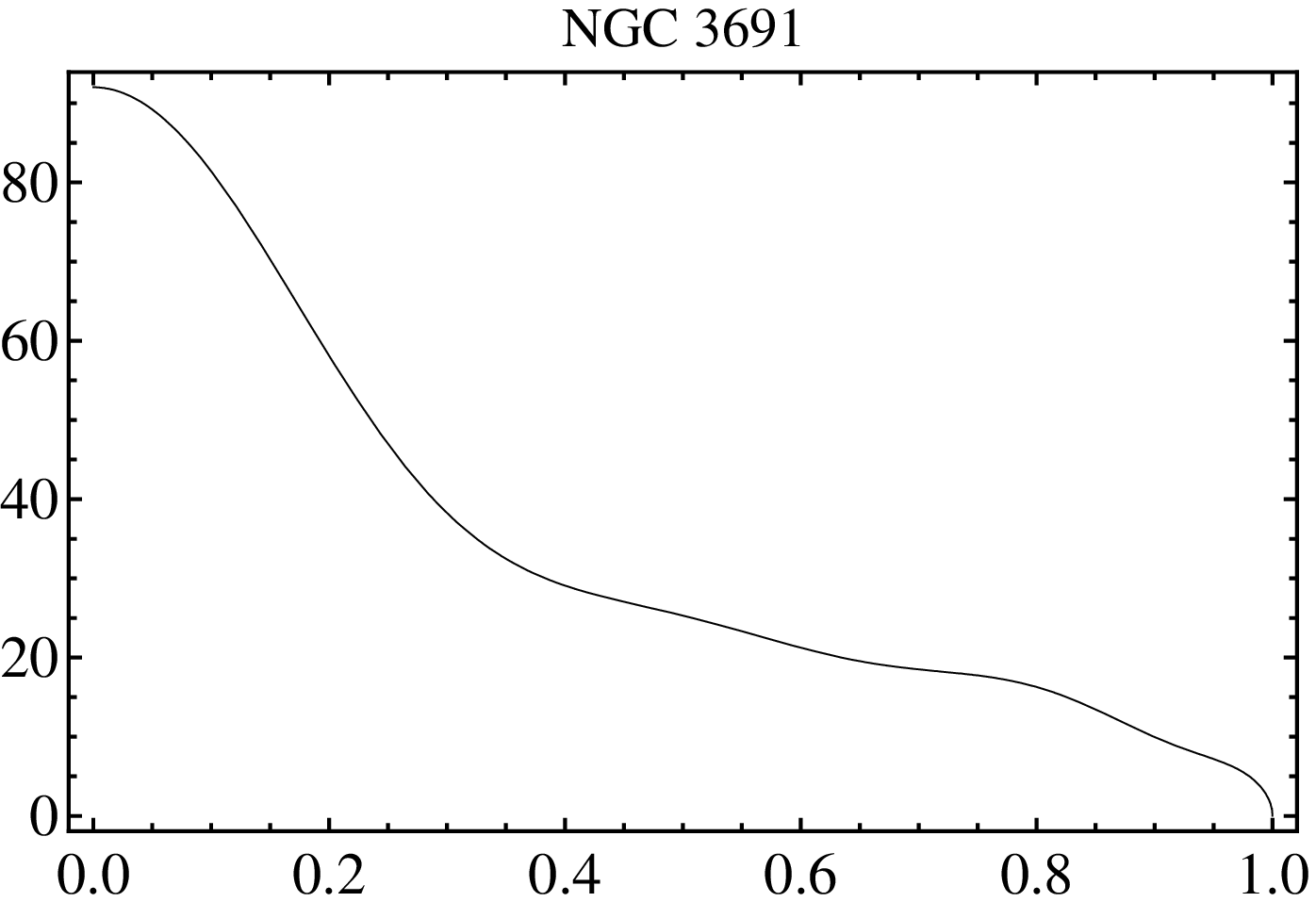} & \epsfig{width=5.4cm,file=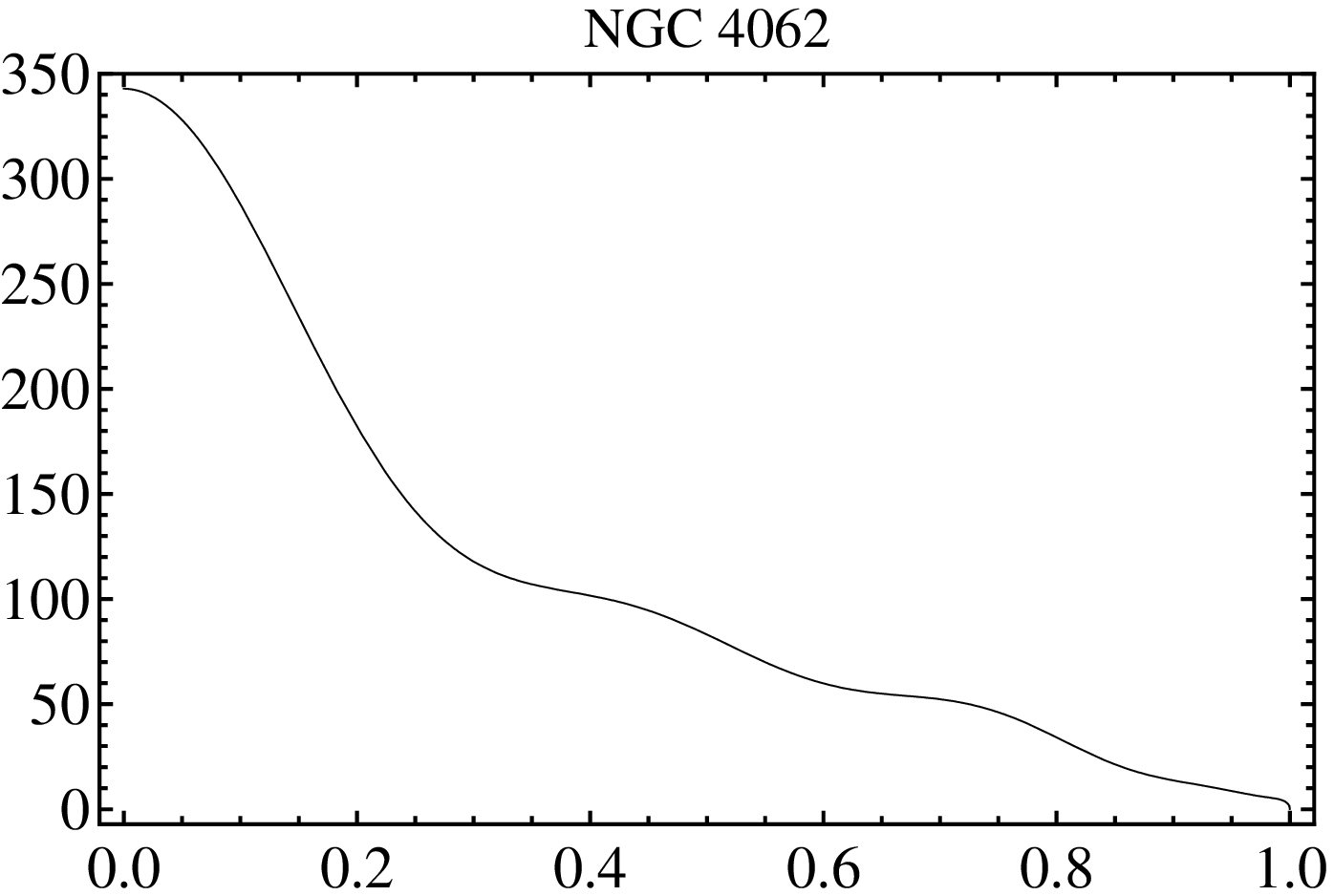} \\
  \epsfig{width=5.4cm,file=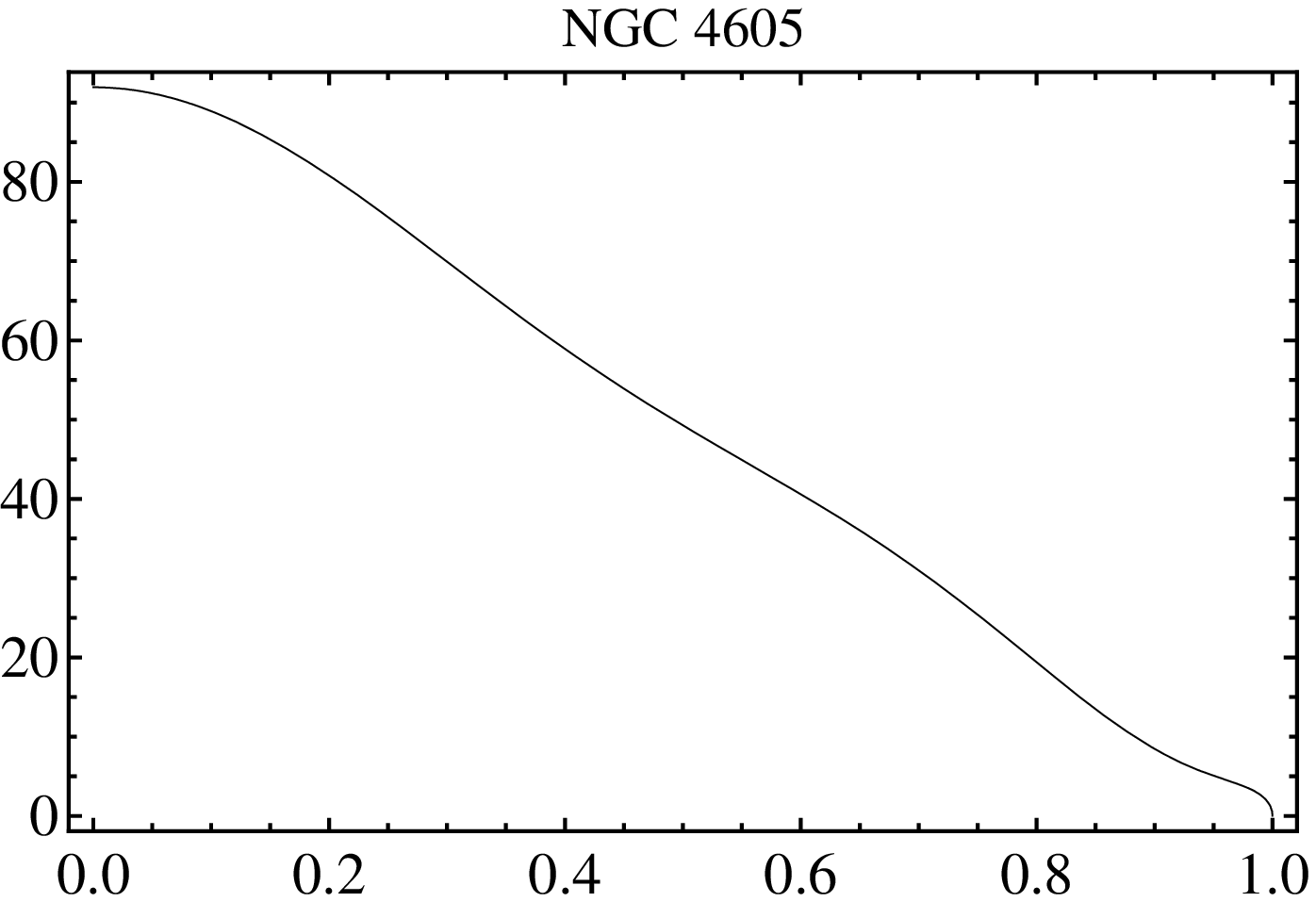} & \epsfig{width=5.4cm,file=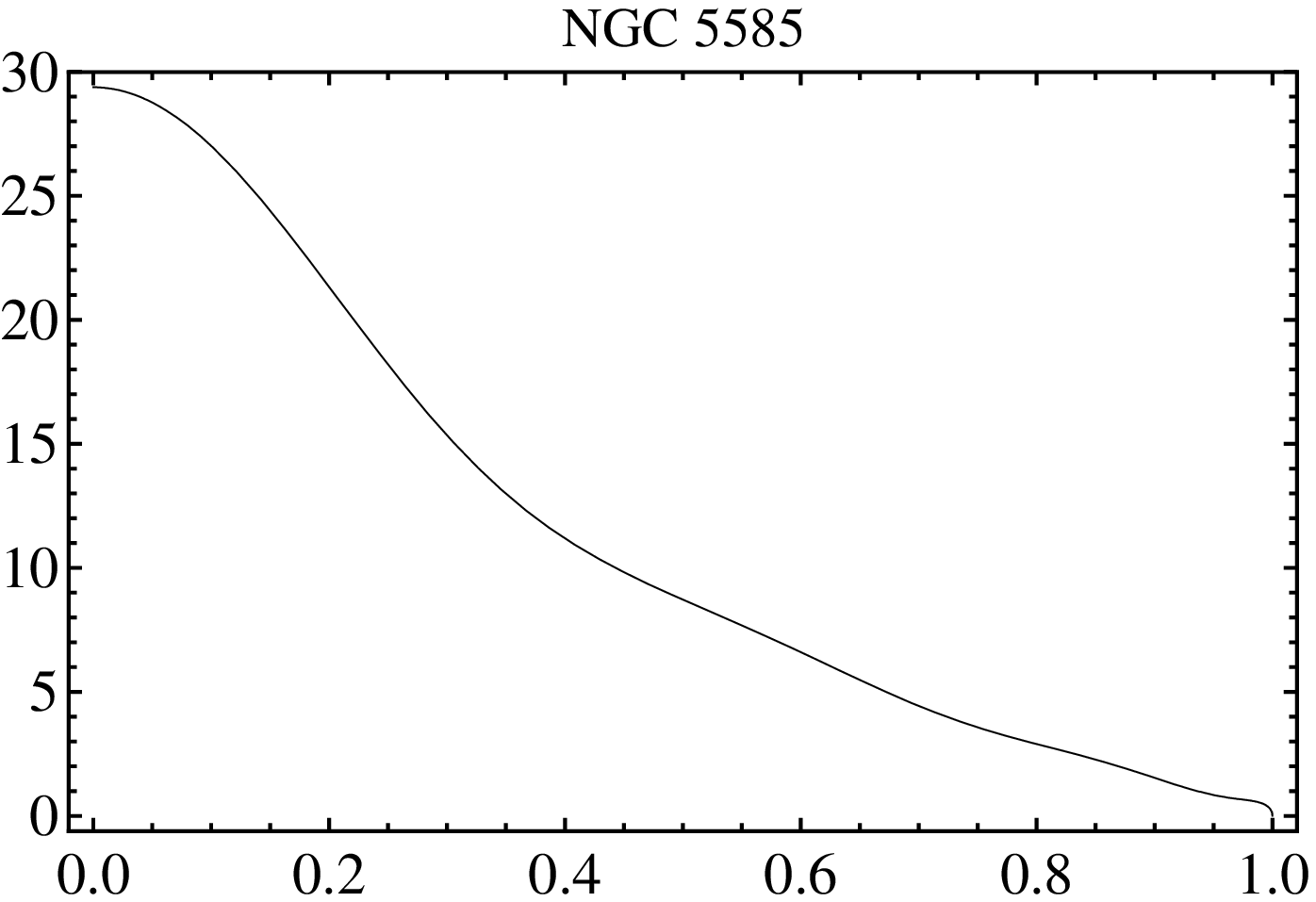} & \epsfig{width=5.4cm,file=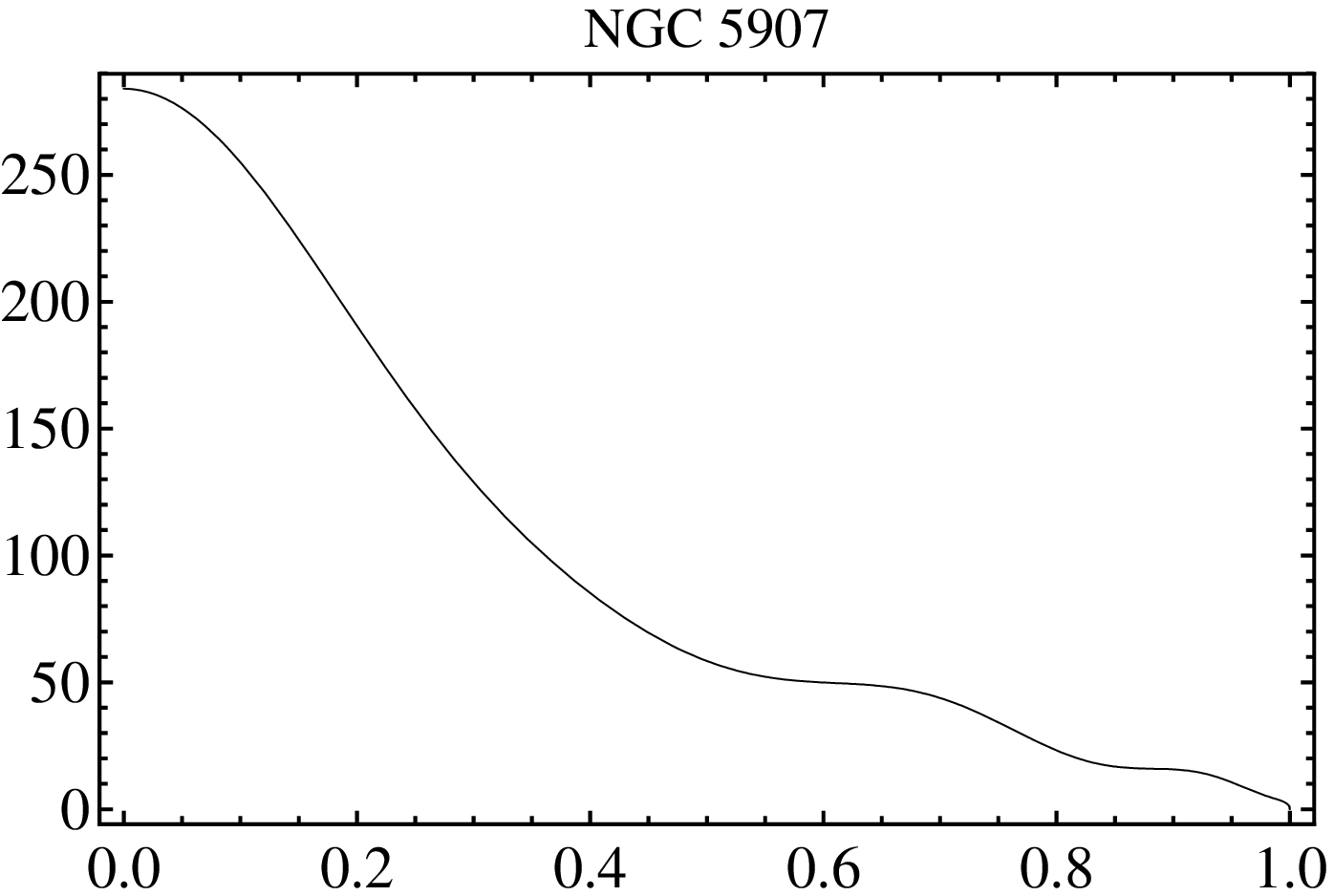}
\end{array}
$$
\caption{Mass density profiles, $\Sigma$ ($10^{10}$kg m$^{-2})$ vs. $R/a$, for the nine spiral galaxies of figure \ref{fig:Vc} via
relation (\ref{massprofile}).} \label{fig:Dens}
\end{figure*}

\begin{figure*}
$$
\begin{array}{ccc}
  \epsfig{width=5.4cm,file=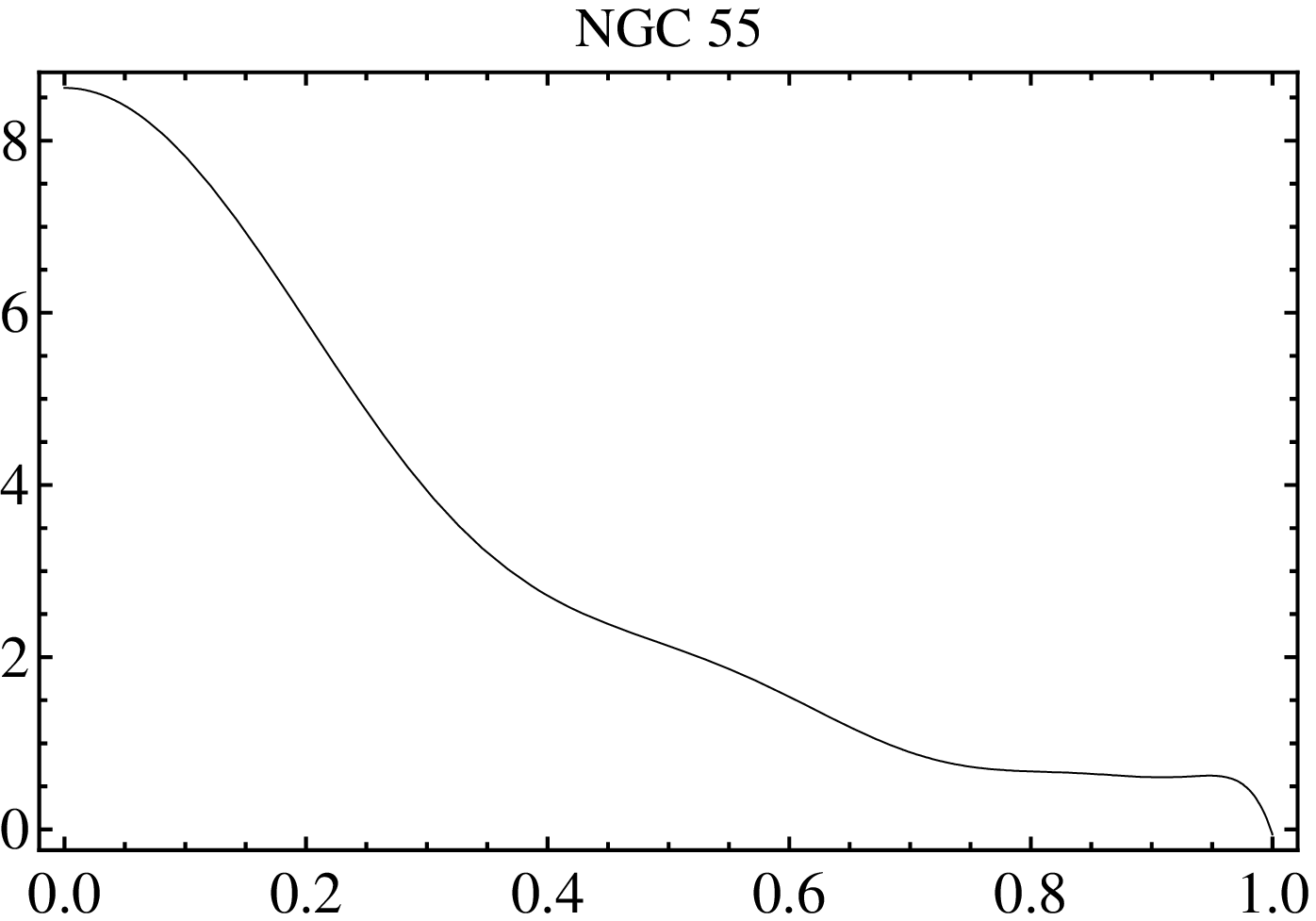} & \epsfig{width=5.4cm,file=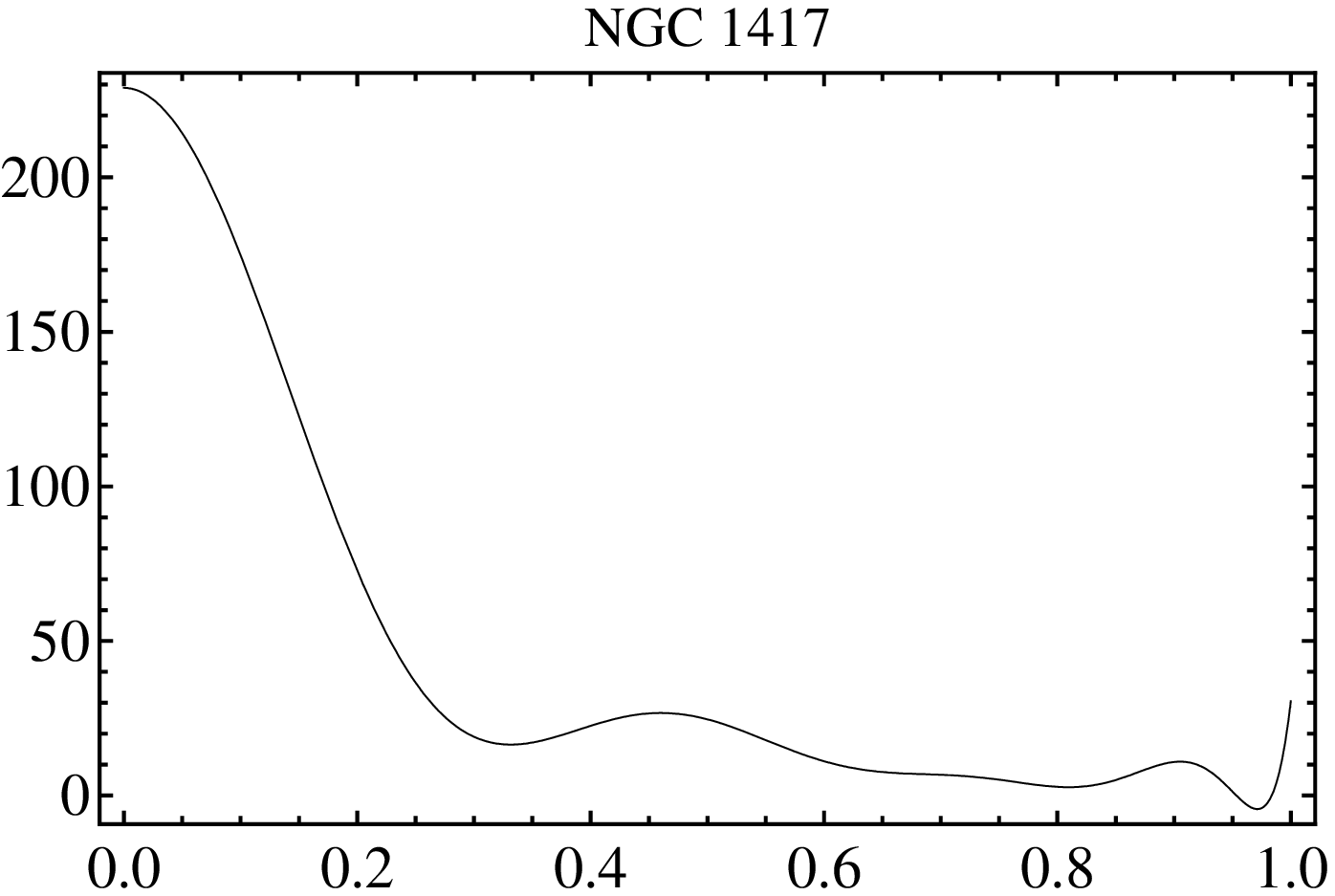} & \epsfig{width=5.4cm,file=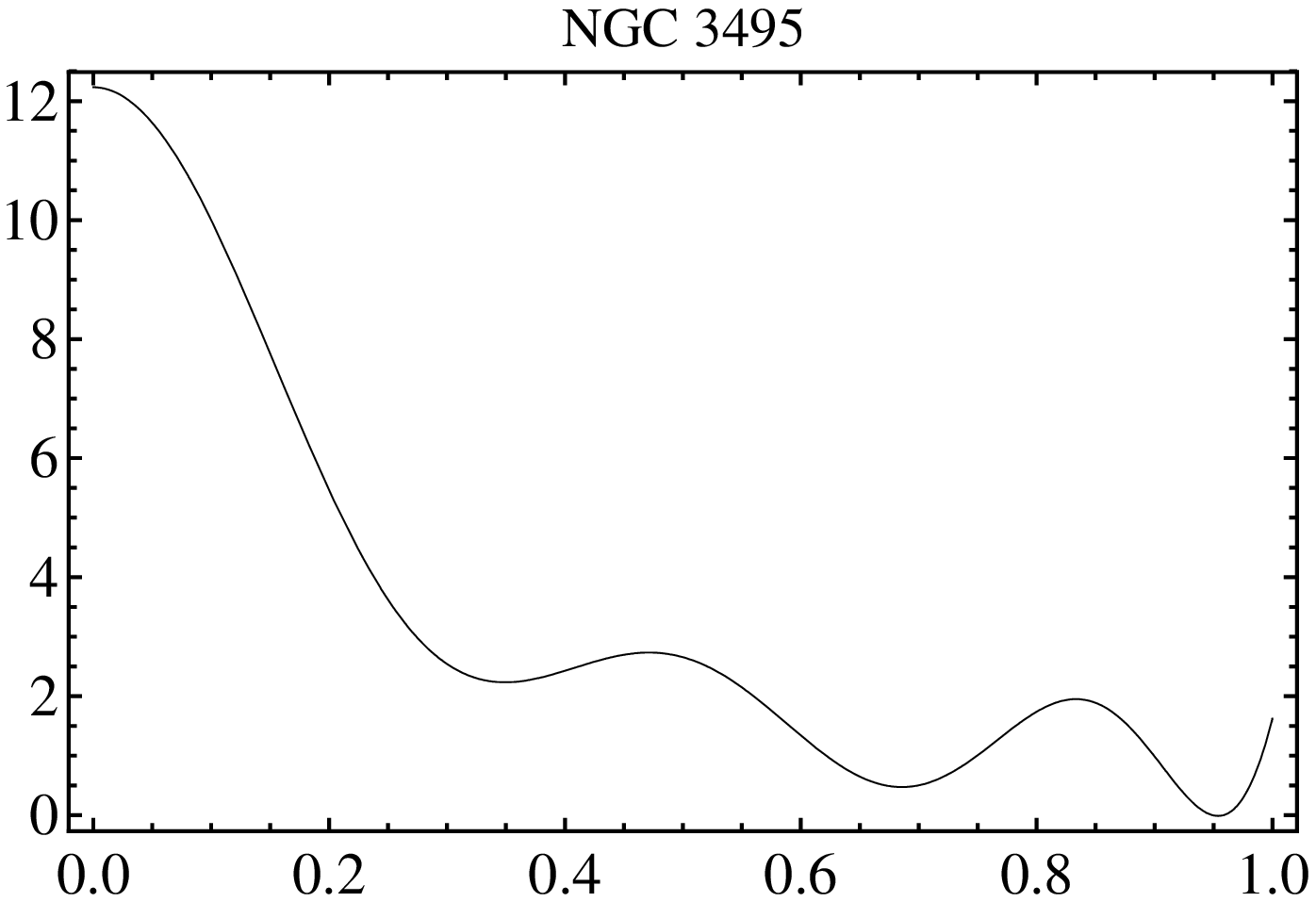} \\
  \epsfig{width=5.4cm,file=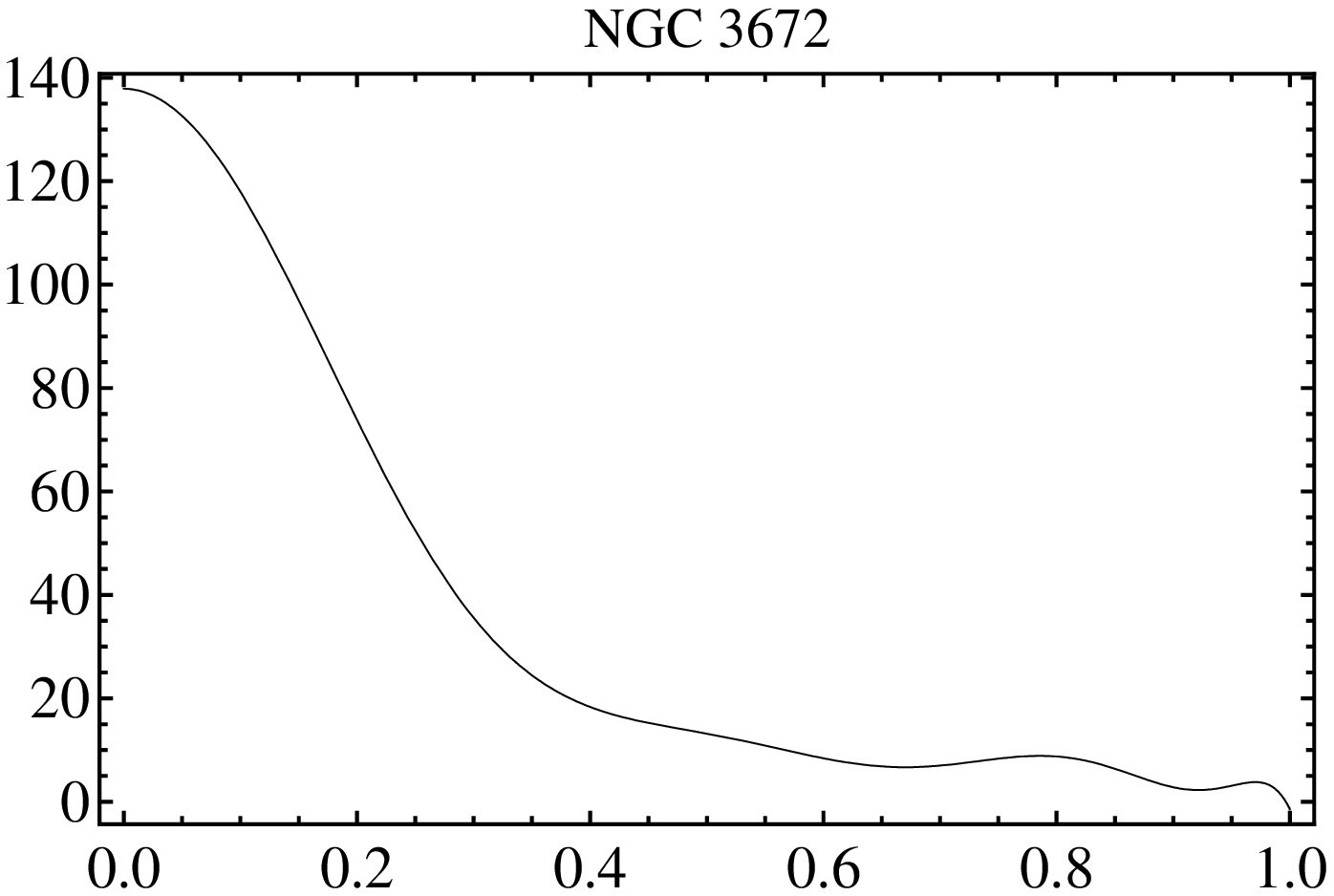} & \epsfig{width=5.4cm,file=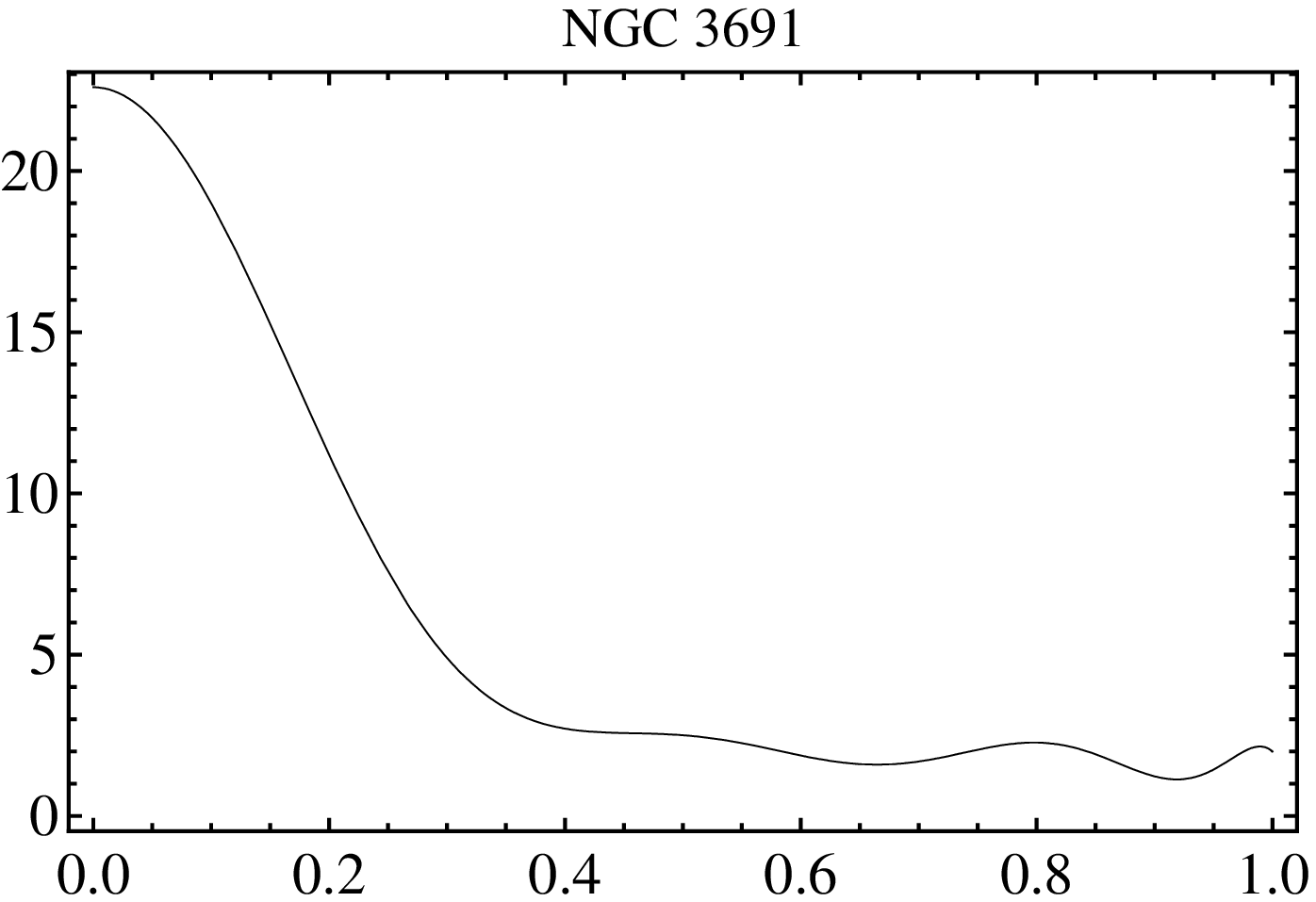} & \epsfig{width=5.4cm,file=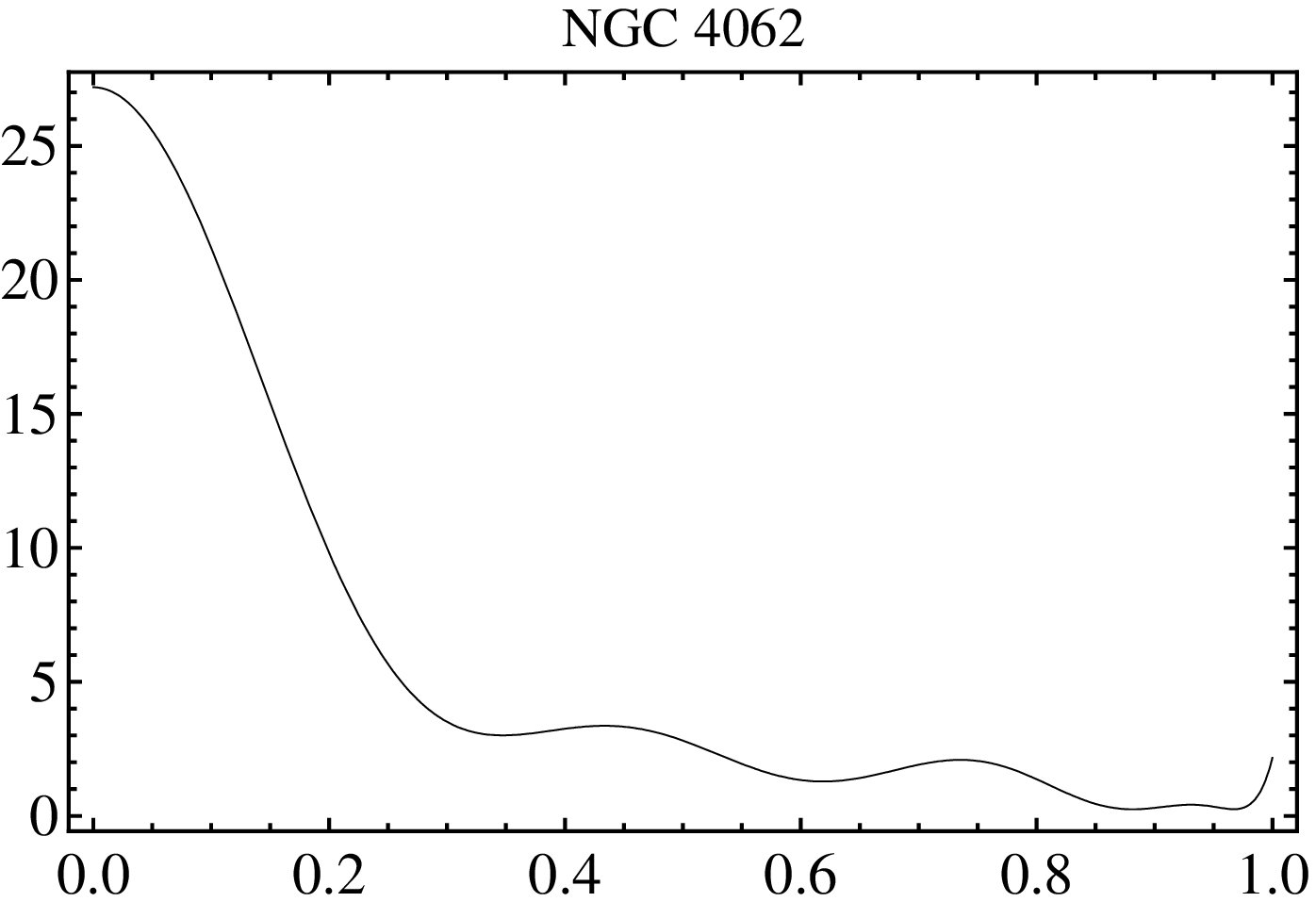} \\
  \epsfig{width=5.4cm,file=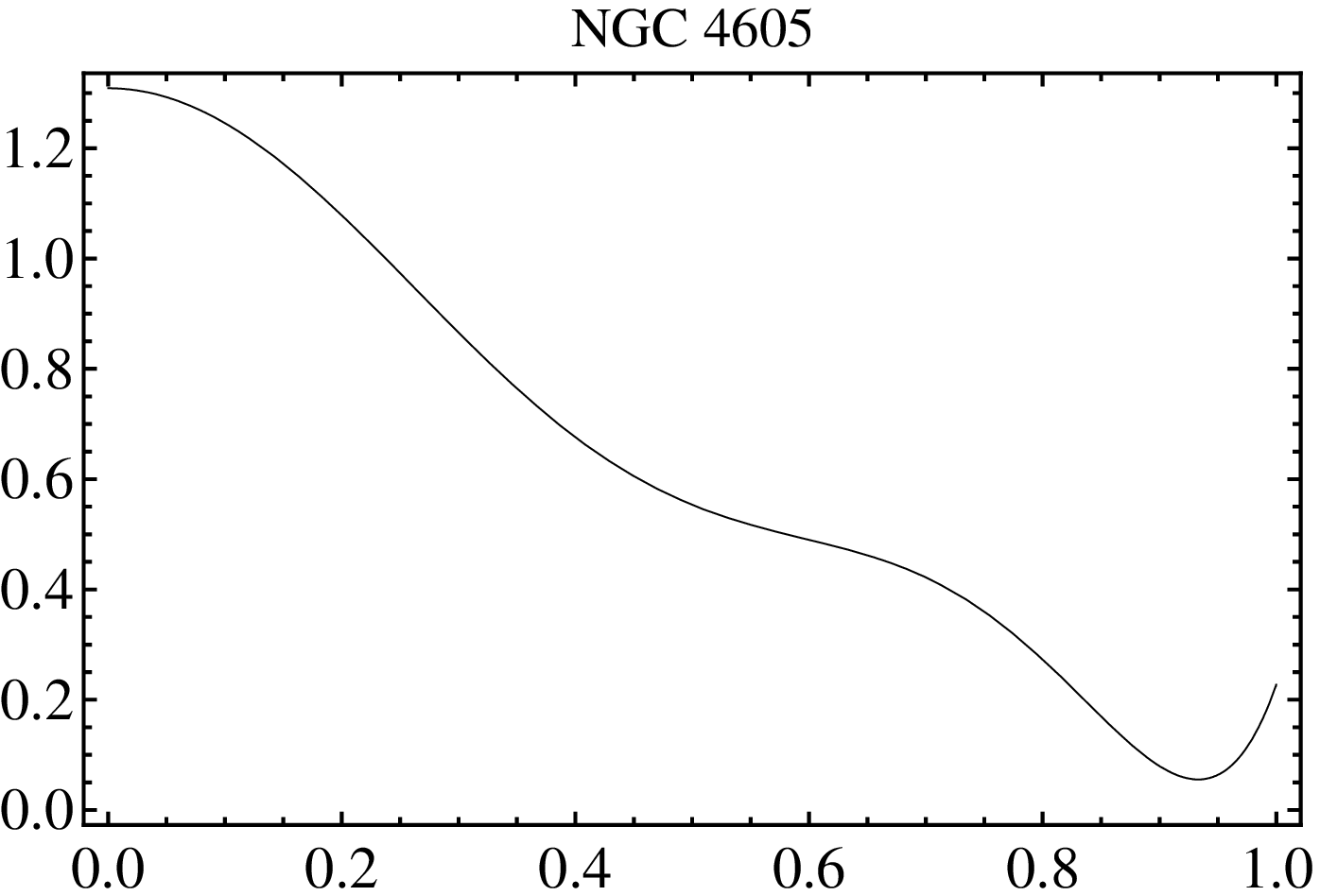} & \epsfig{width=5.4cm,file=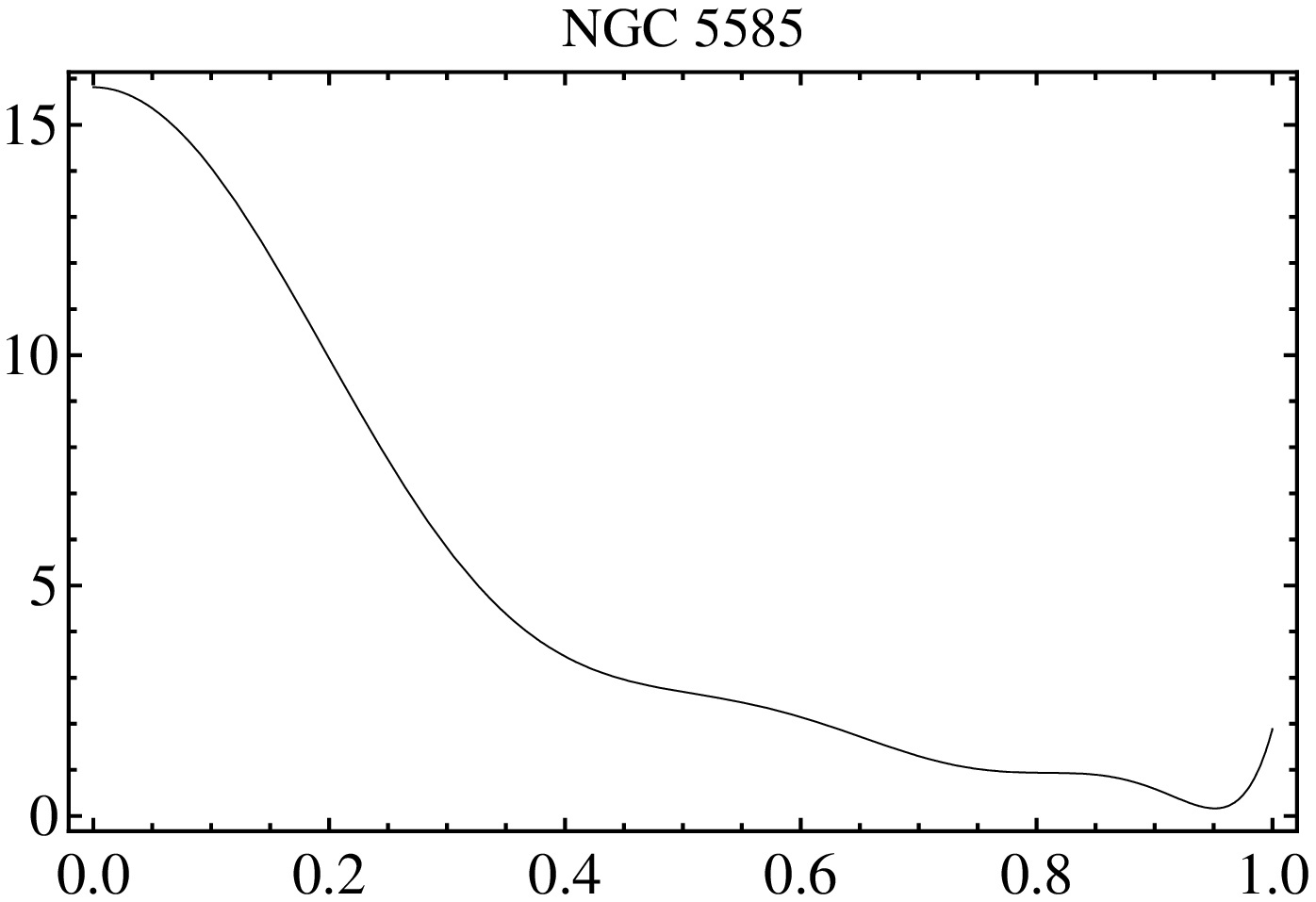} & \epsfig{width=5.4cm,file=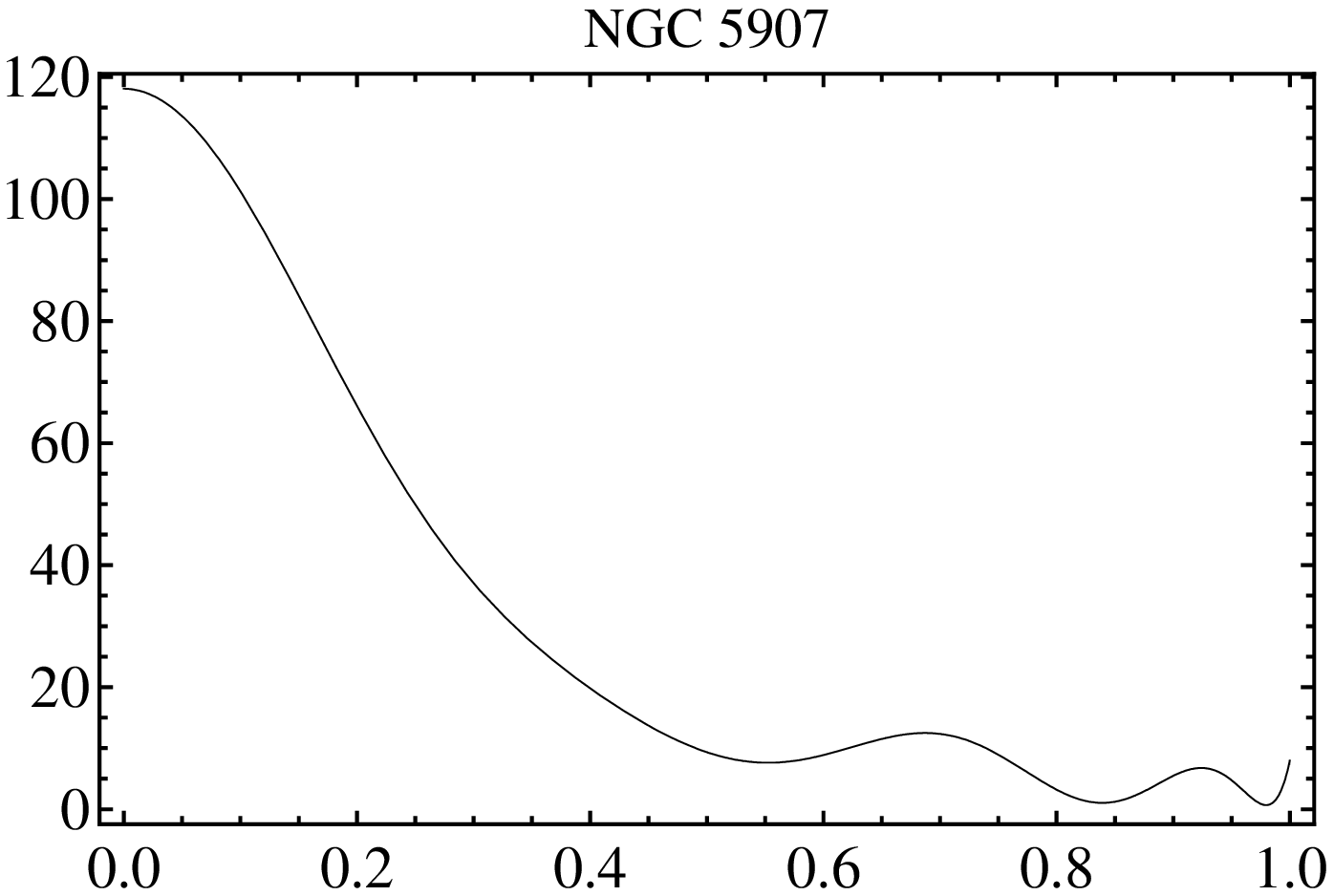}
\end{array}
$$
\caption{Quadratic epicyclic frequency, $\kappa^{2}$ ($10^{-22}$s$^{-2}$) vs. $R/a$, for the sample of figures \ref{fig:Vc} and \ref{fig:Dens}.
In all of these cases we obtain $\kappa^{2}\geq0$, meaning that quasi-circular orbits, on the optical disk region,
are stable under radial perturbations.} \label{fig:epi}
\end{figure*}

In order to illustrate  the above statements, we present additional models for nine spiral galaxies: NGC 55, 1417, 3495, 3672, 3691, 4062, 4605,
5585 and 5907. It can be assumed that all of them obey the rotation law (\cite{ngc})
\begin{equation}\label{velcirchunter}
    v_{c}^{2}=\sum_{n=1}^{m}A_{2n}(R/a)^{2n},
\end{equation}
where $a$ is the radius of the stellar disk, i.e. the optical radius, and $A_{2n}$, along with the integer $m$,
 are constants to be determined by fitting the
observational data corresponding to the circular velocity $v_{c}$.
The surface mass density of the thin disk is given by the relation
\begin{equation}\label{massprofile}
    \Sigma(R)=\frac{1}{2\pi a G \eta}\sum_{n=0}^{m}C_{2n}(2n+1)q_{2n+1}(0)P_{2n}(\eta),
\end{equation}
where $\eta=\sqrt{1-R/a}$, $P_{n}$ are the Legendre polynomials, $q_{n}(\xi)=i^{n+1}Q_{n}(i\xi)$, $Q_{n}$ being the Legendre
Functions of second kind, and $C_{2n}$ are determined from $A_{2n}$ through the relation
\begin{equation}\label{c2n}
    C_{2n}=\frac{4n+1}{4n(2n+1)q_{2n}(0)}\sum_{k=1}^{m}A_{2k}\int_{-1}^{1}x(1-x^{2})^{k}P'_{2n}(x)dx,
\end{equation}
for $n\neq 0$ and
\begin{equation}\label{c0}
    C_{0}=\sum_{n=1}^{m}(-1)^{n+1}C_{2n},
\end{equation}
from which the total mass of the stellar disk can be determined:
\begin{equation}\label{totalmass}
    M= a C_{0}/G.
\end{equation}
In figure \ref{fig:Vc} we show the fits to observational
data, corresponding to rotation curves from \cite{sofue1999}. The results of the fitting,
i.e. the values of $m$ and $A_{2n}$, determine the constants
$C_{2n}$ (see table \ref{tabla1}) and the mass density  profiles (eq.(\ref{massprofile})), which are shown in figure
\ref{fig:Dens}. In addition, we can obtain the quadratic epicyclic frequency (fig. \ref{fig:epi}) from the relation
\begin{equation}\label{epiquadratic}
    \kappa^{2}(R)=\sum_{n=1}^{m}2(n+1)A_{2n}(R/a)^{2n-2},
\end{equation}
which is provided by \cite{ngc}. We can see that in all of these cases it is possible to obtain models whose circular orbits
are all stable.

\begin{deluxetable*}
{c r r r r r r r r r}
  \hline
  \hline
   &  &  &  &  &  &  &  &  & \\
  NGC  &  0055 &  1417 &  3495 &  3672 &  3691 &  4062 &  4605 &  5585 &  5907\\
  Type & SBm & SABb & Sc & Sc & SBb & SABc & SBc & SABc & SABc\\
  a (kpc) & 9.78 & 9.56 & 4.82 & 11.83 & 7.33 & 3.81 & 2.42 & 11.84 & 9.76\\
   &  &  &  &  &  &  &  &  & \\
  \hline
  \hline
  &  &  &  &  &  &  &  &  & \\
  m & 9 & 9 & 8 & 8 & 8 & 9 & 5 & 7 & 9\\
  &  &  &  &  &  &  &  &  & \\
  \hline
   &  &  &  &  &  &  &  &  & \\
  Constants $C_{2n}$(km$^{2}$ s$^{-2}$)\\
  $C_{0}$ & 4403.69 & 38276.21 & 10182.74 & 28122.12 & 10083.08 & 14795.57 & 5183.18 & 4851.66 & 33246.82\\
  $C_{2}$ & 6432.41 & 58035.31 & 13146.72 & 42417.66 & 12615.37 & 21724.96 & 7147.02 & 7406.72 & 50304.17 \\
  $C_{4}$ & 2943.90 & 30770.12 & 4294.24 & 23104.54 & 4497.90 & 9714.98 & 2200.70 & 3702.67 & 26775.38 \\
  $C_{6}$ & 1264.92 & 17058.56 & 2474.02 & 15892.97 & 3617.42 & 5308.24 & 380.94 & 1795.96 & 16739.60 \\
  $C_{8}$ & 525.62 & 10486.79 & 1201.58 & 11299.94 & 2821.16 & 4794.21 & 359.43 & 1098.36 & 12338.09 \\
  $C_{10}$ & 343.56 & 11084.03 & 394.77 & 6320.53 & 1749.81 & 3455.58 & 215.35 & 707.52 & 7730.72 \\
  $C_{12}$ & 247.33 & 12483.17 & 1358.85 & 3771.99 & 1168.16 & 2438.53 & $\cdots$ & 439.27 & 2137.03 \\
  $C_{14}$ & 105.10 & 9164.04 & 1333.32 & 2623.33 & 886.72 & 2516.24 & $\cdots$ & 181.76 & 173.13 \\
  $C_{16}$ & 12.78 & 6018.08 & 311.41 & 955.89 & 299.02 & 1921.78 & $\cdots$ & $\cdots$ & 1456.49 \\
  $C_{18}$ & -12.67 & 2692.43 & $\cdots$ & $\cdots$ & $\cdots$ & 660.05 & $\cdots$ & $\cdots$ & 1006.18 \\
   &  &  &  &  &  &  &  &  & \\
  \hline
  &  &  &  &  &  &  &  &  & \\
  Maximum Disk Mass $(10^{10}M_{\bigodot})$ & 1.00 & 8.51 & 1.14 & 7.74 & 1.72 & 1.31 & 0.29 & 1.34 & 7.54 \\
  &  &  &  &  &  &  &  &  & \\
  \hline
  \hline
  &  &  &  &  &  &  &  &  & \\

\caption{Parameters for the nine spiral galaxies chosen: $a$ is the radius of the optical disk, $m$ is
the number of expansion constants used to obtain the best fit (i.e. the $A_{2n}$ of eq.
(\ref{velcirchunter})), $C_{2n}$ are obtained from the fits according to (\ref{c2n}) and (\ref{c0}).
The estimates for the total mass of the maximum disk are computed from (\ref{totalmass}).}\label{tabla1}
\end{deluxetable*}

\section{Integrability of motion near a stable circular orbit}\label{sec:integrability}

We remarked in the Introduction that motion is integrable near a stable circular orbit.
This fact follows from the separability of the Hamilton-Jacobi
equation near $(R_{o},0)$. There are many evidences of this general behavior for thin disks
in the literature (\cite{ngc, hunter, javier, saa, pn, pedraza}),
where it is found numerically, by means of Poincar\'e sections, that motion is integrable around what appears to be a stable point
of the effective potential, corresponding to a stable circular orbit in the equatorial plane. These evidences also show that
the integrable region goes well beyond the neighborhood of the stable circular orbit, where the approximation of a separable
potential not valid.

As in the case of a smooth density distribution, for density profiles of the form (\ref{totaldensity}) motion near a stable
circular orbit of radius $R_{o}$ in the plane $z=0$ is nearly integrable. This follows from the separability of the
effective potential near the stable point $(R_{o},0)$. In fact, up to first order in $z$, we have
  \begin{eqnarray}
   \Phi_{eff}(R,z) &\approx& \Phi_{eff}(R,0) + \frac{\partial\Phi_{eff}}{\partial |z|}(R,0) |z|\nonumber \\
    &=& \Phi_{eff}(R,0) + 2\pi G\Sigma(R) |z|.\label{potaprox}
  \end{eqnarray}
Since the orbit is radially stable, we can approximate $\Sigma(R)\approx\Sigma(R_o)$, discarding second-order and higher  terms. Thus,
  \begin{equation}\label{Aphiapprox}
   \Phi_{eff}(R,z) \approx \Phi_{eff}(R,0) + 2\pi G\Sigma(R_o) |z|,
  \end{equation}
and the corresponding approximate Hamiltonian, $H = (P_R^2 + P_z^2)/2 + \Phi_{eff}$,
is also separable. Thus, since $|z|$ is continuous,
we can solve the corresponding Hamilton-Jacobi equation assuming that the generating function has the form
  \begin{equation}\label{separableS}
   S(R,z, \vec{J}) = S_R(R, \vec{J}) + S_z(z, \vec{J}),
  \end{equation}
where $\vec{J}$ are the approximate action variables.
In this way, we obtain by quadratures two independent integrals of motion near $(R_{o},0)$: one
for the $R$-coordinate, $J_{R}$, and another for the $z$-coordinate, $J_{z}$. In the next subsection
we study with some detail this idea,
focusing on the $z$-component, which is of special interest here.

\subsection{Shape of nearly equatorial orbits}

Consider an orbit in the equatorial plane and the corresponding vertical perturbation. If the original orbit is circular, motion
will occur on the torus given by the approximate action variables $J_R, J_z$, which are obtained by solving the Hamilton-Jacobi
equation with a separable generating function (\ref{separableS}):
  $$
   E = \frac{1}{2}\Big(\frac{\partial S_R}{\partial R}\Big)^2 + \frac{1}{2}\Big(\frac{\partial S_z}{\partial z}\Big)^2 +
  \Phi_{eff}(R,0) + 2\pi G \Sigma(R_o) |z|.
  $$
We have that
  \begin{equation}\label{Hz}
   H_z \equiv \frac{p_z^2}{2} + \omega |z|
  \end{equation}
is an integral of motion, where
  \begin{equation}\label{omega}
   \omega \equiv 2\pi G \Sigma(R_o).
  \end{equation}
In consequence,
  $$
   \frac{1}{2}\Big(\frac{\partial S_z}{\partial z}\Big)^2 + \omega |z| = E_z,
  $$
with $H_z = E_z$. Thus $S_z$ is given by
  $$
   S_z = \sqrt{2}\int\sqrt{E_z - \omega|z|}dz,
  $$
depending only on $z$. The corresponding action variable is (see \cite{GD} for a discussion)
  $J_z =  \Delta S_z/2\pi$, and in consequence,
 $$
   J_z = \frac{2\sqrt{2}}{\pi} \int_0^Z\sqrt{E_z - \omega|z|}dz,
 $$
where $Z\equiv z_{\max}$ is the maximum value of $z$ along the orbit. From (\ref{Hz}) we have
$Z = E_z/\omega$ and, since $z$ is positive along the range of integration, we can write
  $J_z = 4\sqrt{2} E_z^{3/2}/3\pi \omega$ or, in terms of $Z$,
  \begin{equation}
   J_z = \frac{4\sqrt{2}}{3\pi}\omega^{1/2}Z^{3/2}.
  \end{equation}

The action-angle formalism described above for the $z$-part of the approximate Hamiltonian (\ref{Hz}) gives
exactly the period (\ref{zperiod}) and the amplitude (\ref{zamplitude}) for the vertical oscillations, as expected.

\subsection{Adiabatic invariance of $J_z$}

We now study the effects of adiabatic variations in the approximate potential of eq. (\ref{Hz}) (see section 3.6 of \cite{GD}). If we
consider a slow change in $\omega$ (see eqs. (\ref{Hz}) and (\ref{omega})),
  $$
   \omega ' = s\omega,
  $$
where $s\approx 1$ is a dimensionless quantity that varies slowly with time, adiabatic invariance of $J_z$ gives
  $
   \omega '^{1/2} Z'^{3/2} = \omega^{1/2} Z^{3/2}
  $, and we have
  $
   Z' = Zs^{-1/3}
  $, or equivalently,
  \begin{equation}\label{ZZ'}
   \frac{Z'}{Z} = \bigg(\frac{\omega}{\omega '}\bigg)^{1/3}.
  \end{equation}

Now consider $R$ as a function of time, which induces an ``effective'' time variation in the parameter $\omega$ (defined by (\ref{omega}))
appearing in (\ref{Hz}):
  \begin{equation}
   \omega(t) = 2\pi G \Sigma \big(R(t)\big).
  \end{equation}
Assuming that this effect is an adiabatic variation in the approximate potential of eq. (\ref{Hz}),
corresponding to the perturbed equatorial orbit in the effective potential (\ref{potaprox}),
eq. (\ref{ZZ'}) relates the maximum $z$-amplitudes of a given orbit at different values of $R$ through the following expression:
  \begin{equation}\label{ZZ'sigma}
   \frac{Z(R)}{Z(\tilde{R})} = \bigg(\frac{\Sigma(\tilde{R})}{\Sigma(R)}\bigg)^{1/3}.
  \end{equation}
This relation determines the ``envelope'' of the orbit in the meridional plane.
Therefore we expect that if the characteristic frequencies of the vertical perturbations are much higher than the frequencies of the
$R$-perturbations, then this approximation will be valid and the maximum
$z$-amplitude of motion will be given by eq. (\ref{ZZ'sigma}).
This result can be compared with the ``envelopes'' of (numerically calculated) 3D orbits in a given potential $\Phi_{eff}$ near
its minimum. Eq. (\ref{ZZ'sigma}) must also be compared with the corresponding
relation for smooth potentials (eq. (3.279) of \cite{GD}).

\subsection{Numerical Experiments}\label{sec:numerical}

In order to illustrate the applicability of the ideas sketched above, we perform simulations for the motion
of test particles around mass distributions of the form (\ref{totaldensity}). We shall focus
on the validity of relation (\ref{ZZ'sigma}); for this reason we present a number of situations
in which the orbits integrated are far from being considered as  ``nearly equatorial'' or ``nearly circular''.

\begin{figure*}
$$
\begin{array}{ccc}
  \epsfig{width=8cm,file=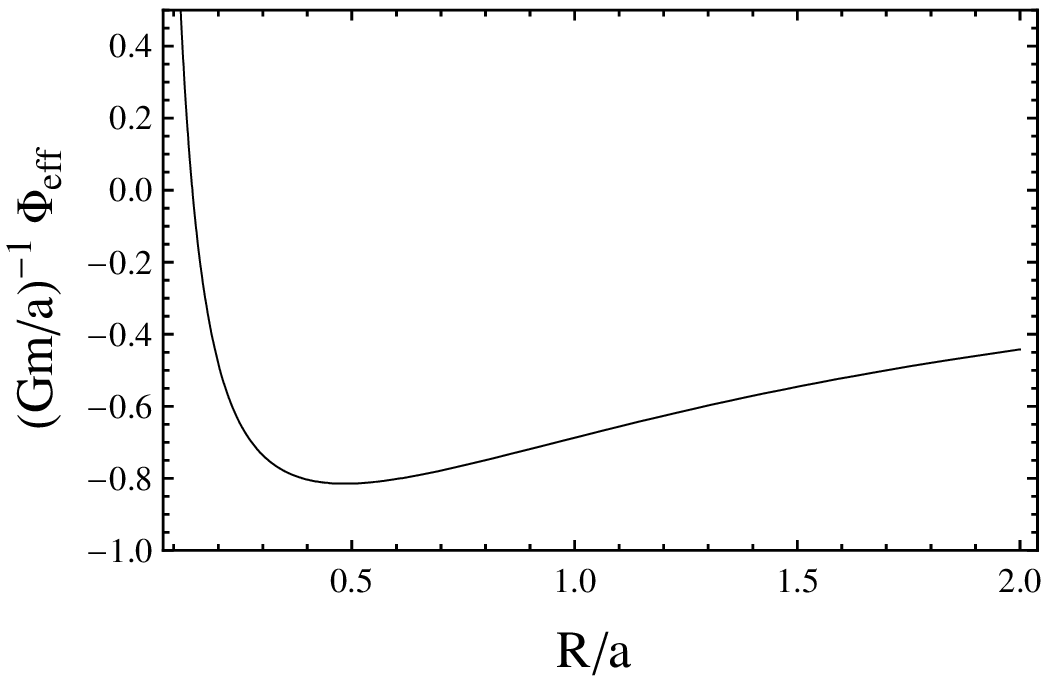} & \epsfig{width=8cm,file=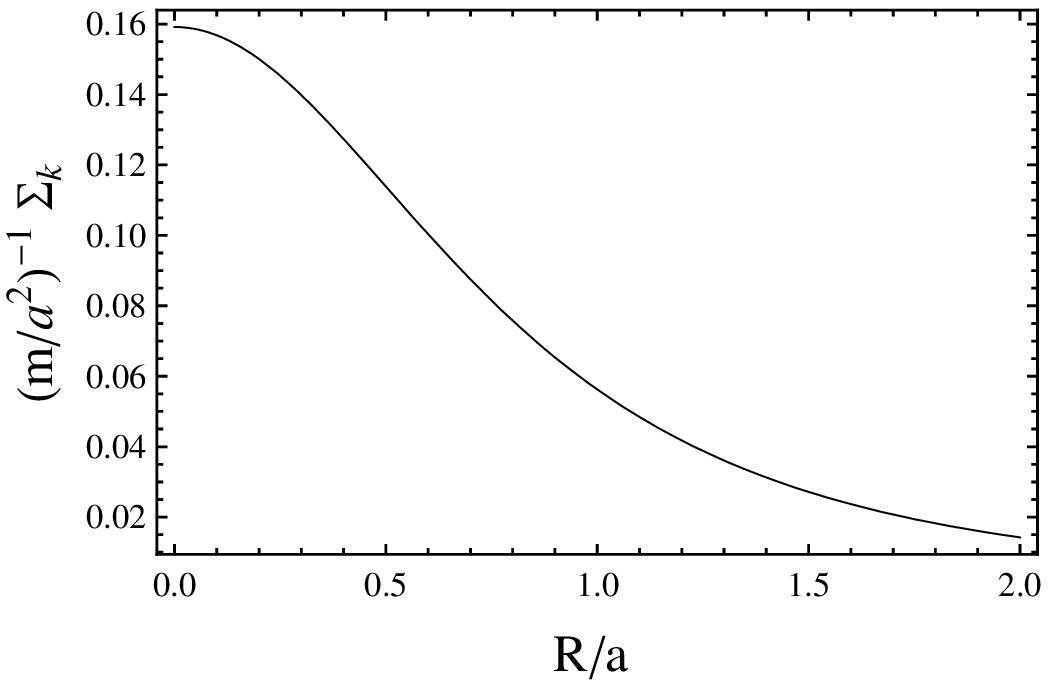}  \\
  \end{array}
$$
\caption{Kuzmin disk: Dimensionless effective potential in the equatorial plane for $\ell/\sqrt{aGm}=0.2$ (left side) and dimensionless surface density
(right side).}
\label{fig:kV}
\end{figure*}

\begin{figure*}
$$
\begin{array}{ccc}
  \epsfig{width=8cm,file=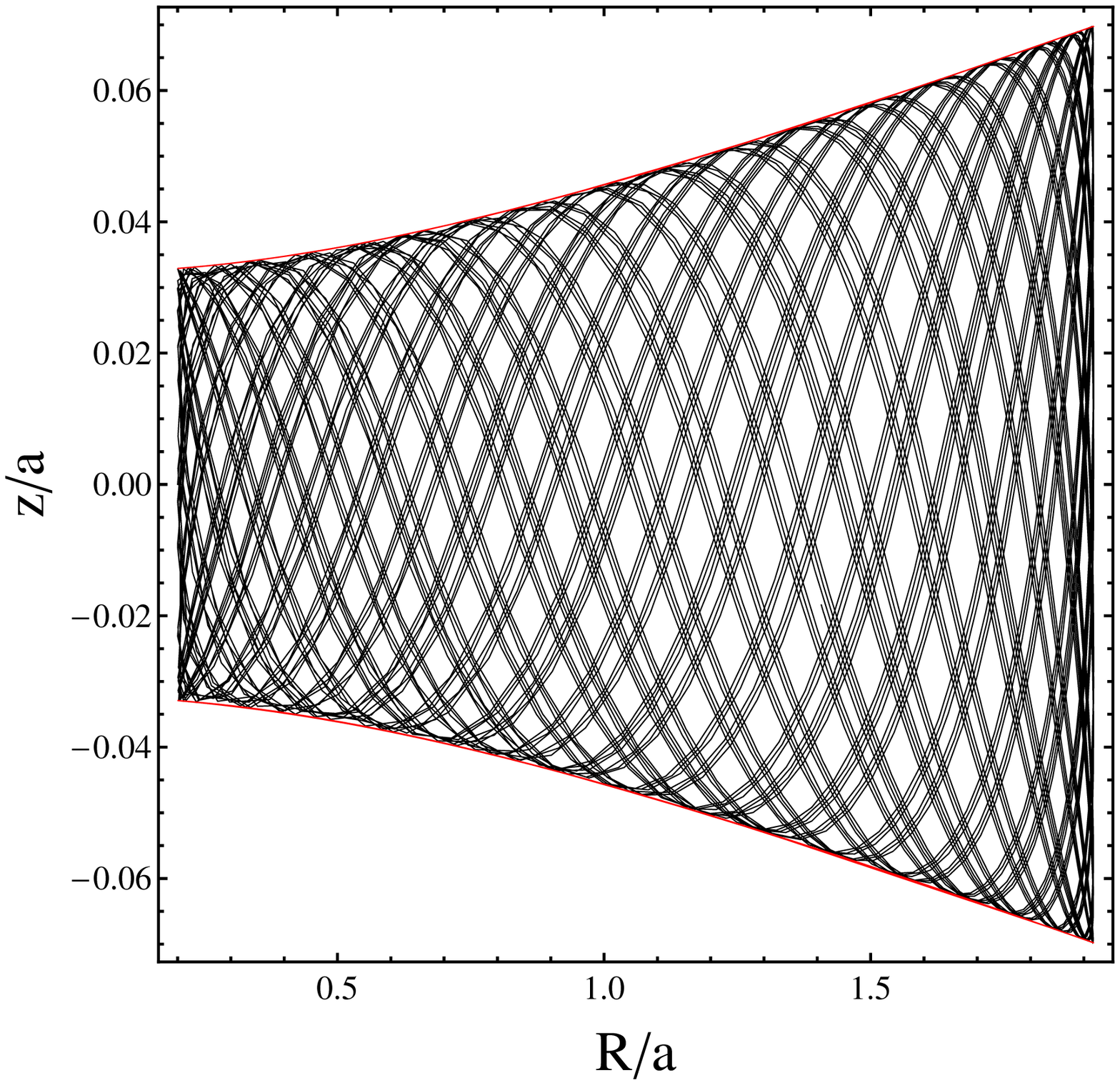} & \epsfig{width=8cm,file=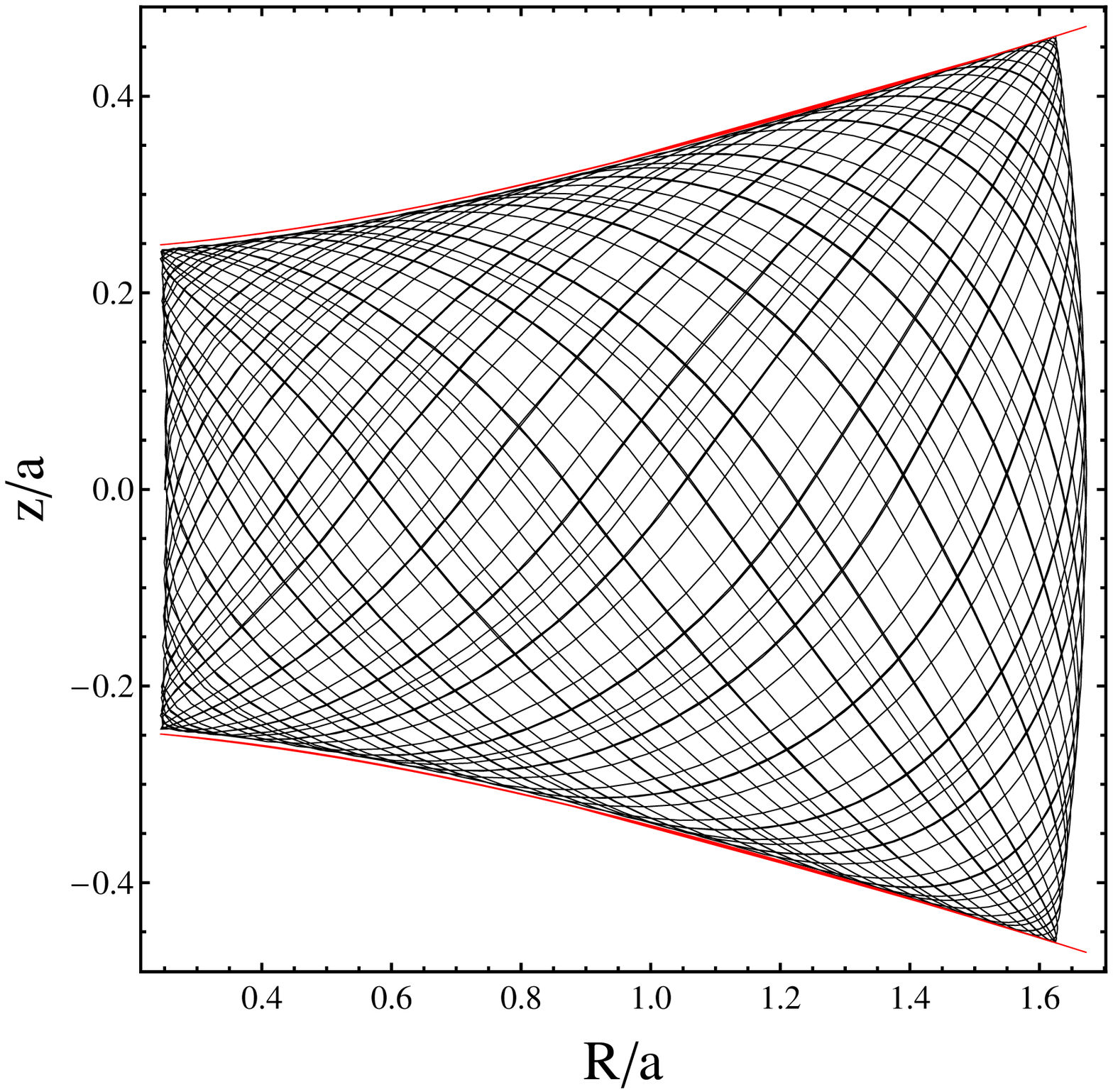}  \\
  \end{array}
$$
\caption{Motion in the meridional plane around a Kuzmin disk with $aE/Gm=-0.45$, $\ell/\sqrt{aGm}=0.2$ and
initial conditions $z/a=10^{-15}$,  $P_{R}=0$ , R/a=0.2 (left), R/a=0.25 (right).
The prediction of eq. (\ref{ZZ'sigma}) is in red and the meridional-plane orbit is in black.
The orbit in the left panel can be considered as a nearly equatorial orbit and its vertical amplitude is very well modeled by the red line.
The same happens with the orbit in the right panel, although its vertical amplitude it is not near the equatorial plane.}
\label{fig:ko}
\end{figure*}

At first, we present the results of numerically calculated orbits in
two cases: (i) the Kuzmin disk (fig. \ref{fig:ko}) and (ii) the Kuzmin disk surrounded by a Plummer halo
(fig. \ref{fig:kpo}). The APDP for the Kuzmin disk is given by
  \begin{equation}
   \Phi_k = - \frac{Gm}{\sqrt{R^2 + (|z|+a)^2}},
  \end{equation}

  \begin{equation}
   \Sigma_k = \frac{m}{2\pi a^{2}}\left(1+\frac{R^2}{ a^2} \right)^{-3/2},
  \end{equation}
and the APDP for the Plummer halo is
  \begin{equation}
   \Phi_p = - \frac{GM}{\sqrt{r^2 + b^2}},
  \end{equation}

  \begin{equation}
   \rho_p = \frac{3M}{4\pi b^3}\left(1+\frac{r^2}{b^2}\right)^{-5/2},
  \end{equation}
where $r^2 = R^2 + z^2$ (\cite{GD}).
In the case (i) we choose $\ell/\sqrt{aGm}=0.2$ and $aE/Gm=-0.45$ for numerical integrations.
For this value of angular momentum the effective potential has a minimum near $R/a=0.5$ (see fig. \ref{fig:kV})
and for this value of energy we can obtain disk-crossing orbits in the region $0<R/a<2$. We find that there is a great
number of orbits obeying the relation (\ref{ZZ'sigma}), even in regions far away from the critical point (i.e. corresponding to
the radius of the circular orbit), where the approximation (\ref{Aphiapprox}) presumably should not be valid.
We remark that in all of these cases, the test particle passes through (or very near) the critical point, at least one time. Figure \ref{fig:ko} shows
two examples of this fact by plotting the motion in the meridional plane of two orbits along with the orbit's envelope computed from
(\ref{ZZ'sigma}). Note that the prediction of (\ref{ZZ'sigma}) holds well beyond the vicinity of
the thin disk, as it can be viewed in the right panel of fig. \ref{fig:ko}.

\begin{figure*}
$$
\begin{array}{ccc}
  \epsfig{width=8cm,file=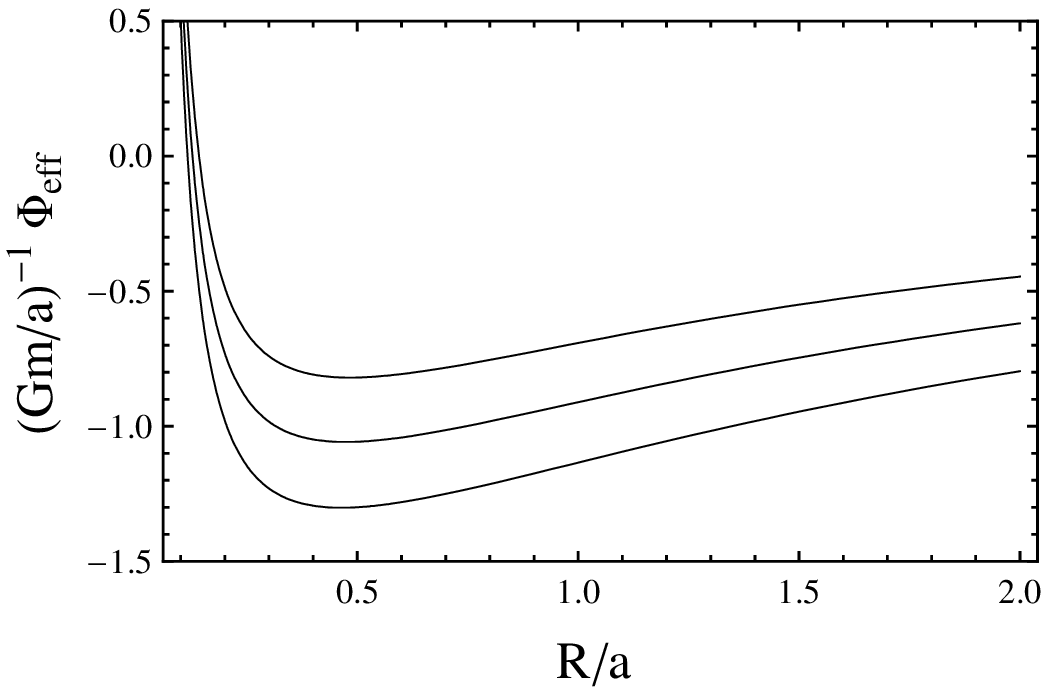} & \epsfig{width=8cm,file=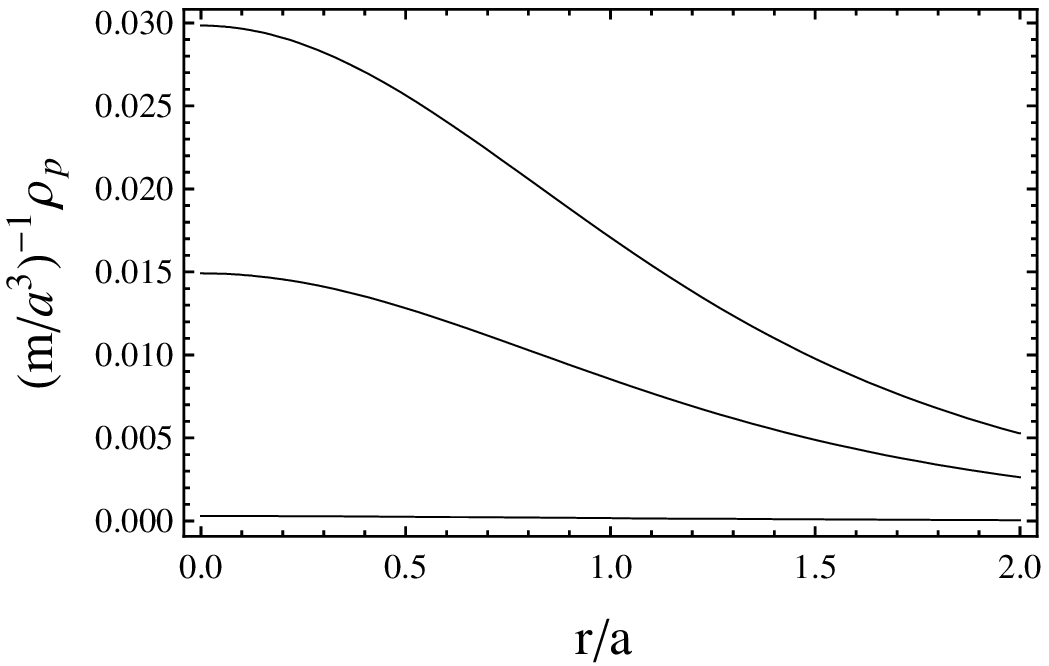}  \\
  \end{array}
$$
\caption{Kuzmin disk + Plummer halo with $b/a=2$: a) Effective potential in the equatorial plane, using $\ell/\sqrt{aGm}=0.2$ and
$M/m=0.01,0.5,1$ (from top to bottom). b) Density of the Plummer halo for $M/m=0.01,0.5,1$ (from bottom to top).
The surface density of the Kuzmin disk is given in the right panel of fig. \ref{fig:kV}.}
\label{fig:kpV}
\end{figure*}

\begin{figure*}
$$
\begin{array}{ccc}
  \epsfig{width=5.4cm,file=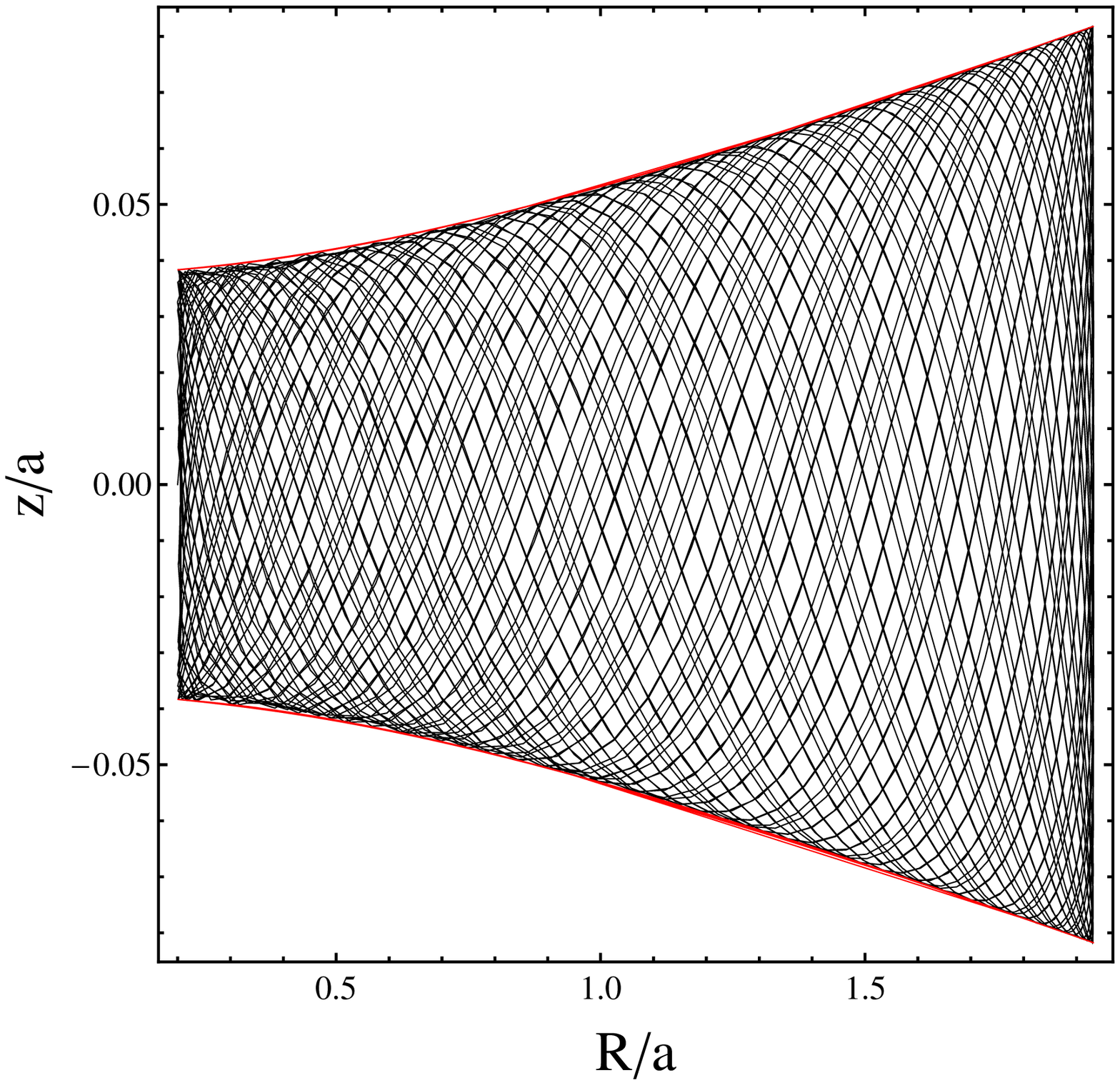} & \epsfig{width=5.4cm,file=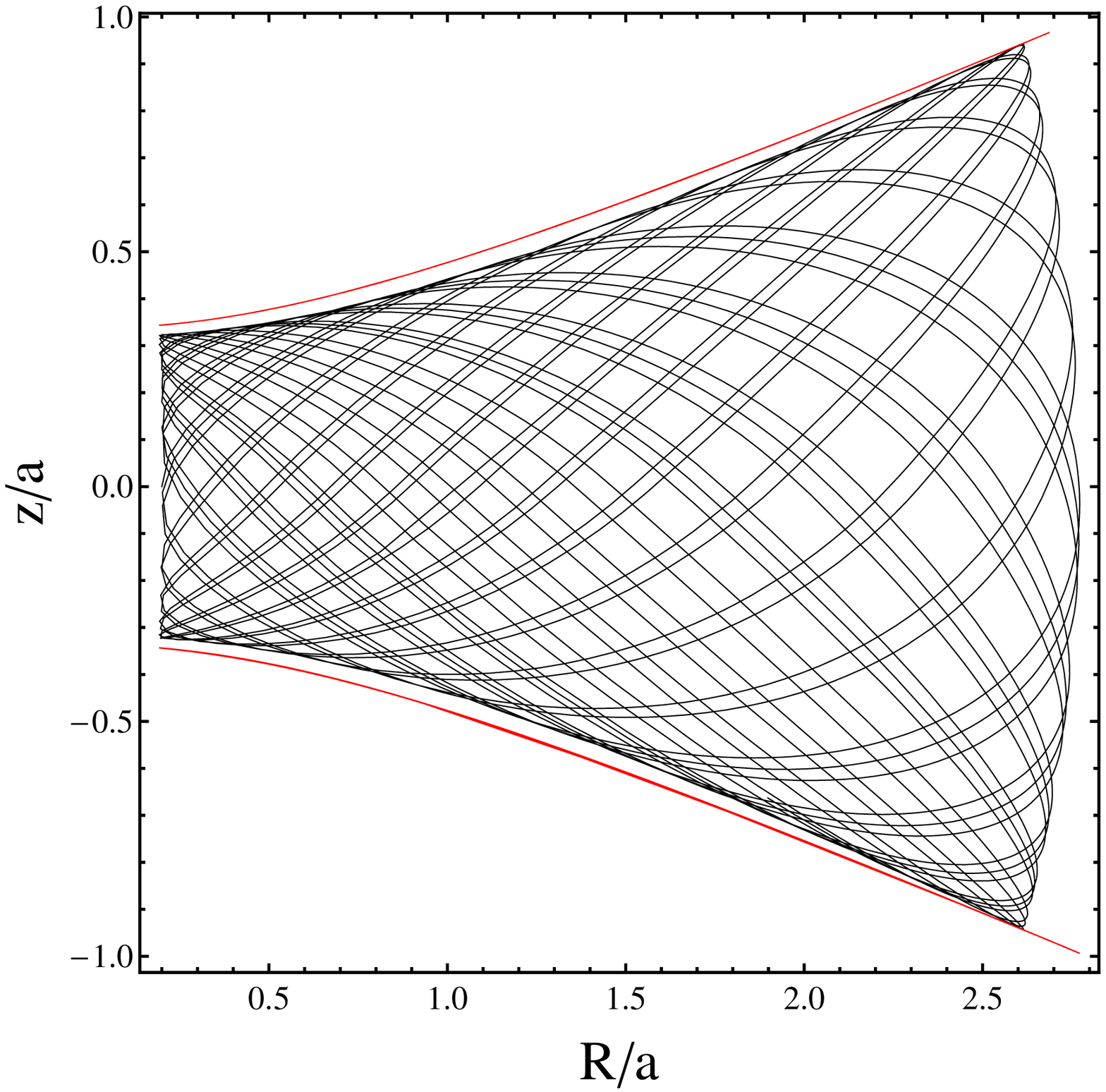} \epsfig{width=5.4cm,file=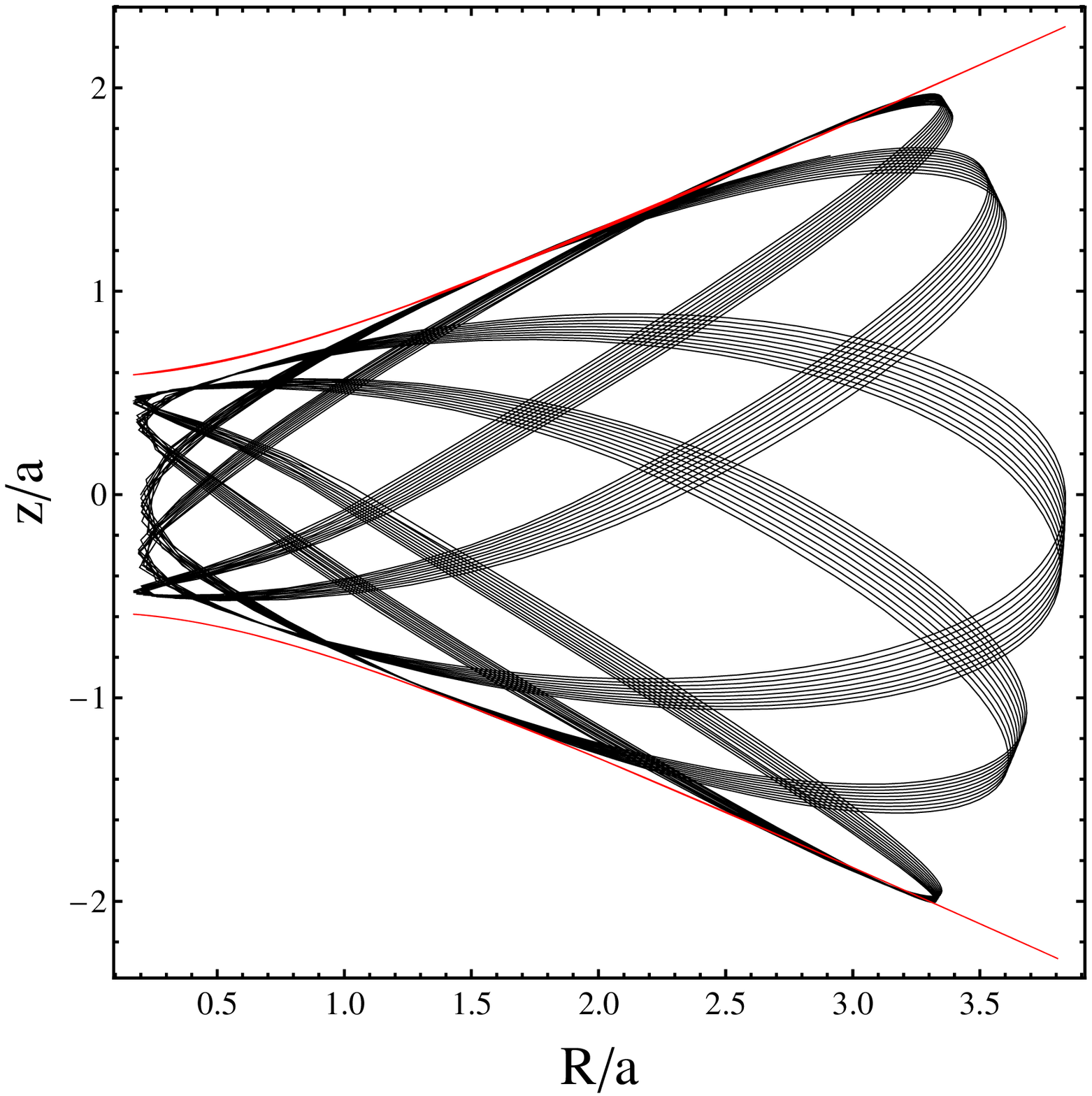} \\
  \end{array}
$$
\caption{Orbits in Kuzmin + Plummer potential, with $aE/Gm=-0.45$, $\ell/\sqrt{aGm}=0.2$, initial conditions
$R/a=0.2$, $z/a=10^{-15}$, $P_{R}=0$
and using the same parameters of Fig. \ref{fig:kpV}.
The prediction of eq. (\ref{ZZ'sigma}) is in red for $M/m=0.01,0.5,1.0$, from the left to the right.}
\label{fig:kpo}
\end{figure*}

\begin{figure*}
$$
\begin{array}{ccc}
  \epsfig{width=8cm,file=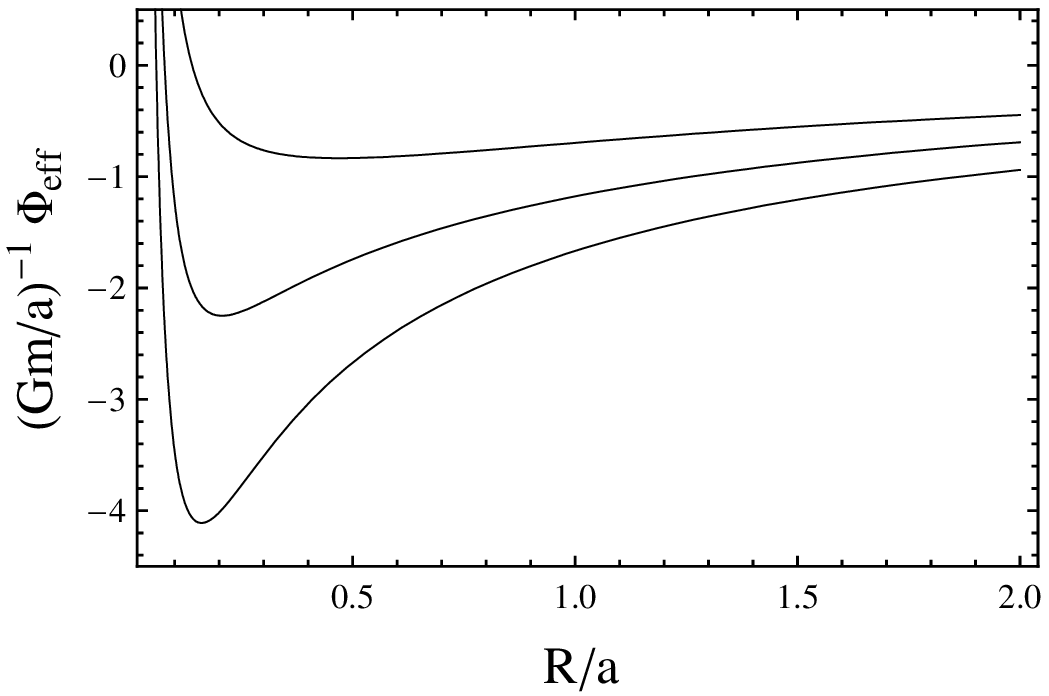} & \epsfig{width=8cm,file=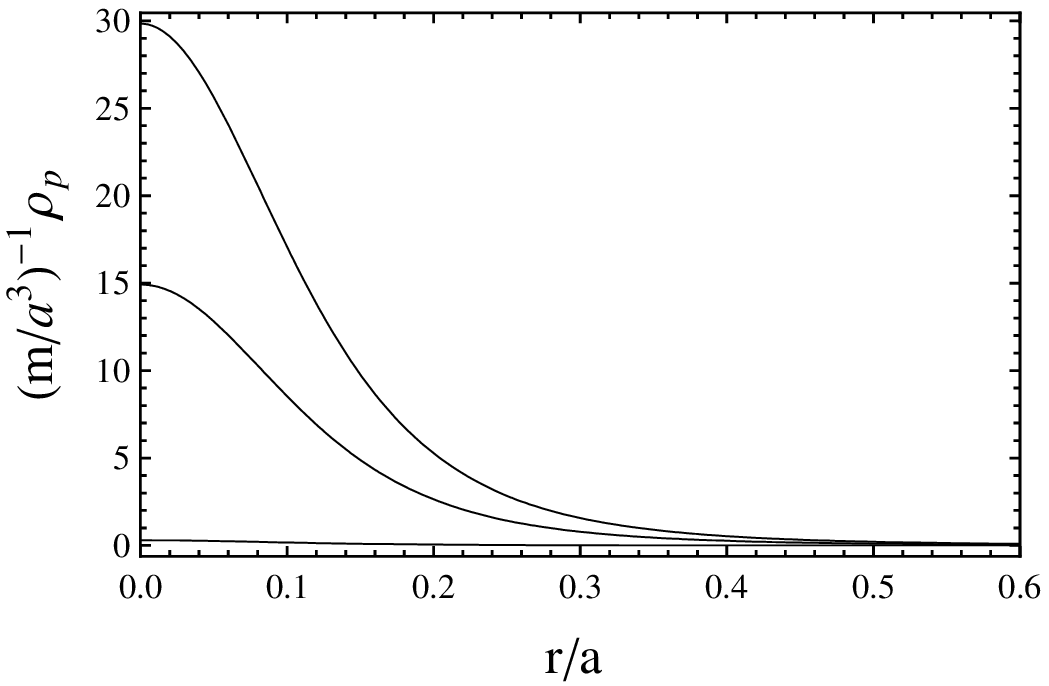}  \\
  \end{array}
$$
\caption{Kuzmin disk + Plummer halo with $b/a=0.2$: a) Effective potential in the equatorial plane, using $\ell/\sqrt{aGm}=0.2$ and
$M/m=0.01,0.5,1$ (from top to bottom). b) Density of the Plummer halo for $M/m=0.01,0.5,1$ (from bottom to top).
The surface density of the Kuzmin disk is given in fig. \ref{fig:kV}b.}
\label{fig:kpV2}
\end{figure*}

In the case (ii), with the addition of the Plummer halo, we consider three situations:  $M/m=0.01,0.5,1$, in order to account gradually
the contribution of this component in the orbits' behavior. First we choose an extended halo ($b/a=2$) and
then a more concentrated one ($b/a=0.2$), as it is shown in figures \ref{fig:kpV} and \ref{fig:kpV2}, respectively. By maintaining the same values
of energy and angular momentum as in the
above case, we obtain the meridional-plane orbit for each ratio $M/m$. The results of the computation for $b/a=2$ are shown in fig \ref{fig:kpo}, from
which we can see small deviations from the predictions of (\ref{ZZ'sigma}) when $M/m=0.5$ and $M/m=1$
(for $M/m=0.01$ the test particle describes a  nearly equatorial orbit which is very well modeled by (\ref{ZZ'sigma})).
It is remarkable that such deviations are not
very significant when the halo mass is of the order of the disk mass and the amplitude of vertical oscillations
is comparable with the effective size of the disk.

\begin{figure*}
$$
\begin{array}{ccc}
  \epsfig{width=5.4cm,file=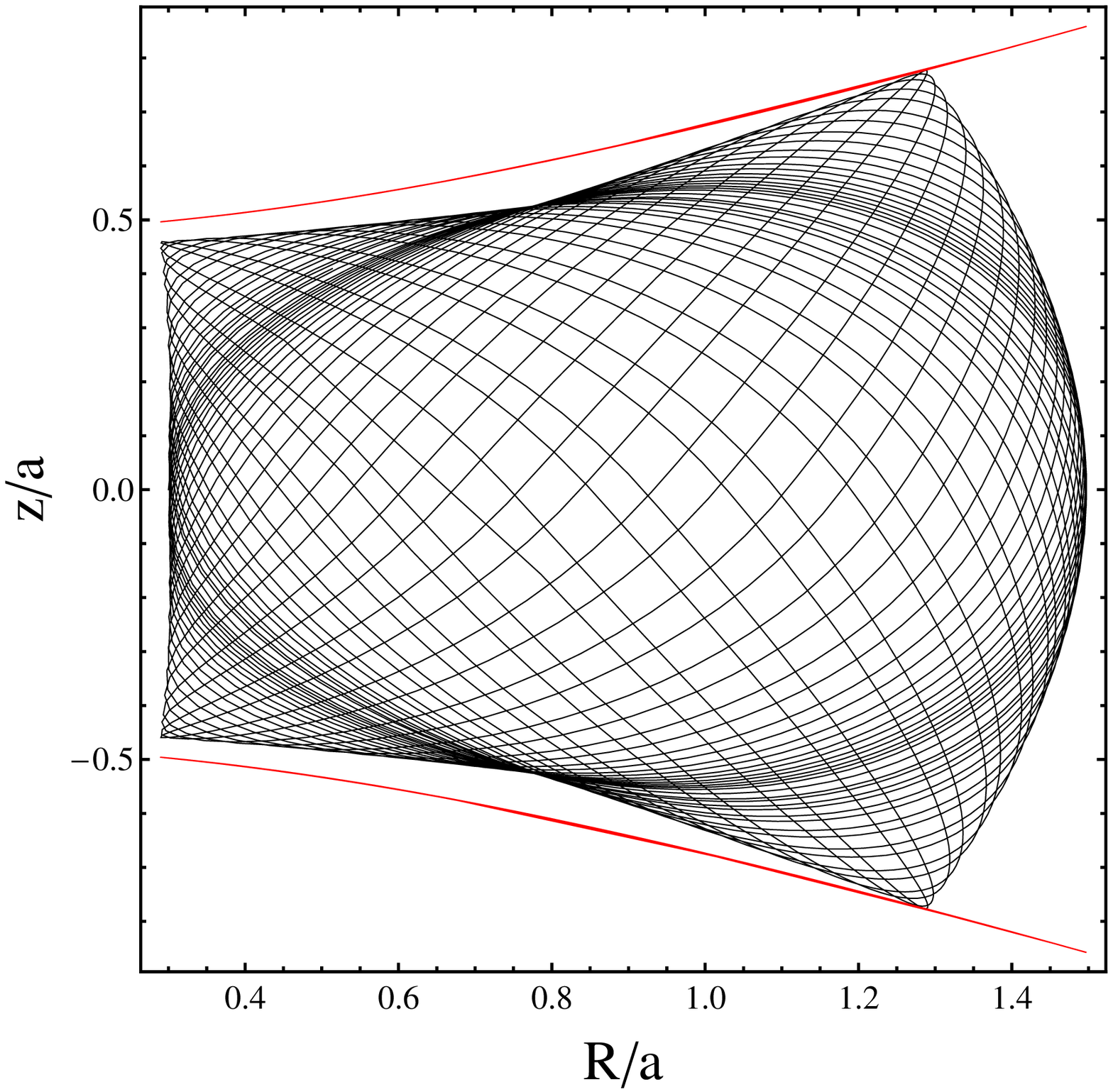} & \epsfig{width=5.4cm,file=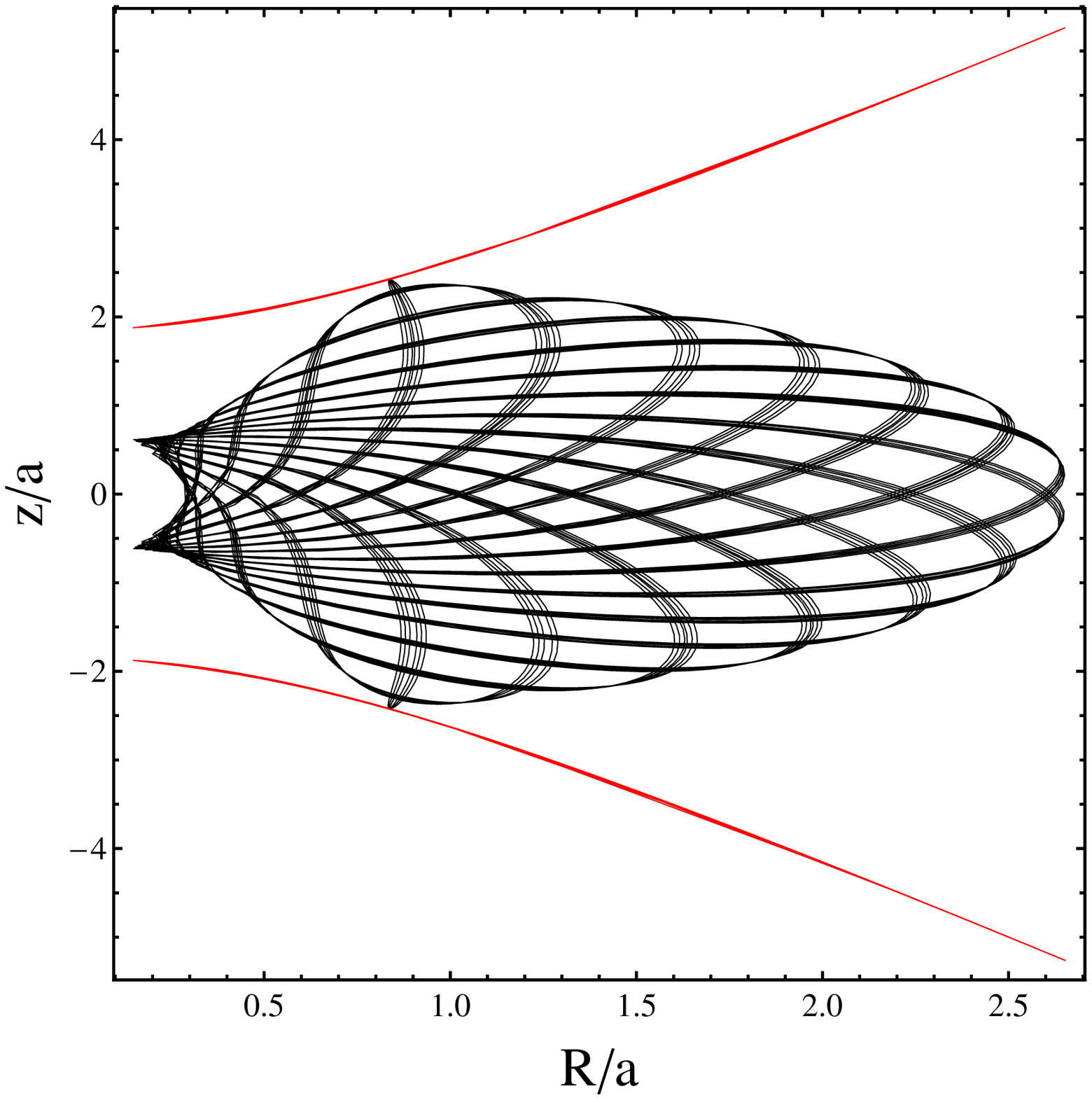} \epsfig{width=5.4cm,file=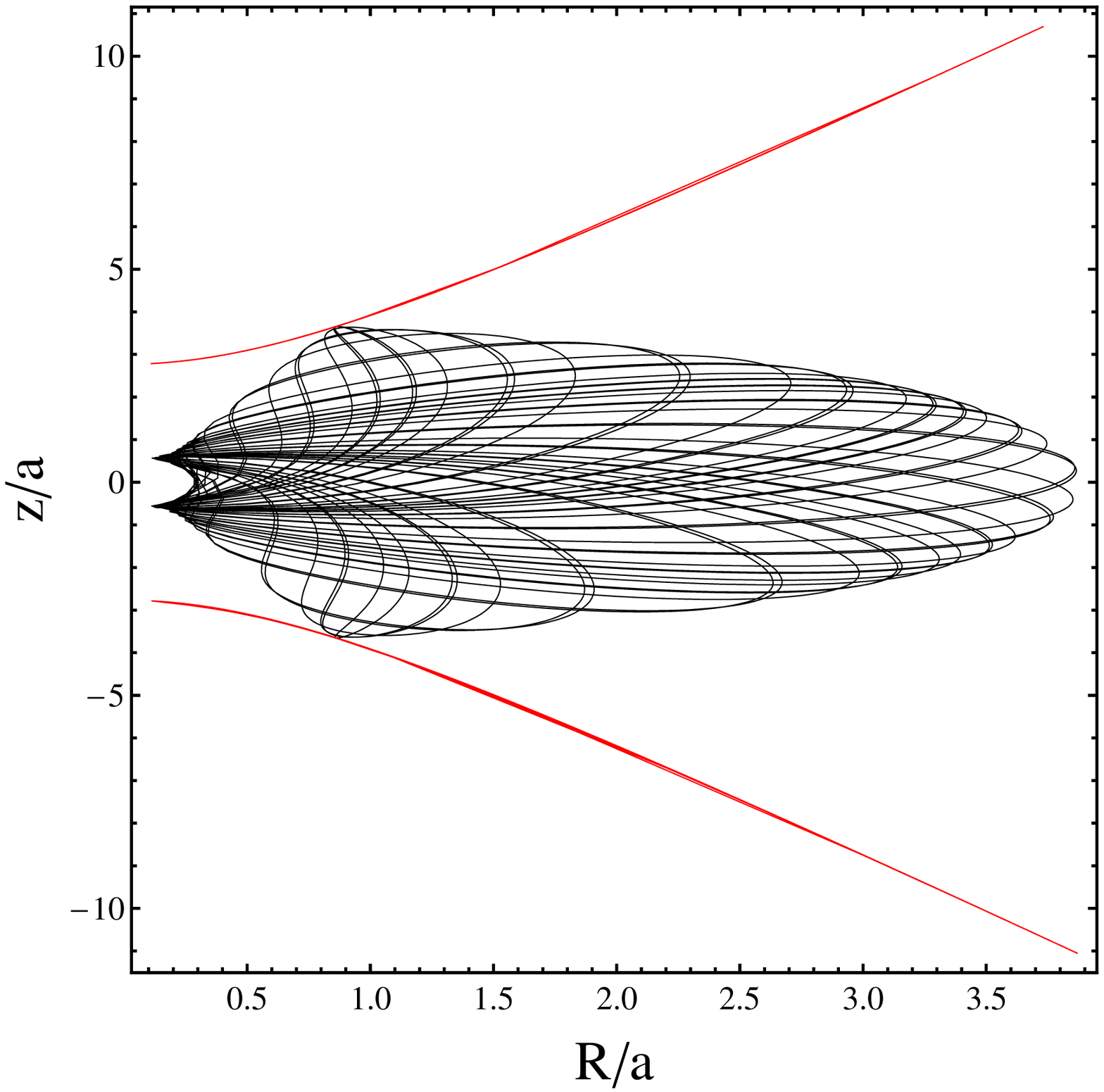} \\
  \end{array}
$$
\caption{Orbits in Kuzmin + Plummer potential, with
$aE/Gm=-0.45$, $\ell/\sqrt{aGm}=0.2$, initial conditions $R/a=0.3$, $z/a=10^{-15}$, $P_{R}=0$ and using the same
parameters of Fig. \ref{fig:kpV2}.
The prediction of eq. (\ref{ZZ'sigma}) is in red for $M/m=0.01,0.5,1$, from the left to the right.}
\label{fig:kpo2}
\end{figure*}

In contrast, the results of the computation for $b/a=0.2$, which are shown in fig. \ref{fig:kpo2}, reveal significant deviations from the prediction of
(\ref{ZZ'sigma}). This is due to the fact that, in all of the situations illustrated, the orbit does not pass through the critical point
of the effective potential (see fig. \ref{fig:kpV2}). However, note that for $M/m=0.01$ the orbit passes (say) near
the critical point and the deviation is significantly smaller than in the other two cases.


\begin{figure*}
$$
\begin{array}{cc}
  \epsfig{width=8cm,file=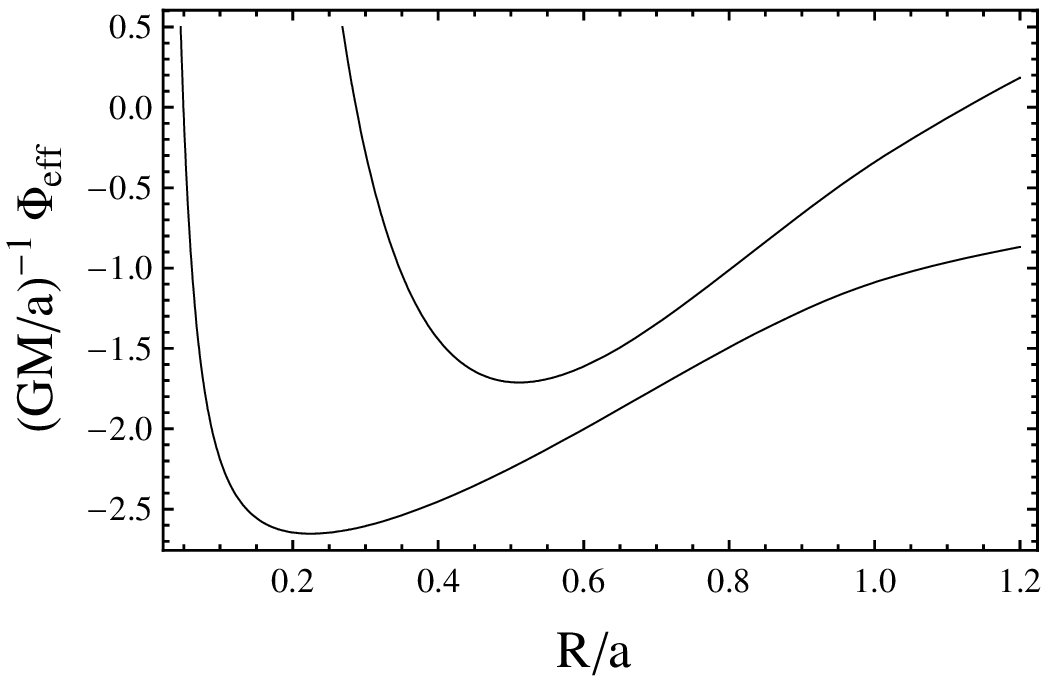} & \epsfig{width=8cm,file=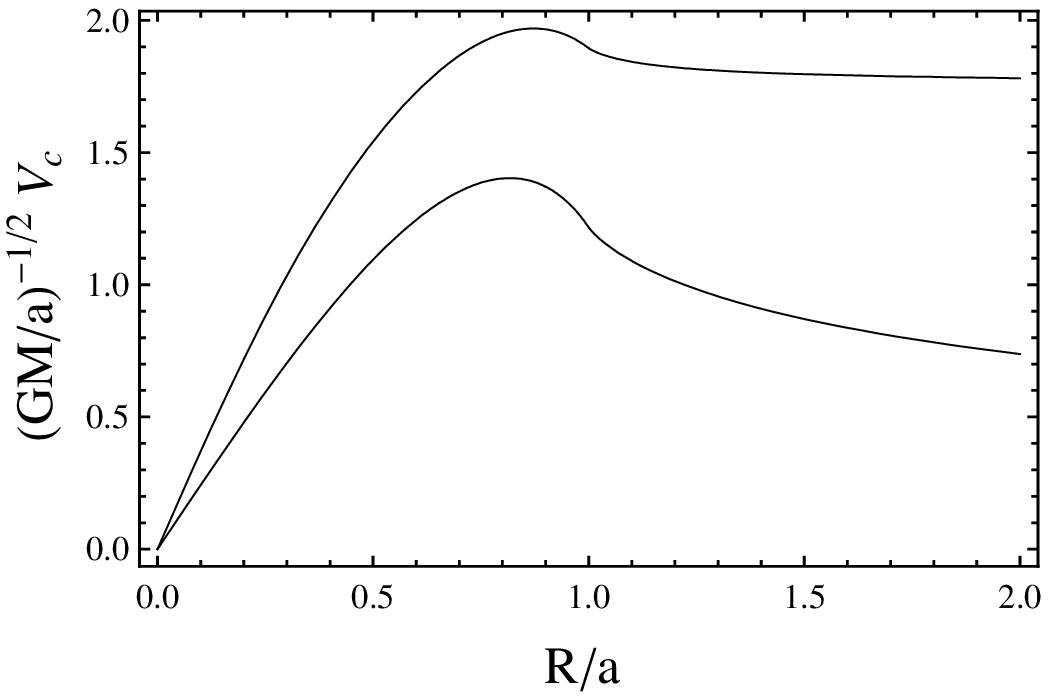}
\end{array}
$$
\caption{Model $\Phi_{K2}+\Phi_{H}$.Left panel: Dimensionless effective potential for $\ell/\sqrt{aGM}=0.12$, $V\sqrt{a/GM}=0.1$, $d/a=1.2$ (lower curve)
and $\ell/\sqrt{aGM}=0.8$, $V\sqrt{a/GM}=1.2$,
$d/a=0.6$ (upper curve). Right panel: Dimensionless Circular velocity for $\ell/\sqrt{aGM}=0.12$, $V\sqrt{a/GM}=0.1$, $d/a=1.2$ (lower curve)
and $\ell/\sqrt{aGM}=0.8$, $V\sqrt{a/GM}=1.2$, $d/a=0.6$ (upper curve).} \label{fig:potVcK2halo}
\end{figure*}

As a second example, we perform  numerical calculations with a mathematically  more involved model:
the second member of the generalized Kalnajs disks (\cite{gonzalez-reina}) immersed in a
 halo's spherical logarithmic potential. The surface mass density of the thin disk is a monotonically
decreasing function,
$\Sigma = 5M/(2\pi a^{2})( 1 - R^{2}/a^{2})^{3/2}$, where $a$ is the radius of the disk (it is a disk with
finite extension, as in subsection \ref{sec:FTDModels}) and $M$ its total mass. The corresponding
gravitational potential can be cast in oblate spheroidal coordinates, $\xi=a^{-1}\mbox{Re}[\sqrt{R^{2}+(z-\mbox{i}a)^{2}}]$,
$\eta=-a^{-1}\mbox{Im}[\sqrt{R^{2}+(z-\mbox{i}a)^{2}}]$, through the relation
\begin{eqnarray}
\Phi_{K2} &=& - \frac{GM}{a} \left[ \cot^{-1} \xi + A
(3\eta^{2} - 1)\right. \nonumber \\
 & & \left.\quad \quad \quad \quad \quad \quad\quad + B ( 35 \eta^{4} - 30 \eta^{2} + 3)\right], \label{eq:4.23}
\end{eqnarray}
with
\begin{subequations}\begin{align}
A &= \frac{5}{14} \left[(3\xi^{2} + 1) \cot^{-1} \xi - 3 \xi \right], \\
B &= \frac{3}{448} \left[ (35 \xi^{4} + 30 \xi^{2} + 3) \cot^{-1} \xi - 35 \xi^{3} -
\frac{55}{3} \xi \right].
\end{align}\end{subequations}
This potential leads to a Keplerian rotation curve, in contrast with
the first member of the family (the well known Kalnajs disk), which describes a configuration rotating as a rigid body.
On the other hand, the potential modeling the spherical halo  is given by
\begin{equation}\label{pothalo}
    \Phi_{H}=V^{2}\ln(R^{2}+z^{2}+d^{2})
\end{equation}
where $V$ and $d$ are  parameters to be determined according to the observational data (for example, the galactic rotation curve).
This component was used by \cite{johnston} to simulate the motion of dwarf galaxies around the Milky Way. Here we present the results of numerical
experiments of the test-particle motion in the potential $\Phi_{K2}+\Phi_{H}$,
by considering two situations: (i) A case in which the disk component is dominant, where $V\sqrt{a/GM}=0.1$, $d/a=1.2$, for the halo, and
$\ell/\sqrt{aGM}=0.12$, $Ea/GM=-1$ and $Ea/GM=-2$, for the orbits;
(ii) a case characterized by a dominant halo component, with $V\sqrt{a/GM}=1.2$, $d/a=0.6$, and  $\ell/\sqrt{aGM}=0.8$, $Ea/GM=-0.5$ for orbits.
The effective potential and rotation curve corresponding to both cases are shown in fig. \ref{fig:potVcK2halo}.

\begin{figure*}
$$
\begin{array}{cc}
  \epsfig{width=8cm,file=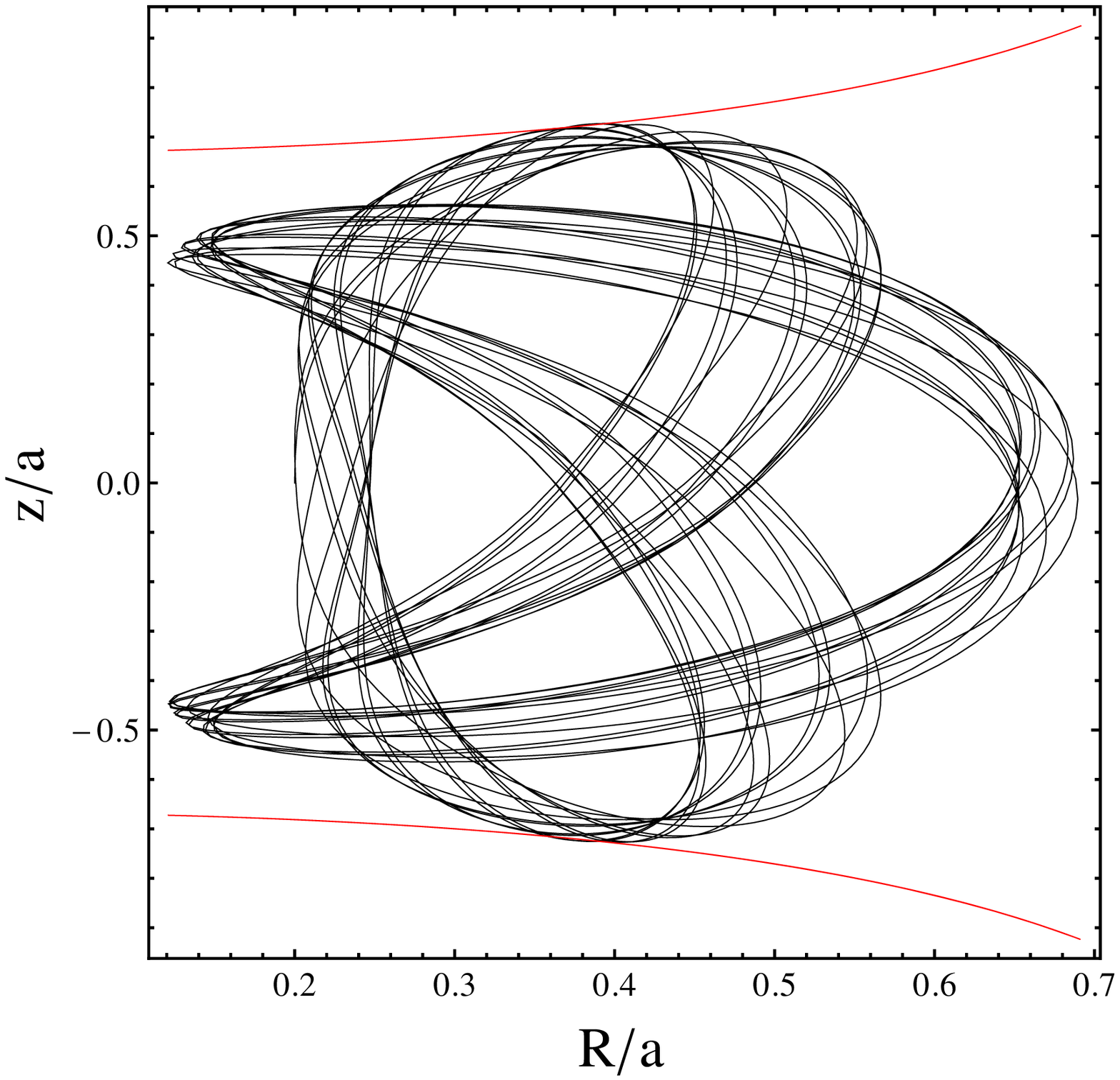} & \epsfig{width=8cm,file=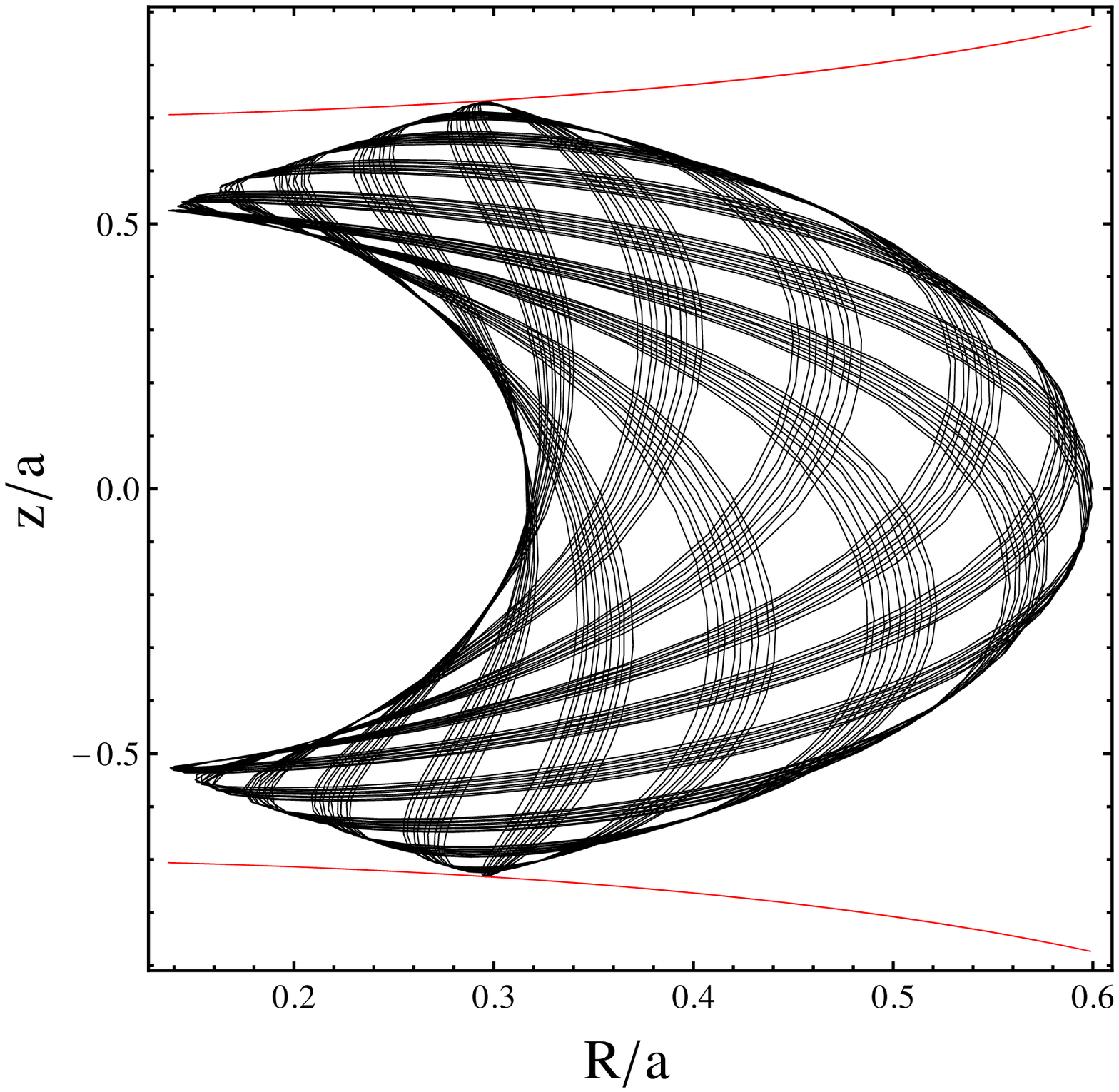}\\
  \epsfig{width=8cm,file=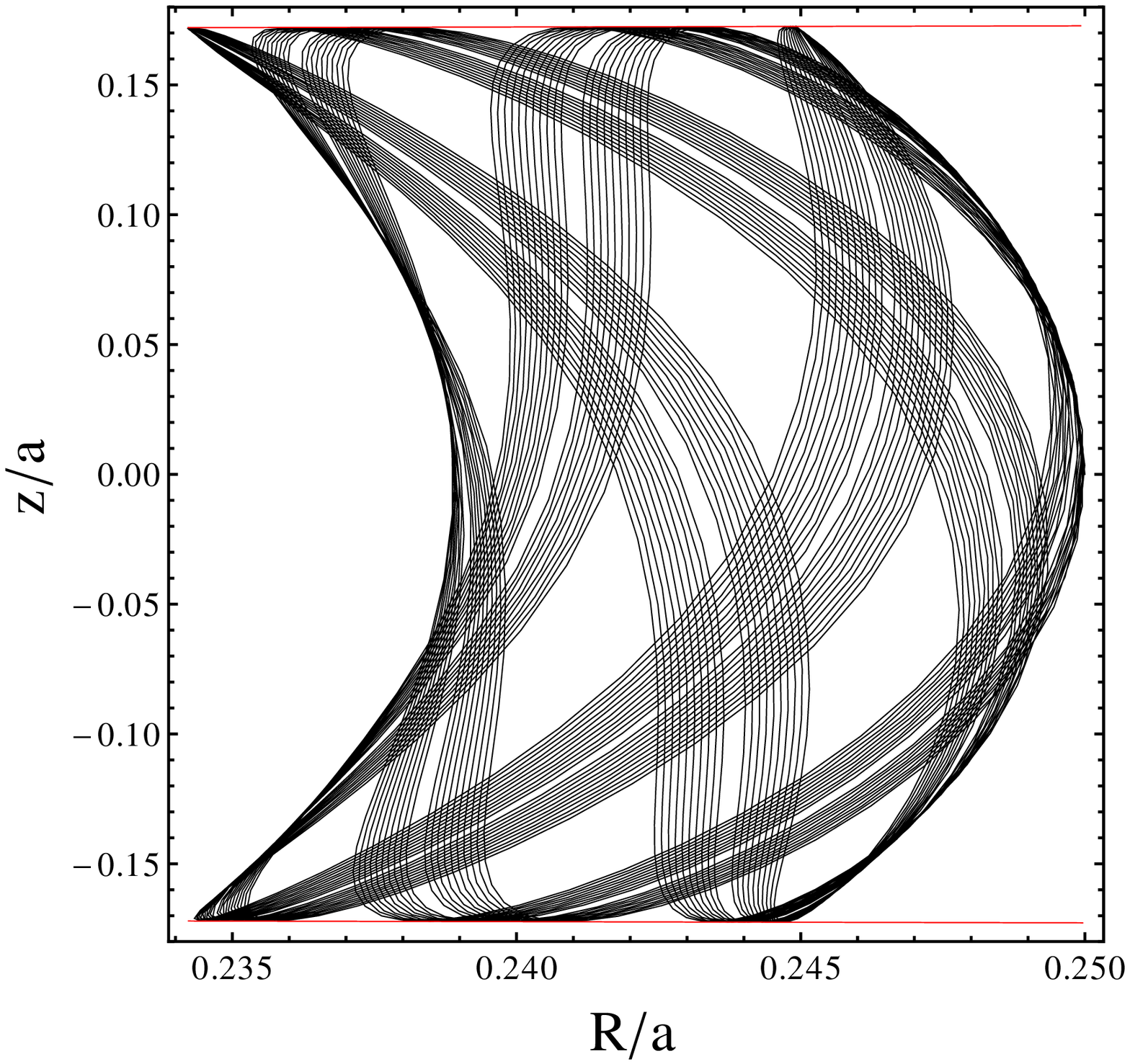} & \epsfig{width=8cm,file=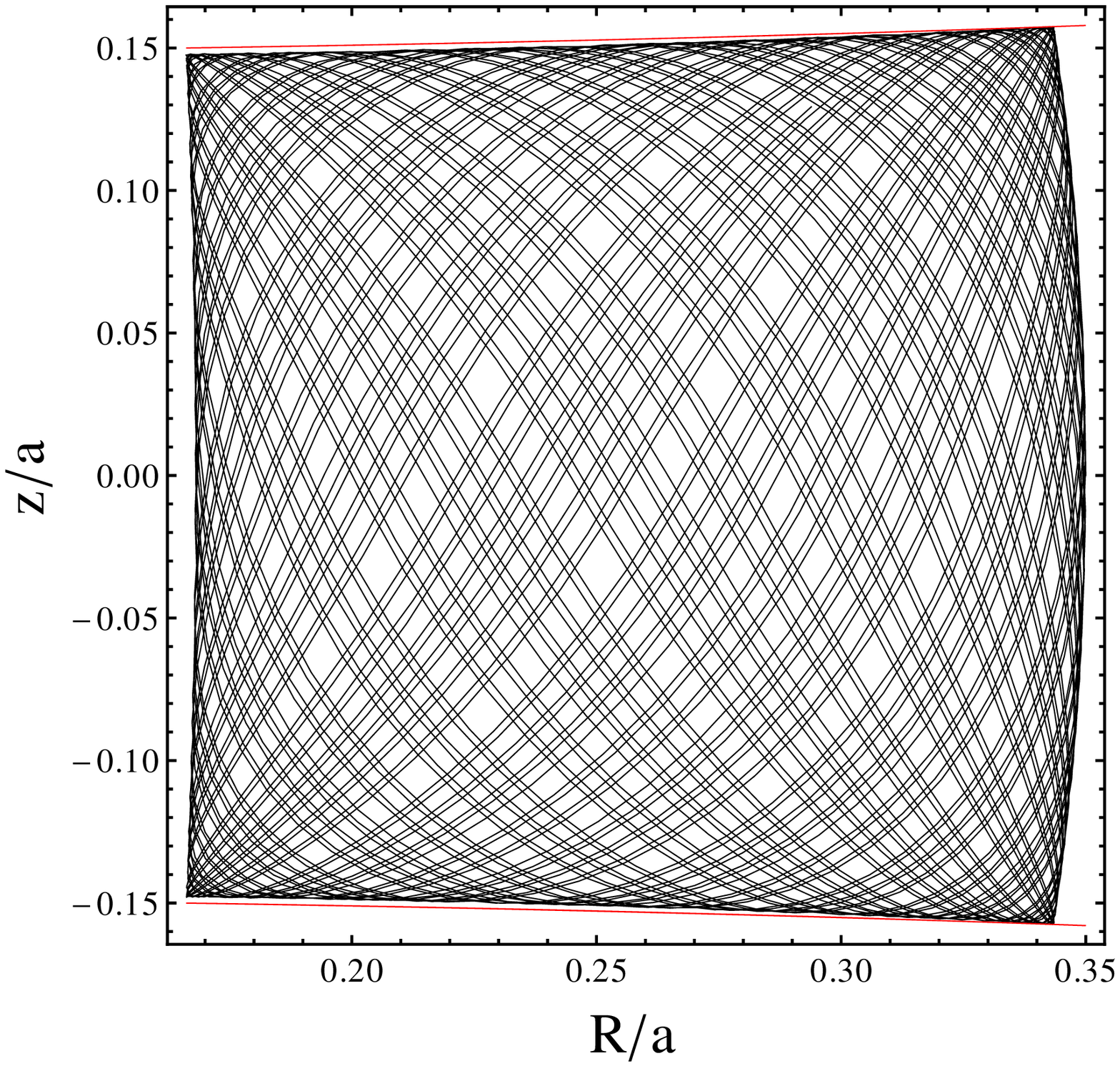}
\end{array}
$$
\caption{Orbits in the meridional plane  for the model $\Phi_{K2}+\Phi_{H}$ with $\ell/\sqrt{aGM}=0.12$, $V\sqrt{a/GM}=0.1$, $d/a=1.2$.
Top panels: $Ea/GM=-1$, $R/a=0.2$ (left) and $R/a=0.6$ (right). Bottom panels $Ea/GM=-2$, $R/a=0.25$ (left) and $R/a=0.35$ (right).
In all panels, $z/a = 10^{-15}$. } \label{fig:orbitK2halo}
\end{figure*}

In the case (i) we obtain results similar to the reported ones for the Kuzmin-Plummer potential. Working with  $Ea/GM=-1$, we find orbits which
do not pass through the critical point of the effective potential and, in consequence, do not follow the prediction of eq. (\ref{ZZ'sigma}) (top pannels
of fig. \ref{fig:orbitK2halo}). In contrast, we find that for $Ea/GM=-2$, there is a great number of orbits whose vertical amplitude is described
by (\ref{ZZ'sigma}) with high precision. In the bottom panels of fig.  \ref{fig:orbitK2halo} we show two of them.

\begin{figure*}
$$
\begin{array}{cc}
  \epsfig{width=8cm,file=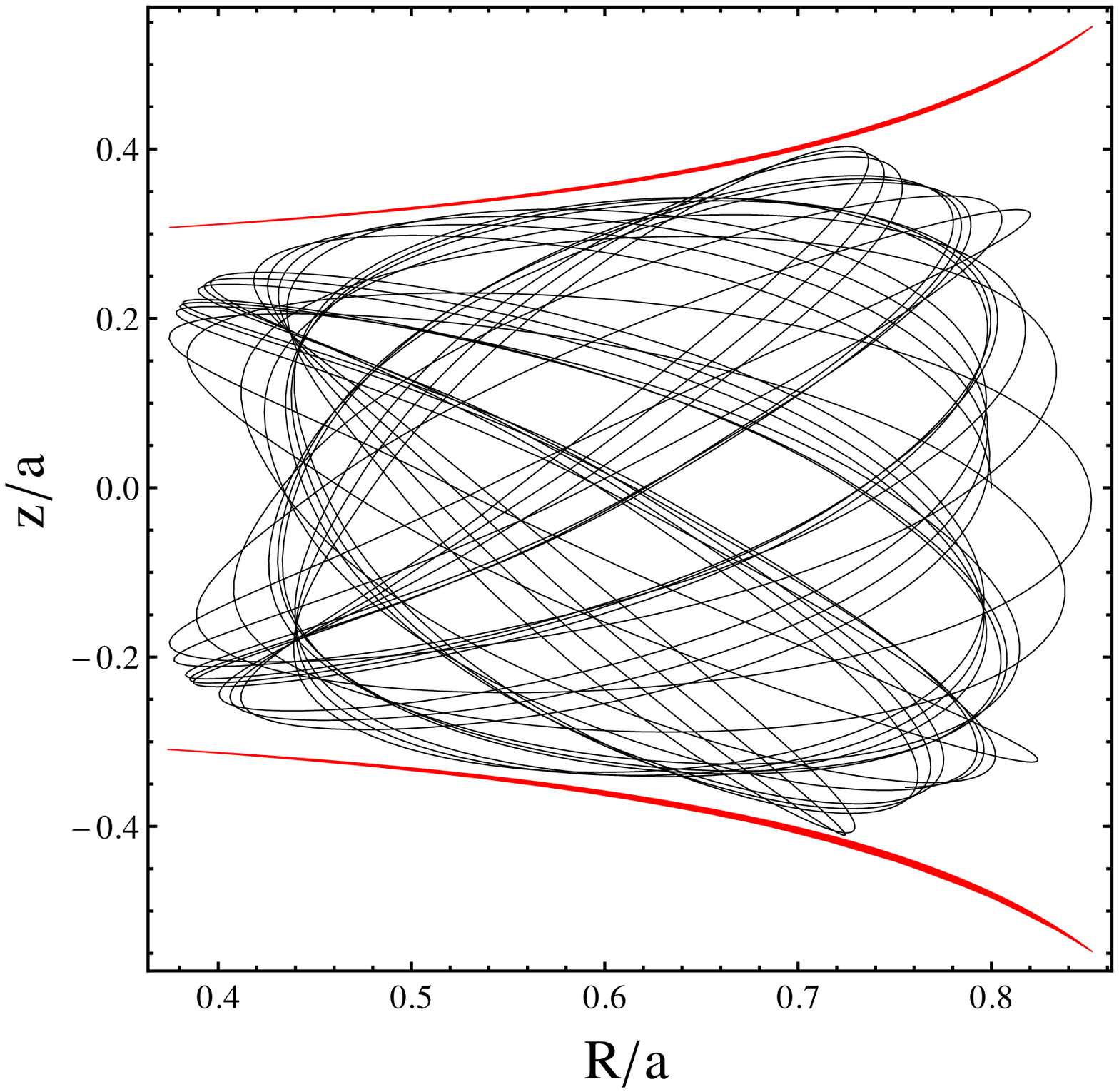} & \epsfig{width=8cm,file=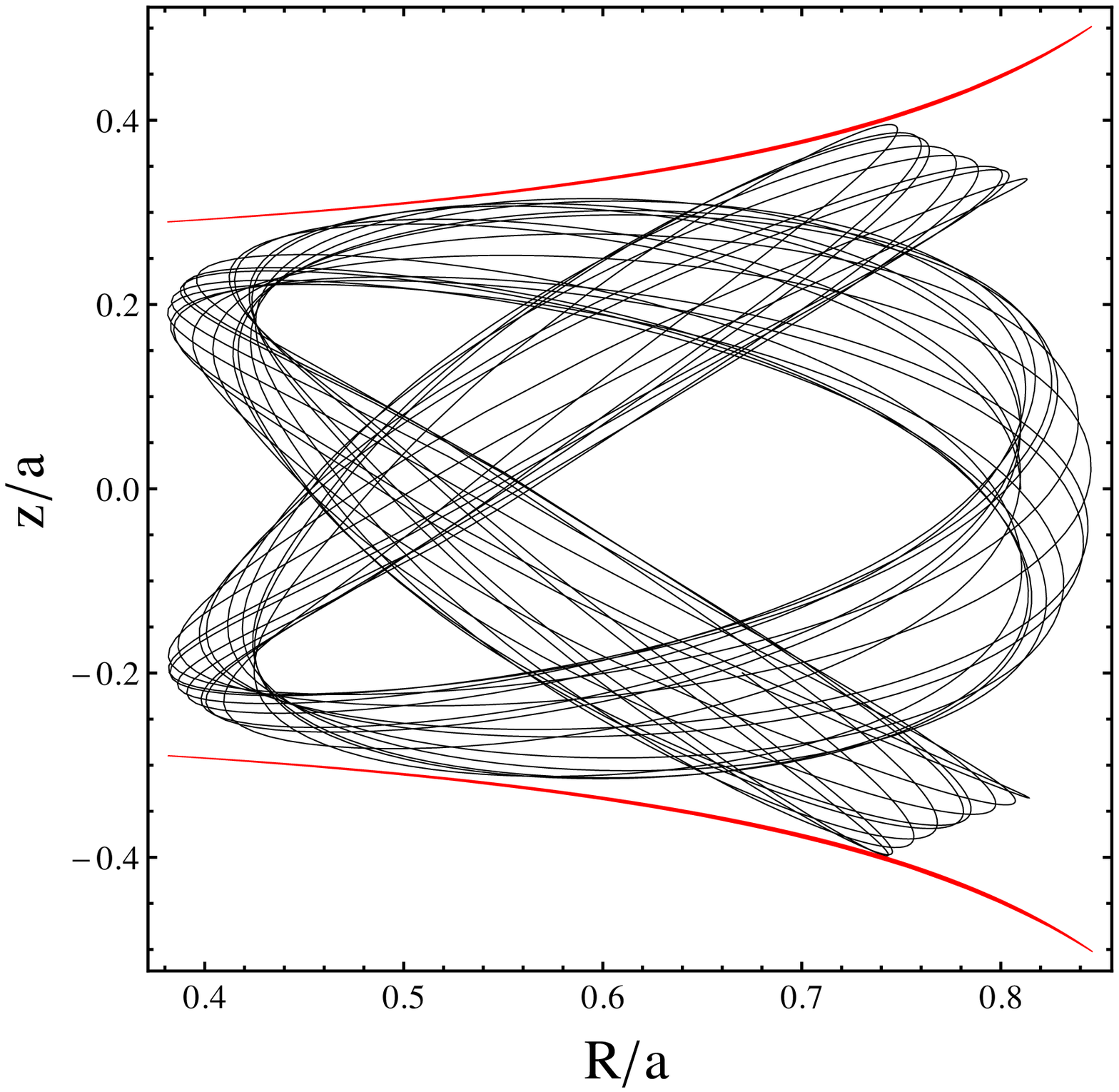}\\
  \epsfig{width=8cm,file=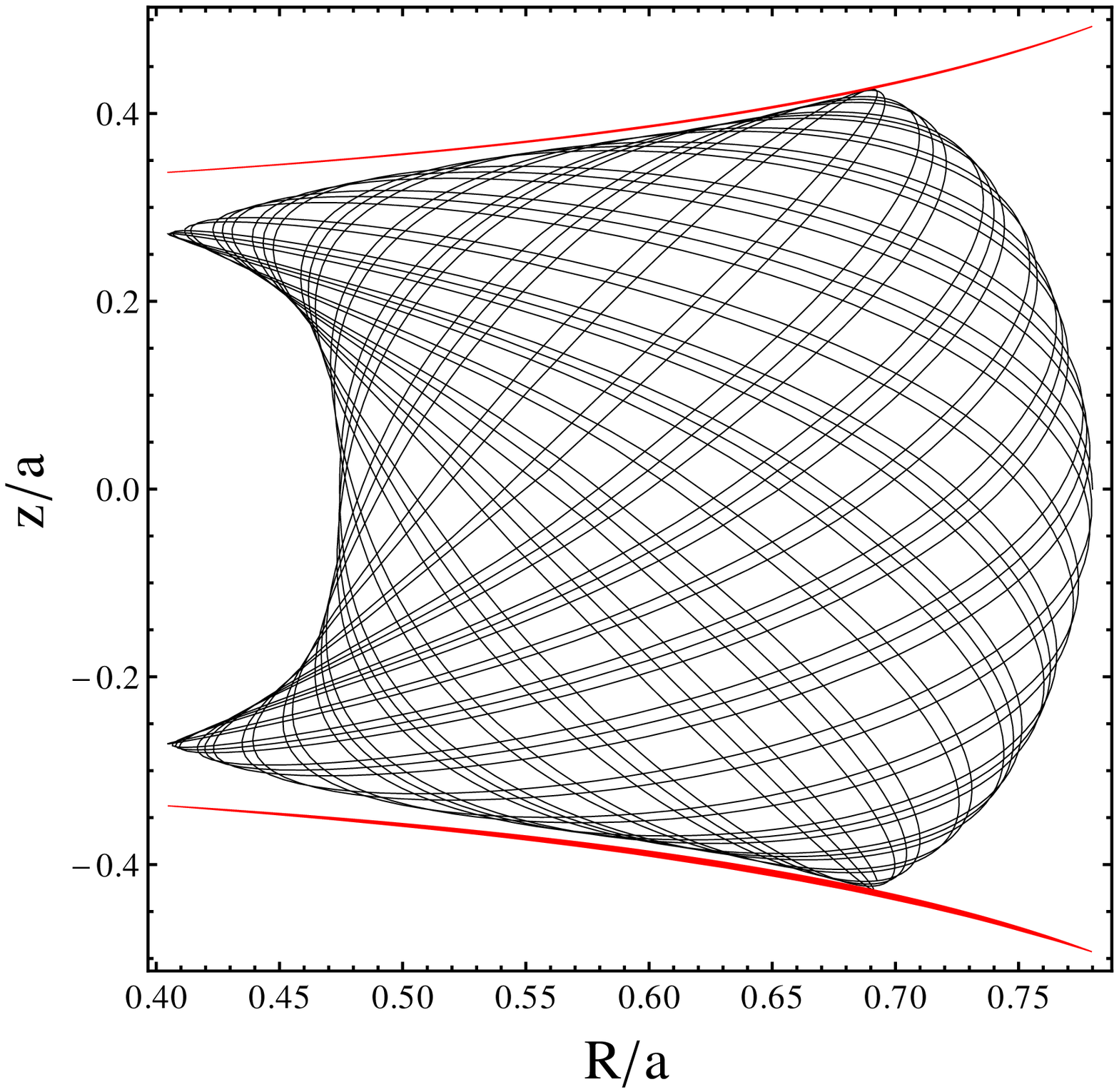} & \epsfig{width=8cm,file=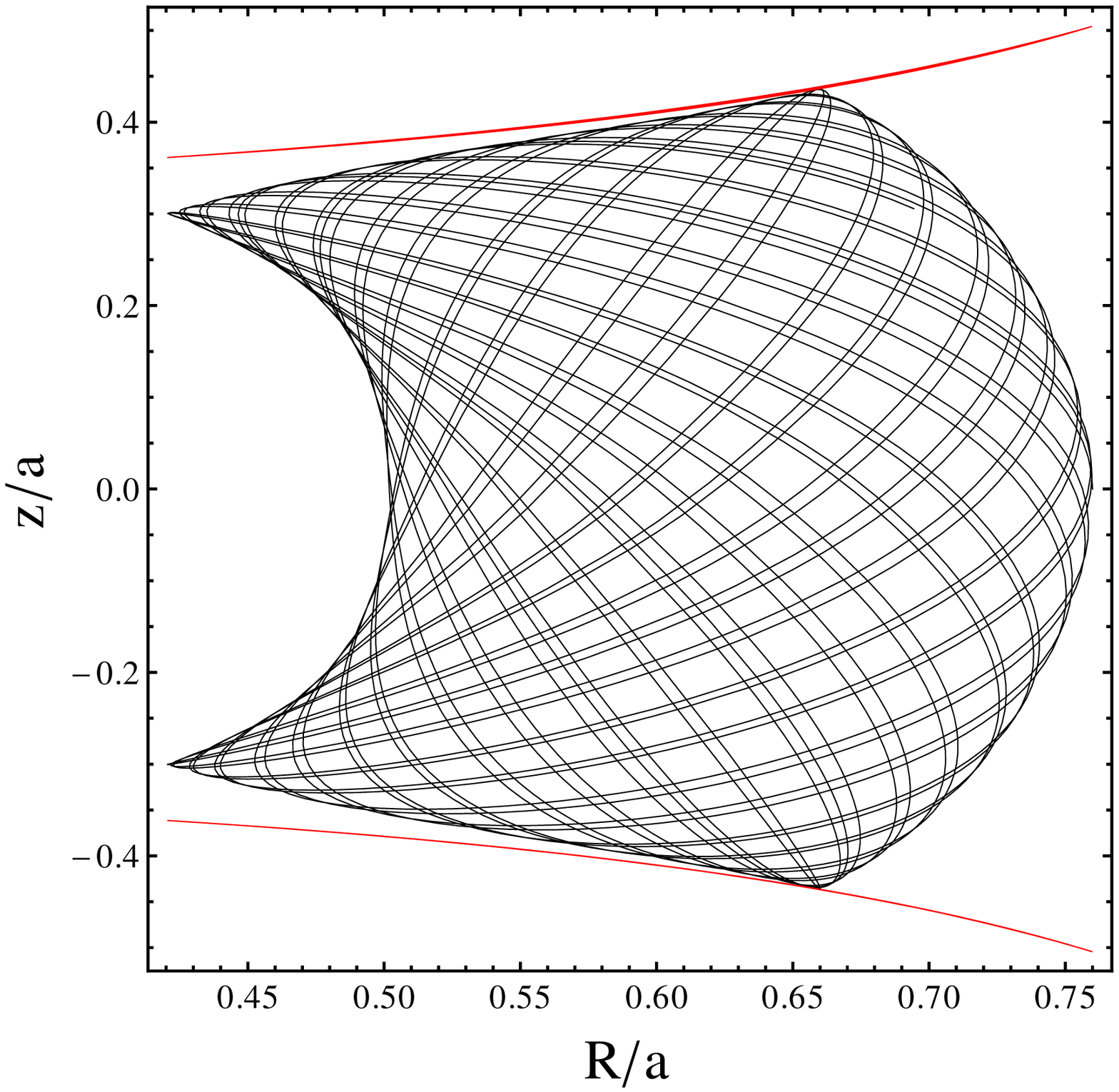}
\end{array}
$$
\caption{Orbits in the meridional plane  for the model $\Phi_{K2}+\Phi_{H}$ with $\ell/\sqrt{aGM}=0.8$, $V\sqrt{a/GM}=1.2$, $d/a=0.6$, $Ea/GM=-0.5$,
 and initial conditions $R/a=0.8$, $R/a=0.81$, $R/a=0.78$, $R/a=0.76$ ($z/a = 10^{-15}$ in all panels). Note the chaotic orbit which is bounded by
 the red line with a reasonably good precision.} \label{fig:orbitK2halo3}
\end{figure*}

\begin{figure}
\epsfig{width=7.5cm,file=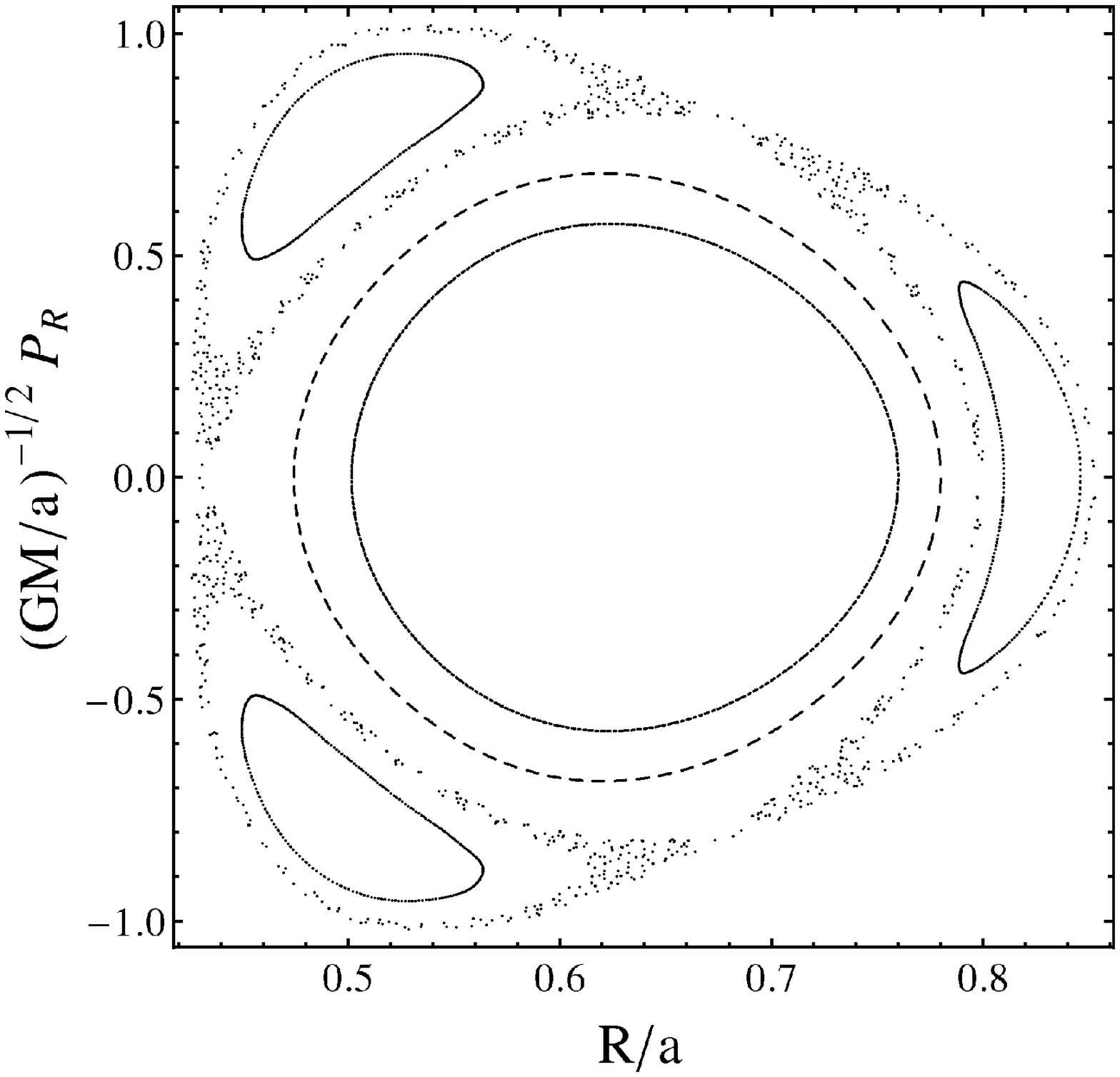}
\caption{Surface of section ($z=0$, $P_{z}>0$) formed by the consequents of the four orbits in fig. \ref{fig:orbitK2halo3}.} \label{fig:poincareK2halo3}
\end{figure}

In situation (ii) we note an interesting phenomenon. Here we find orbits which, in general, do not follow eq. (\ref{ZZ'sigma}) but there is a number of them
which pass near the red line where the particle reaches the maximum amplitude. This is the case illustrated in the top panels of
fig. \ref{fig:orbitK2halo3} where we present a chaotic orbit (left side) and a loop orbit (right side), which pass through
the critical point of the effective potential (near $R=0.5$) and also approach the red line. Note that the $z$-amplitude of the oscillations
is comparable with the $R$-amplitude. The bottom panels of this figure show two examples of regular orbits which do not obey the prediction of (\ref{ZZ'sigma}).
The chaoticity  or regularity of the aforementioned orbits can be visualized in the surface of section of fig. \ref{fig:poincareK2halo3}.

We point out that the accurateness of eq. (\ref{ZZ'sigma}) for numerically integrated orbits in the above examples (specially when dealing with
nearly equatorial orbits)
can also be seen as an indirect verification of the stability criterion
(\ref{condition}), since the derivation of eq. (\ref{ZZ'sigma}) depends on the formalism introduced to obtain condition (\ref{condition}).

We performed numerical integrations with the
Runge-Kutta method of fourth order with variable time step. Conservation of energy was checked with a
precision characterized by a maximum relative error of about $10^{-6}$. All the computations were performed in
the \textit{Laborat\'{o}rio de Computa\c{c}\~{a}o Paralela Patricio Letelier} at the  Instituto de Matem\'{a}tica,
Estat\'{i}stica e Computa\c{c}\~{a}o Cient\'{i}fica of UNICAMP.

\section{Stability criteria in Modified theories of gravity}\label{sec:MOND}

The results obtained for the vertical stability of equatorial circular orbits
(as well as the predictions about the amplitude  of nearly circular orbits) are valid
for Newtonian gravity, but there is no guarantee that this also happens in other theories.
However, most theories of modified gravity have the property that,
in the Newtonian limit,  test particles do follow a
Hamiltonian flow determined by a modified potential. This makes it  possible to formulate, in a similar fashion as in Newtonian gravity,
 a  criterion for the Liapunov stability of circular orbits and an adequate description of
the oscillations of nearly circular orbits (assuming adiabatic invariance of the approximate vertical action). In this section we
illustrate this fact by addressing two examples: MOND (\cite{bekenstein1984}) and RGGR (\cite{davi2010}).

MOND can be formulated as a potential theory, with a potential $\Psi$ satisfying the field equation (\cite{bekenstein1984})
  \begin{equation}
   \nabla\cdot\big[\mu(|\nabla\Psi|/a_o)\nabla\Psi\big] = 4\pi G \rho,
  \end{equation}
where $\mu(x)$ is the MOND interpolating function, $a_o$ is a characteristic acceleration scale and $\rho$ is the baryonic matter
density. For a thin disk of matter in the plane $z=0$ we obtain from Gauss's theorem
  \begin{equation}
   \frac{\partial\Psi}{\partial|z|}\bigg|_{z=0} = \frac{2\pi G \Sigma(R)}{\mu(|\nabla\Psi|/a_o)},
  \end{equation}
which enables us to conclude that the  condition for vertical stability of equatorial circular orbits is $\Sigma>0$, the same as in Newtonian gravity.
Moreover, it can be verified that the condition for radial stability is $\kappa^2>0$, where $\kappa^2$ is calculated from the MOND potential $\Psi$.

On the other hand, we find that the adiabatic invariance of the approximate vertical action leads to the relation
 $$
   \frac{Z(R)}{Z(\tilde R)} =
\left[\frac{\frac{\partial\Psi}{\partial|z|}(\tilde R,0)}{\frac{\partial\Psi}{\partial|z|}(R,0)}\right]^{1/3},
  $$
which in terms of the baryonic surface mass density of the thin disk reads
  \begin{equation}\label{ZMOND}
   \frac{Z(R)}{Z(\tilde R)} = \left(\frac{\mu_R\Sigma(\tilde R)}{\mu_{\tilde R}\Sigma(R)}\right)^{1/3},
  \end{equation}
where $\mu_R \equiv \mu(|\nabla\Psi(R,0)|/a_o)$.

For RGGR we have the modified potential (\cite{davi2010})
\begin{equation}
 \tilde\Phi = \Phi_N + \frac{c^2}{2G_0}G,
\end{equation}
where $\Phi_N$ is the Newtonian potential, $G_0$ is the ``standard'' gravitational constant and $G = G(\Phi_N)$ takes into account
renormalization group corrections. The phenomenologically adopted form of G gives us (see \cite{davi2010})
  \begin{equation}
   \nabla\tilde\Phi = \bigg[1- \frac{V^2_\infty}{\Phi_N}\bigg]\nabla\Phi_N,
  \end{equation}
where $V^2_\infty$ is the asymptotic circular velocity and $\Phi_N \to 0$ as $|\vec{x}|\to \infty$. Therefore we have
  \begin{equation}
   \frac{\partial \tilde\Phi}{\partial |z|}\bigg|_{z=0} = 2\pi G\bigg[1- \frac{V^2_\infty}{\Phi_N}\bigg]\Sigma(R),
  \end{equation}
and we see that the vertical stability condition for circular orbits is again $\Sigma> 0$, and the radial stability condition is also $\kappa^2> 0$
(computed with the potential $\tilde\Phi$). Adiabatic invariance of $J_z$ for nearly circular orbits gives us
(see eq. (\ref{ZMOND}))
  \begin{equation}\label{ZRGGR}
   \frac{Z(R)}{Z(R')} =
\left[\frac{\left(1- \frac{V^2_\infty}{\Phi_N(R',0)}\right)\Sigma(R')}{\left(1- \frac{V^2_\infty}{\Phi_N(R,0)}\right)\Sigma(R)}\right]^{1/3}.
  \end{equation}

We see that modified theories of gravity predict deviations from the Newtonian behavior of nearly equatorial orbits. For orbits
of astrophysical objects with a considerable radial amplitude these different predictions
could, in principle, be compared with astronomical data.

\section{Conclusions}

We analyzed the stability of equatorial circular orbits in
thin disks surrounded by smooth axisymmetric structures. The presence
of the thin disk does not allow us to proceed in the same way as in
the smooth case because of the delta-like singularity. In particular, the
vertical stability criterion for smooth potentials, $\nu^{2}>0$, is not applicable
anymore. We developed a consistent vertical stability criterion
for circular orbits, which together with the (unchanged) radial stability
criterion, ensures Liapunov stability of the corresponding circular
orbit. Based on this new formalism, we find that nearly
equatorial orbits have a third integral of motion,
which is given by eq. (\ref{ZZ'sigma}). This is supported by numerical simulations, which additionally reveal
that orbits with great vertical amplitude can be described approximately
by this integral (sec. \ref{sec:numerical}). It would be interesting to see if
this dependence on the surface density is present in more realistic models, not described by a razor-thin disk, but
incorporating a stellar distribution described by a thickened disk.

The introduction of the new stability criterion leads to the
conclusion that all of the thin disk models  presented in \cite{gonzalez-reina,ngc,pedraza}
are stable in a first approximation, contrary to the statements shown in such references. This fact urged us to
obtain additional models in sec. \ref{sec:FTDModels} in order to show that Hunter's method,
taking into account the stability criterion constructed here,
is a powerful tool to model the maximum disk of a number of flat galaxies.
Having tested the stability of orbits in this class of models, it would be
interesting to carry out more conclusive stability analyses based on
statistical mechanics considerations (i.e. perturbed solutions of Boltzmann equation,
Toomre's criterion, etc). We  also have to point out that the stability analyses performed in
references \cite{pn} and \cite{javier} need to be corrected.

We also briefly addressed the problem of the stability criterion in modified theories of gravity, such
as MOND and RGGR, in sec. \ref{sec:MOND} (the same problem, in the realm of general relativity theory, is being studied
and the results will be shown in a next paper).
 We point out  that whenever is possible establish a
Hamiltonian formulation of the motion, it is also possible to perform an analysis similar to
the presented here, in the Newtonian gravity realm. In particular, for the two examples studied here, we
find that the stability criteria are also given by the relations $\Sigma>0$ and $\kappa^{2}>0$ and that
the assumption of adiabatic invariance leads to eqs. (\ref{ZMOND}) and (\ref{ZRGGR}), introducing
deviations from the Newtonian relation (\ref{ZZ'sigma}). This fact can be used as an additional test
of these theories, once we have at disposal the required observational data.

\section*{Acknowledgements}
The work of  R.S.S.V. and J.R.-C. is supported by Funda\c{c}\~{a}o de Amparo \`{a} Pesquisa do
Estado de S\~{a}o Paulo (FAPESP), grants 2010/00487-9 and 2009/16304-3, respectively.
We acknowledge the usage of the HyperLeda database (http://leda.univ-lyon1.fr).

\appendix

\section{Proof of the vertical stability condition}

\subsection{The z-equation}

Consider a circular orbit of radius $R_0$ in the plane $z=0$, with $z$-component of angular momentum
given by $l$. We define a vertical perturbation
on this orbit as an instantaneous increase in the $z$-component of the velocity of the particle (at time $t=0$, say):
  \begin{equation}
   \vec{v}(0) \mapsto \vec{v}(0) + v_{0z}\hat{z}.
  \end{equation}
This ``shift'' in velocity does not change the value of $l$, but increases the energy of the orbit by an amount
$\frac{1}{2}(v_{0z})^2$.

We are interested in small vertical perturbations to the circular orbit. The term ``small'' will become clearer in the course of the proof.
The questions we want to answer are the following:

\begin{enumerate}

\item\textit{For sufficiently small $v_{0z}$, what are the conditions that make the particle come back to the plane of the disk
in a finite time?}

\item\textit{If the particle comes back, is the z-amplitude of motion compatible with the assumption of a small perturbation?
In this case, is the $R$-variation negligible?}

\item\textit{After coming back and crossing the disk, what happens to the particle? Does it stay near the original circular orbit for
$t \to \infty$ ?}

\end{enumerate}

\subsubsection{Does the particle come back to the disk?}

Without loss of generality, assume $v_{0z} > 0$. Then the particle goes to the region $z > 0$, and
$\ddot{z} = - \partial\Phi/\partial z$ implies
  \begin{equation}\label{Azdot}
   \dot{z}(t) = \dot{z}(0) - \int_0^t \frac{\partial \Phi}{\partial z}(R(t'), z(t'))dt',
  \end{equation}
while the equation for R gives
  \begin{equation}
   \dot{R}(t) - \dot{R}(0) = - \int_0^t \frac{\partial\Phi_{eff}}{\partial R} (R(t'), z(t')) dt'.
  \end{equation}
Since $\dot{R}(0) = 0$ and $\partial\Phi_{eff}/\partial R(R_0,0) = 0$, the continuity of
$\partial\Phi_{eff}/\partial R$ implies that
for sufficiently small $z(t')$ we will have $\dot{R}(t)$ negligible, and then the assumption of constant R
will be reasonable. This condition is satisfied for small enough $v_{0z}$, at least for a short time. Let us assume for now
$R\approx R_0$ and study its consequences. We will come back to this issue later on.
Equation (\ref{Azdot}) reduces to
  \begin{equation}\label{AzdotR0}
   \dot{z}(t) = v_{0z} - \int_0^t \frac{\partial \Phi}{\partial z}(R_0, z(t'))dt'.
  \end{equation}

Since $\partial \Phi/\partial |z|$ is continuous and $\partial \Phi/\partial |z|(R_o,0)>0$, take $a > 0$ such that
$\partial \Phi/\partial |z|(R_o,z) > 0$ for $z\in [-a,a]$. If we define
  \begin{equation}
   \alpha_{R_0} = \min_{z\in [-a,a]} \bigg\{\frac{\partial \Phi}{\partial |z|}(R_0,z) \bigg\},
  \end{equation}
we have that $\alpha_{R_0} > 0$ by compacity.

If $z(t') < a$ for $t'\in [0, t]$, it follows that $\ddot{z}(t') < 0$ (since $\ddot{z} = - \partial \Phi/\partial z
= - \partial \Phi/\partial |z|$), and the vertical velocity of the particle tends to decrease. Eq. (\ref{AzdotR0})
implies
  \begin{equation}\label{Ainequalityzdot}
   \dot{z}(t) \leq v_{0z} - \alpha_{R_0}t.
  \end{equation}
In particular, $\dot{z}(t) \leq v_{0z}$, which implies
  \begin{equation}\label{Ainequalityz}
   z(t) \leq v_{0z}t.
  \end{equation}
Then, defining
  \begin{equation}
   t_c \equiv \frac{a}{v_{0z}},
  \end{equation}
we have that $z(t) < a$ for $t < t_c$.
Integrating (\ref{Ainequalityzdot}), we also have the inequality
  \begin{equation}\label{Ainequalityz2}
   z(t) \leq v_{0z}t - \frac{\alpha_{R_0}}{2}t^2.
  \end{equation}
The first instant of time in which the right-hand side of eq. (\ref{Ainequalityz2}) is equal to $a$ is
  \begin{equation}
   t_c' = \frac{1}{\alpha_{R_0}}\bigg[v_{0z} - \sqrt{(v_{0z})^2 - 2 \alpha_{R_0} a} \bigg]
  \end{equation}
if the discriminant is positive. We also have $z(t) < a$ for $t < t_c'$.

Since $a$ and $\alpha_{R_0}$ depend only on the potential and not on the particular trajectory of the particle,
we can fix these quantities by imposing the condition that $a$ is sufficiently small to keep $\dot{R}$ small enough, in such a way that
the $R$-variation of the perturbed orbit can be neclected at least for a short time
(we will see below that this is guaranteed for long times by the radial stability
of the circular orbit). Doing this, we have from the definition of $t_c$ and from eq. (\ref{Ainequalityzdot}) that there is a
critical value $\overline{v_{0z}}$ such that if $v_{0z} < \overline{v_{0z}}$ then there will be a time $\tilde{t}\in [0, t_c]$
(depending on $v_{0z}$) such that $\dot{z}(\tilde t) < 0$. Since $\alpha_{R_0} > 0$ and $z(\tilde t) < a$ by construction,
the particle will have $\dot{z}(t) < 0$ for $t > \tilde t$, and since the upper estimate of $\dot{z}(t)$,
eq. (\ref{Ainequalityzdot}), decreases with $t$, $\dot{z}(t)$ will not approach zero for larger $t$, which implies the particle will
hit the disk in a finite time.

We can obtain from eq. (\ref{Ainequalityz2}) an upper estimate for the total time interval the particle stays in the $z > 0$
region. If we make the right hand side of eq. (\ref{Ainequalityz2}) equal to zero (which implies $z(t) \leq 0$), we find
  \begin{equation}\label{Adeltat}
   \Delta t = \frac{2 v_{0z}}{\alpha_{R_0}}.
  \end{equation}
We can also obtain an estimate for $\overline{v_{0z}}$: from eq. (\ref{Ainequalityz2}), a sufficient condition to have $z(t) < a$
during the whole oscillation is
  \begin{equation}
   v_{0z}t - \frac{\alpha_{R_0}}{2}t^2 < a.
  \end{equation}
This condition will be satisfied for all $t$ if, and only if, $v_{0z} < \sqrt{2 \alpha_{R_0} a}$. Thus, we have that a lower limit
for the critical value $\overline{v_{0z}}$ is
  \begin{equation}
   \overline{v_{0z}} \geq \sqrt{2 \alpha_{R_0} a}.
  \end{equation}
That is, if $v_{0z} < \sqrt{2 \alpha_{R_0} a}$ the particle will hit again the plane of the disk within the time interval
$\Delta t$ given by (\ref{Adeltat}) and the oscillation around the original circular orbit will have a vertical amplitude smaller
than $a$. In this way, we have also answered the second question: the amplitude of the oscillation is less than $a$ (given the
above conditions), where $a$ can be taken arbitrarily small.
In fact, we can obtain an estimate for this amplitude: given $v_{0z}$, it follows from eqs. (\ref{Ainequalityz2}) and
(\ref{Adeltat}) that the amplitude of the perturbation will be smaller than the value of the right-hand side of eq.
(\ref{Ainequalityz2}) evaluated at $\Delta t/2$:
  \begin{equation}\label{Azmax}
   z_{max} \leq \frac{(v_{0z})^2}{2 \alpha_{R_0}}.
  \end{equation}
The case $v_{0z} < 0$ is analogous, because of the $z$-symmetry of the system. Thus we have answered the first question:
For small enough $|v_{0z}|$, the condition for the particle come back to the plane of the disk in finite time is
$\Sigma(R_0) > 0$.

\subsubsection{Disk-crossing and asymptotic behavior}

Assume that the $R$-variation of the vertically perturbed orbit is negligible. This implies that the projection of the perturbed
orbit on the $z = 0$ plane is the original circular orbit, and then by
conservation of the mechanical energy, the particle will hit the disk with a velocity with vertical component $-v_{0z}$.

The equation of motion for $z$ implies a discontinuity in the particle's acceleration while crossing the disk, but its velocity
is continuous. This implies that the particle will have a velocity $-v_{0z}$ just after crossing the disk, and since the
system is symmetric with respect to the $z = 0$ plane the particle will strike again the disk (now from the
other side). Motion after crossing the disk will be analogous to the oscillation before crossing it, because of the $z$-symmetry
of the potential and the nature of the new initial conditions. Thus, it follows that for sufficiently small $|v_{0z}|$ the
vertically perturbed orbit will remain oscillating around the original circular orbit for $t\to\infty$,
with characteristic period given by
$2 \Delta t$ (see eq. (\ref{Adeltat})) and characteristic amplitude given by (\ref{Azmax}).

In this sense, the vertical stability condition $\Sigma(R_0) > 0$ derived in this section has, for thin disks represented
by density distributions
of the form (\ref{totaldensity}), the same status as the condition $\nu^2 > 0$ for smooth axisymmetric potentials (\cite{GD, ngc}).
Both conditions assume small perturbations and neglect the $R$-variation. We now analyze the effect of this variation and obtain
a stability condition under general small perturbations.

\subsection{Effects on the R-coordinate of the trajectory}

In order to be able to neglect the $R$-variation of the perturbed orbit, a sufficient condition is that the original circular
orbit is stable under small radial perturbations, as we shall see in the following. This translates into the inequality
  \begin{equation}
   \kappa^2 (R_0) > 0,
  \end{equation}
where $\kappa^2 (R_0) = \frac{\partial^2\Phi_{eff}}{\partial R^2}(R_0,0)$ is the quadratic epicyclic frequency of the perturbation
(\cite{GD, ngc}). Indeed, assuming $\Sigma(R_0) > 0$ and $\kappa^2 (R_0) > 0$ (with $R_0$ not on the border of the thin surface
distribution if this distribution is finite in size), we have
  \begin{eqnarray}
   \frac{\partial^2\Phi_{eff}}{\partial R^2}(R_0,0) &>& 0, \label{stab1}\\
   \frac{\partial\Phi_{eff}}{\partial |z|}(R_0,0) &>& 0, \label{stab2}
  \end{eqnarray}
which imply $(R_0, 0)$ is a (strict) local minimum of the potential $\Phi_{eff}(R, z)$: there is a neighborhood of $(R_0, 0)$
in which $\Phi_{eff}(R, z) > \Phi_{eff}(R_0, 0)$ if $(R,z)\neq (R_0,0)$. Thus, for small enough $v_{0z}$ (corresponding to
small enough $(E - \Phi_{eff}(R_0, 0))$), the orbit will oscillate around $(R_0,0)$ in the meridional plane with an amplitude
that can be made arbitrarily small.

It also follows from conditions (\ref{stab1}) and (\ref{stab2}) that, if we neglect changes in $l$ due to small radial perturbations,
the circular orbit will be Liapunov stable under
small perturbations in any direction of the meridional plane. In fact, Liapunov stability of $(R_0, 0)$ depends only on the
continuity of $\Phi_{eff}$ and not on its smoothness (see \cite{arnold}, chap. 5, p. 99). This result is the generalization to thin disks represented
by density distributions of the form (\ref{totaldensity}) of the general stability condition $\nu^2>0, \kappa^2>0$ for circular
orbits (\cite{GD, ngc, pn}).

Finally we note that, while the condition $\Sigma(R_0) > 0$ depends only on the thin disk, condition $\kappa^2 (R_0) > 0$ depends also
on the smooth 3D distribution, in such a way that this spatial distribution can affect the radial stability of the circular orbit
and, as a consequence, the behavior of vertically perturbed orbits for large enough time.

\end{document}